\documentclass{pas}

\usepackage{enumitem}
\usepackage{gensymb}
\usepackage{multirow}
\usepackage{amsmath}

\usepackage{graphicx}
\usepackage{float}

\usepackage[T1]{fontenc}

\begin{document}

\lefttitle{Publications of the Astronomical Society of Australia}
\righttitle{Jan K\'{a}ra et al.}

\jnlPage{1}{4}
\jnlDoiYr{2021}
\doival{10.1017/pasa.xxxx.xx}

\articletitt{Research Paper}

\title{A study of transients from ground-based surveys reveals new ultra-compact accreting white dwarf binaries}

\author{\gn{Jan} \sn{K\'{a}ra},$^{1}$ 
        \gn{Liliana} \sn{Rivera Sandoval},$^{1,2}$ 
        \gn{Wendy} \sn{Mendoza},$^{1}$
        \gn{Thomas J.} \sn{Maccarone},$^{3}$
        \gn{Manuel} \sn{Pichardo Marcano},$^{4}$
        \gn{Luis E.} \sn{Salazar Manzano},$^{5}$
        \gn{Ryan J.} \sn{Oelkers},$^{1}$ and 
        \gn{Jan} \sn{van Roestel}$^{6}$
        }

\affil{$^1$Department of Physics and Astronomy, University of Texas Rio Grande Valley, Brownsville, TX 78520, USA\\
        $^2$South Texas Space Science Institute, University of Texas Rio Grande Valley, Brownsville, TX 78520, USA\\
        $^3$Department of Physics \& Astronomy, Texas Tech University, Box 41051, Lubbock, TX, 79409-1051, USA\\
        $^4$Universidad Nacional Aut\'onoma de M\'exico, Instituto de Astronom\'ia, Ciudad Universitaria, 04510 Ciudad de M\'exico, Mexico\\
        $^5$Department of Astronomy, University of Michigan, Ann Arbor, MI 48109, USA\\
        $^6$Institute of Science and Technology Austria, Am Campus 1, 3400 Klosterneuburg, Austria
        }

\corresp{J. K\'{a}ra, Email: jan.kara@utrgv.edu}



\begin{abstract}
AM~CVn stars are ultra-compact semi-detached binaries consisting of a white dwarf primary and a hydrogen-depleted secondary. In this paper we present spectroscopic and photometric results of 15 transient sources pre-classified as AM~CVn candidates. Our analysis confirms 9 systems of the type AM~CVn, 3 hydrogen-rich cataclysmic variables (accreting white dwarfs with near-main-sequence stars for donors) and 3 systems that could be evolved cataclysmic variables. 
Eight of the AM~CVn stars are analysed spectroscopically for the first time, which increases the number of spectroscopically confirmed AM CVns by about $10\%$.
TESS data revealed the orbital period of the AM~CVn star ASASSN-20pv to be $P_\mathrm{orb}=27.282\,\mathrm{min}$, which helps to constrain the possible values of its mass ratio.
TESS also helped to determine the superhump periods of one AM~CVn star (ASASSN-19ct, $P_\mathrm{sh}=30.94\,\mathrm{min}$) and two cataclysmic variables we classify as WZ Sge stars ($P_\mathrm{sh}=90.77\,\mathrm{min}$ for ZTF18aaaasnn and $P_\mathrm{sh}=91.6\,\mathrm{min}$ for ASASSN-15na).
We identified very different abundances in the spectra of the AM~CVns binaries ASASSN-15kf and ASASSN-20pv (both $P_\mathrm{orb}\sim 27.5$ min), suggesting different type of donors.
Six of the studied AM~CVns are X-ray sources, which helped to determine their mass accretion rates.
Photometry shows that the duration of all the superoutbursts detected in the AM~CVns is consistent with expectations from the disc instability model. 
Finally, we provide refined criteria for the identification of new systems using all-sky surveys such as LSST.
\end{abstract}


\maketitle

\section{Introduction}

Accreting white dwarf binaries are semi-detached binary systems in which a white dwarf primary is accreting matter from a Roche-lobe-filling secondary star. These system can belong to two different types of object: cataclysmic variables (CVs) and AM Canum Venaticorum (AM~CVn) stars.

CVs \citep[e.g.][]{1995cvs..book.....W} have orbital periods typically between 80 minutes and 10 hours and their donors are hydrogen-rich red dwarfs. There are, however, also examples of CVs with longer orbital periods and sub-giant evolved stars as donors. The transferred matter forms an accretion disc around a white dwarf when no strong magnetic field is present. The accretion disc can undergo transitions between low and high-temperature states, which results in events called dwarf nova outbursts, during which the disc's brightness increases by several magnitudes. 
The dwarf nova outbursts typically display bi-modality in their duration which is most evident in the case of SU UMa type of CVs. This type of CVs exhibits normal dwarf nova outbursts and also longer and more energetic superoutbursts.
The magnetic CVs can be divided in polars (disc-less systems with accretion through the magnetic field lines) and intermediate polars,  where there is a disc formed near the donor star and the matter is later transferred to the WD through magnetic lines. Outbursts in intermediate polars have been observed but are not common \citep[e.g.][]{2017A&A...602A.102H}.

The evolutionary models for CVs with donors resembling a main sequence star at the onset of the accretion predict a period minimum $P_\mathrm{min}\simeq82\,\mathrm{min}$ which agrees with the observed population of such CVs \citep{2011ApJS..194...28K}, even though the evolutionary models predict more CVs around the period minimum than is observed \citep{2023MNRAS.525.3597I, 2024A&A...687A.305M}. There are also CVs with orbital period periods below the period minimum \citep[see for example][]{2012MNRAS.425.2548B, 2020MNRAS.496.1243G, 2025A&A...699A..81K, 2025A&A...700A.107G} which are thought to originate from systems with evolved donors at the onset of the mass accretion. While spectra of these systems show hydrogen lines, they exhibit also strong helium lines.

AM~CVns \citep[e.g.][]{2018A&A...620A.141R, 2025A&A...700A.107G} have orbital periods typically between 5 and 70 minutes and their donors are hydrogen-poor and helium rich white dwarfs or semi-degenerate stars. Their spectra show helium lines and are devoid of hydrogen lines. The only two systems classified as AM~CVns which show optical hydrogen lines are HM Cnc ($P_\mathrm{min}=5.35\,\mathrm{min}$) and 3XMM~J0510-6703 ($P_\mathrm{min}=23.6\,\mathrm{min}$). These two systems are considered to be direct progenitors of AM~CVns in a short-lived phase of accretion of the donor's thin hydrogen shell \citep{2025A&A...700A.107G}. HM Cnc is also one of two known AM CVns showing Lyman $\alpha$ absorption line \citep{2023MNRAS.518.5123M}, the other system is CP Eri \citep{2006ApJ...636L.125S}.

Similarly to CVs, AM~CVns can host an accretion discs around the white dwarfs which can exhibit outbursts of diverse behaviour caused by transitions between low and high-temperature states \citep{2021MNRAS.502.4953D}. Many AM~CVns show normal outbursts and superoutburst, similar to CVs of the type SU UMa. Superoutbursts of AM~CVns can also be followed by a series of normal outbursts called rebrightenings and a slow decline to the system's quiescence brightness, which are properties typical also for CVs of the type WZ Sge, a subtype of SU UMa CVs \citep{2015PASJ...67..108K}. Due to the compactness of AM~CVn systems which limits the size of their disc, the duration of superoutbursts and normal outbursts tends to be short and they show rapid changes of brightness. These properties have posed a challenge to the ground-based detection of AM~CVns. 
This was demonstrated by \cite{2021MNRAS.508.3275P} who also showed how continuous photometry from space-based observatories can be used to study the characteristics of AM~CVn outbursts in detail. 

The current sample of known AM~CVn contains about 100 systems \citep{2025A&A...700A.107G} which is only a small fraction when compared to the known CVs \citep{2001PASP..113..764D, 2006yCat.5123....0D, 2003A&A...404..301R, 2011yCat....102018R, 2020RNAAS...4..219J}. \cite{2025A&A...700A.107G} showed that currently known space density of AM CVns predicts about 50 of these systems within $500\,\mathrm{pc}$ while only about 30 systems are known in this region. 
AM~CVns are key laboratories for study of the accretion physics \citep[e.g.][]{2012A&A...544A..13K}, they are potential progenitors of sub-luminous supernovae .Ia \citep{2007ApJ...662L..95B}, and they are expected to be important sources of low frequency gravitational waves \citep[e.g.][]{2004MNRAS.349..181N, 2023MNRAS.525L..50S}. Hence, it is important to understand their evolution, characterise their mass transfer and mass accretion and the physical processes which govern their outbursts. For that, it is essential to first identify the largest number of AM~CVns and thus, create statistically significant sample which will explore full parameter space of these systems and probe their diverse behaviour. While binary systems can be classified as AM~CVn candidates based on photometric observations, spectroscopy is the only way to unambiguously determine whether the donor is a hydrogen-poor and helium-rich star and therefore spectroscopic studies are necessary for identification and/or confirmation of new AM~CVns, as was done, for example, by \cite{2014MNRAS.439.2848C, 2021AJ....162..113V, 2022MNRAS.512.5440V, 2025MNRAS.537.3078A}. 

Here we present a spectroscopic and photometric study of AM~CVn candidates with the aim of confirming these systems and thus increasing the population number. The available photometric observations obtained from various ground-based observatories and from the Transiting Exoplanet Survey Satellite (TESS) mission were used to characterise the outbursts of the targets and to search for periodic variations.

\section{Observations} 

\subsection{Target selection}

\begin{figure*}
    \centering
    \includegraphics[width=0.99\linewidth]{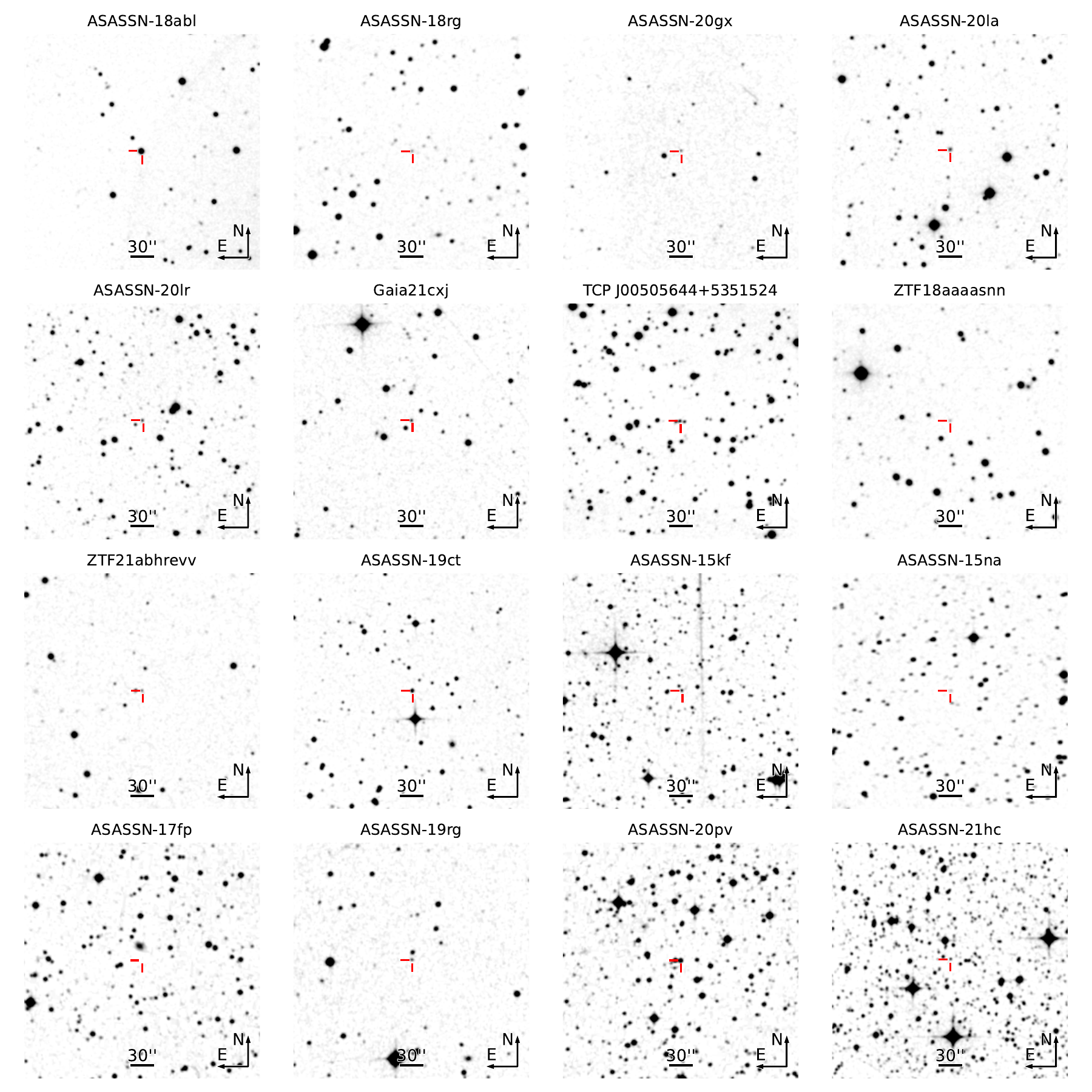}

    \caption{Finding charts for observed targets, the images are blue bands from the Digitized Sky Survey 2 \citep[DSS2,][]{1996ASPC..101...88L}.}
    \label{F:FindingCharts}
\end{figure*}

The sample of selected targets consists of transient sources identified with 
the All-Sky Automated Survey for Supernovae\footnote{\url{https://asas-sn.osu.edu/}} \citep[ASAS-SN;][]{2014ApJ...788...48S, 2017PASP..129j4502K}, 
the Zwicky Transient Facility (ZTF) survey\footnote{\url{https://www.ztf.caltech.edu/ztf-public-releases.html}}, the database of American Association of Variable Star Observers\footnote{\url{https://www.aavso.org}} \citep[AAVSO;][]{AAVSO:ONLINE} and the Variable Stars Network (VS-NET)\footnote{\url{http://www.kusastro.kyoto-u.ac.jp/vsnet/}}. They were classified as AM~CVn candidates based on the properties of their light curves such as short duration of outbursts and superoutbursts, rapid luminosity rise and decline, blue colour of the target, and brightness variations with short periodicity. Multiple targets were observed by amateur astronomers who managed to detect superhump variations during the superoutbursts, which allowed to constrain orbital periods of the targets, as superhump periods are typically a few percent longer than  orbital periods. These estimates were typically reported as an alert in the VSNET.

Figure~\ref{F:FindingCharts} shows the finding charts of the selected targets.
During the realisation of this work four of the targets (ASASSN-15kf, ASASSN-19ct, ASASSN-20pv, and ASASSN-21hc) were classified as confirmed AM~CVn system by \cite{2025A&A...700A.107G} based on the presence of outbursts and short orbital period, but no spectroscopic observations were presented prior to our study. We also included two AM~CVn systems which have been previously spectroscopically analysed. One of them is ASASSN-17fp, which was observed by \cite{2017ATel10334....1C} during its superoutburst. We included this target with the aim to obtain its spectrum during quiescence. The second target is a well-studied V744 And, also known as Gaia21cxj or SDSSJ0129+3842, \citep[e.g.][]{2012MNRAS.419.2836R, 2013MNRAS.432.2048K} which we included for comparison purposes.

\subsection{Ground-based photometric data}
For all targets we obtained available archival photometry from the ZTF, ASAS-SN, AAVSO and the Asteroid Terrestrial-impact Last Alert System (ATLAS) survey \citep{2018PASP..130f4505T}. 
We created light curves to look for signs of outburst activity. We used the Modified Julian Date reference system for all ground based datasets.\looseness=-10

For analysis of the targets and their comparison with other stars we also used photometry from the Sloan Digital Sky Survey \citep[SDSS;][]{2015ApJS..219...12A}, the Panoramic Survey Telescope and the Rapid Response System \citep[Pan-STARRS;][]{2016arXiv161205560C, 2020ApJS..251....7F}, and the SkyMapper Southern Sky Survey \citep{2024PASA...41...61O, 2024yCat.2379....0O}.

\subsection{Gemini observatory spectroscopy}

\begin{table*}
    \centering
    {\tablefont\caption{Log of Gemini observations listing observed targets, their type,  J2000 coordinates (RA, DEC), spectrograph used for observation, date of observation, exposure time of individual spectra $t_\mathrm{exp}$, and number of exposures $N_\mathrm{exp}$.}
    \label{T:LIST}
    \begin{tabular}{l r r r c l c r}
    \toprule
         Name           & Type        &  \multicolumn{1}{c}{RA} &  \multicolumn{1}{c}{DEC}  & Instrument & \multicolumn{1}{c}{Date} &  $t_{\mathrm{exp}}$ [s] & $N_{\mathrm{exp}}$ \\
         \hline
ASASSN-15kf            &           AM~CVn &      15:38:38.20 &     -30:35:50.21 &       GMOS-S &        2024, Apr 1  &          550  &            3 \\
ASASSN-15na            &               CV &      19:19:08.84 &     -49:45:41.00 &       GMOS-S &        2024, Apr 4  &         1000  &            4 \\
ASASSN-17fp            &           AM~CVn &      18:08:51.10 &     -73:04:04.20 &       GMOS-S &        2024, Apr 4  &          300  &            2 \\
ASASSN-18abl           &          Evolved CV candidate &      03:00:54.95 &      18:02:28.92 &       GMOS-N &        2023, Nov 4  &          100  &            3 \\
ASASSN-18rg            &           AM~CVn &      21:17:42.16 &     -02:22:28.52 &       GMOS-N &       2023, Jul 27  &          900  &            5 \\
ASASSN-19ct            &           AM~CVn &      11:33:15.36 &     -37:10:19.96 &       GMOS-S &        2024, Mar 7  &          150  &            2 \\
ASASSN-19rg            &           AM~CVn &      13:25:58.13 &     -14:52:26.31 &       GMOS-S &       2024, Mar 31  &          680  &            7 \\
ASASSN-20gx            &           AM~CVn &      23:49:30.17 &      22:01:29.64 &       GMOS-N &       2023, Nov 12  &          900  &            3 \\
ASASSN-20la            &          Evolved CV candidate &      01:38:51.95 &      46:34:48.90 &       GMOS-N &    2023, Nov 10-11  &          940  &           10 \\
ASASSN-20lr            &          AM~CVn  &      04:22:20.06 &      50:07:12.82 &       GMOS-N &       2023, Nov 12  &          600  &            3 \\
ASASSN-20pv            &           AM~CVn &      10:40:19.50 &     -49:51:29.70 &       GMOS-S &        2024, Mar 6  &          150  &            2 \\
ASASSN-21hc            &           AM~CVn &      16:05:25.27 &     -38:12:11.14 &       GMOS-S &        2024, Mar 2  &          850  &            2 \\
Gaia21cxj              &           AM~CVn &      01:29:40.06 &      38:42:10.45 &       GMOS-N &       2023, Nov 11  &         1100  &            3 \\
TCP J00505644+5351524  &          Evolved CV candidate &      00:50:56.44 &      53:51:51.90 &       GMOS-N &        2023, Nov 9  &          300  &            3 \\
ZTF18aaaasnn           &               CV &      06:11:06.13 &      57:37:34.00 &       GMOS-N &       2023, Nov 22  &          950  &           10 \\
ZTF21abhrevv           &               CV &      00:50:55.88 &      35:58:55.60 &       GMOS-N &       2023, Nov 12  &          900  &            3 
\botrule
    \end{tabular}}
\end{table*}

Spectroscopic data were obtained with the 8.1-m telescopes at the international Gemini Observatory located on Maunakea in Hawai'i and Cerro Pachón in Chile. The observations were taken under the programs GN-2023B-Q-310 and GS-2024A-Q-311 in years 2023 and 2024 (PI-Rivera Sandoval), a log of observations is presented in Table \ref{T:LIST}. All spectra were obtained with GMOS spectrograph \citep{2004PASP..116..425H, 2016SPIE.9908E..2SG} equipped with GG455 broad band filter and R400 grating ($R \sim 1900$) in case of target ASASSN-18abl and R150 low-resolution grating ($R \sim 600$) for the other systems. Targets were observed in $4\times1$ binning, except for ASASSN-18abl, which was observed in $2\times1$ binning. Observation of each targets consist of multiple exposures, which allowed us to use spectral dithering to minimise the effects of the gaps between individual CCDs on the final combined spectra whose typical signal-to-noise ratio is $\sim 20$. 
The spectra cover wavelengths between $4800\,\mathrm{\text{\AA}}$ and $10\,000\,\mathrm{\text{\AA}}$, apart from ASASSN-15na and ASASSN-18abl, for which the upper bound is about $9100\,\mathrm{\text{\AA}}$ due to low signal at longer wavelengths in case of the former target and observational configuration in case of the later target. 

All spectra were reduced using the DRAGONS data reduction software \citep{2023RNAAS...7..214L, DRAGONS}. All spectra were flux-calibrated but low flux at the ends of the wavelength ranges rendered the flux-calibration for some of the targets unreliable. The flux-calibrated spectra are presented in 
Figure~\ref{F:SPEC_ALL}
in the appendix.

\begin{table}
    \centering
    \caption{Summary of the four targets with available TESS photometry
    with detected superoutbursts (SOs), outbursts (OBs), periods and the uncertainties derived from half-widths at half-maximum (FWHM/2) of corresponding peak in Lomb-Scargle periodogram.}
    \label{tab:tesstargetslist_v2}
    {\tablefont\begin{tabular}{@{\extracolsep{\fill}} l l c c c c}
        \toprule
        Sector & Cadence & SOs & OBs & Period & FWHM/2\\
               &         &     &     & [min]  & [min]   \\
        \hline
        \multicolumn{6}{c}{ASASSN-19ct}\\
        \hline
        10 & 30 min & 1 & 18 &       & \\
        36 & 120 s & -  & - &       & \\
        37 & 120 s & 1 & 1  & 30.94$^a$ & 0.23510 \\
        63 & 120 s &  - & - &     &   \\
        90 & 120 s & - & -  &     &   \\
        \hline
        \multicolumn{6}{c}{ASASSN-20pv}\\
        \hline
        9  & 30 min & - & - &       & \\
        10 & 30 min & - & - &       & \\
        36 & 10 min & - & - &       & \\
        37 & 10 min & - & - &       & \\
        63 & 120 s  & - & - & 27.282$^b$ & 0.00952 \\
        90 & 200 s  & - & - & 27.282$^b$ & 0.00833 \\
        \hline
        \multicolumn{6}{c}{ZTF18aaaasnn}\\
        \hline
        19 & 30 min  &  - & - &       & \\
        20 & 30 min  &  - & - &       & \\
        60 & 200 s  &  - & - &       & \\
        73 & 120 s  & 1  & - & 91.21$^a$ & 0.20 \\
        \hline
        \multicolumn{6}{c}{ASASSN-15na}\\
        \hline
        13 & 30 min  & 1  & - & 91.87$^a$  & 0.64908 \\
        27 & 10 min  & -  & - &       & \\
        67 & 200 s  & -  & - &      &  \\
        94 & 200 s & - & - & - & 
        \botrule
    \end{tabular}}
    
    \begin{tabnote}
        $^{(a)}$ - superhump period; $^{(b)}$ - orbital period
    \end{tabnote}
\end{table}  

\subsection{TESS data}
In this work, we use Full-Frame Images (FFIs) and Target Pixel Files (TPFs) of TESS \citep{2015JATIS...1a4003R} to perform a high-cadence investigation of the outburst activity in our targets. TESS FFIs provide continuous photometric monitoring across multiple observational sectors in various cadence modes (30 minutes, 10 minutes, and 200 seconds) and TPFs provide cutouts of FFIs for selected targets and their light curves with cadences of 2 minutes and in some cases also 20 seconds.
Both FFIs and TPFs can be used for detailed light curves  analysis \citep{2015JATIS...1a4003R}. The observational data were queried and downloaded using the Python package \textit{Lightkurve} \citep{2018ascl.soft12013L}, light curves extracted from FFIs were obtained using aperture photometry method. To ensure high quality of the data, we excluded all measurements with the TESS quality flag $q > 0$ to remove instrumental and observational artefacts. Times of observations were converted from Barycentric TESS Julian Date (BTJD) to Modified Julian Date (MJD) using the \textit{astropy.time} package, the flux measurements are given in electrons per second (e/s) \citep{astropy:2022}. 

Table~\ref{tab:tesstargetslist_v2} gives the overview of TESS data available for targets from our study.

\begin{figure*}
    \centering
    \includegraphics[width=0.89\linewidth]{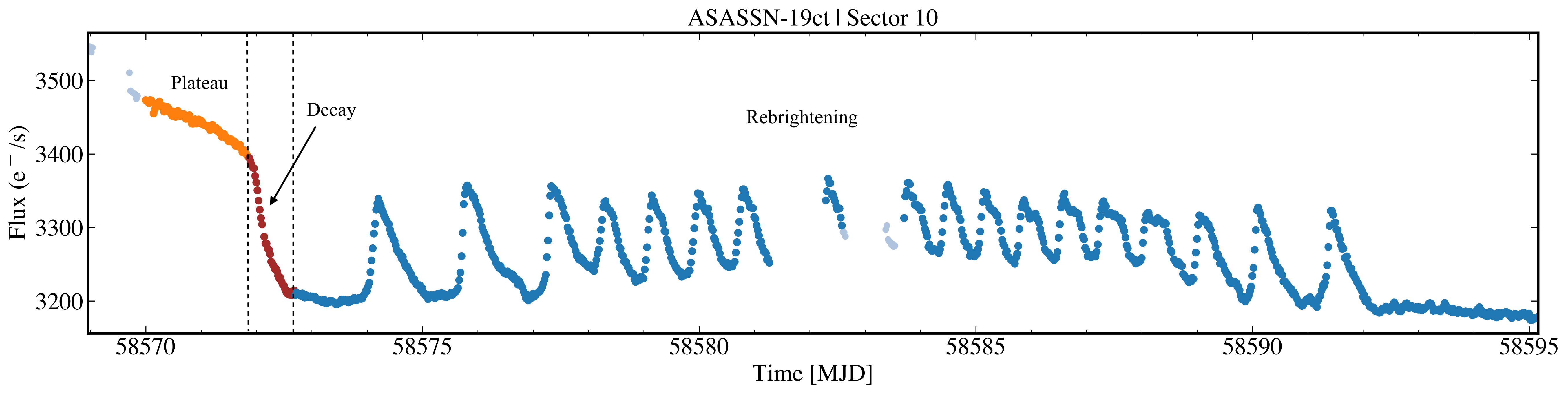}
    \includegraphics[width=0.89\linewidth]{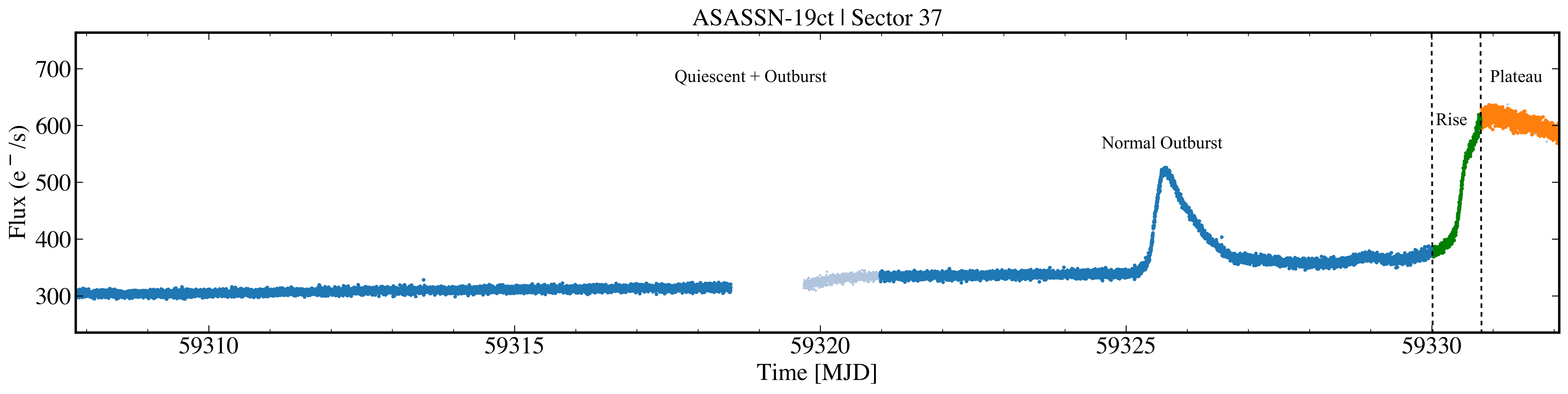}
    \includegraphics[width=0.89\linewidth]{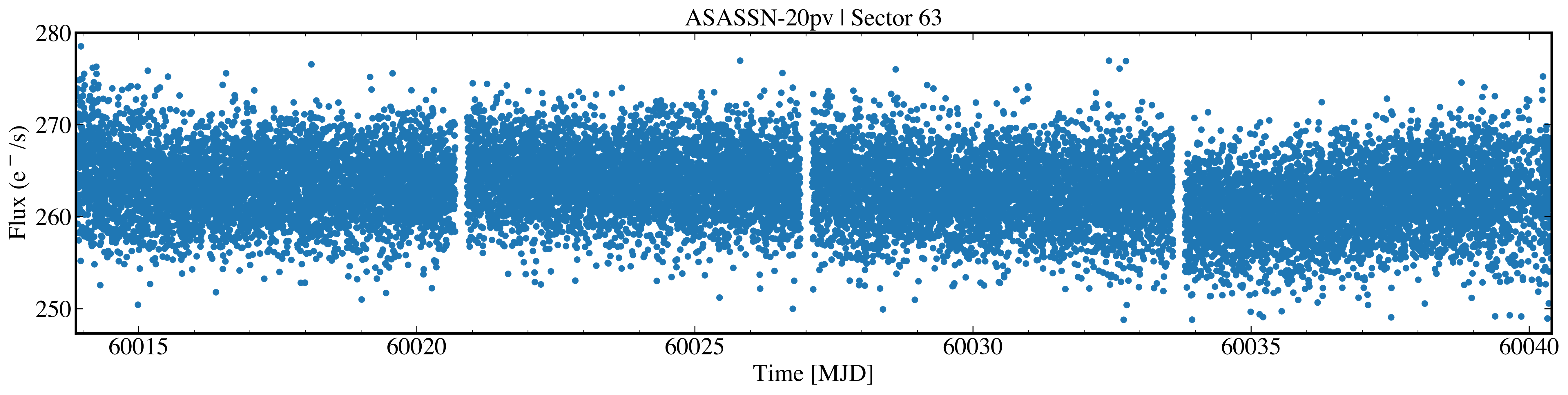}
    \includegraphics[width=0.89\linewidth]{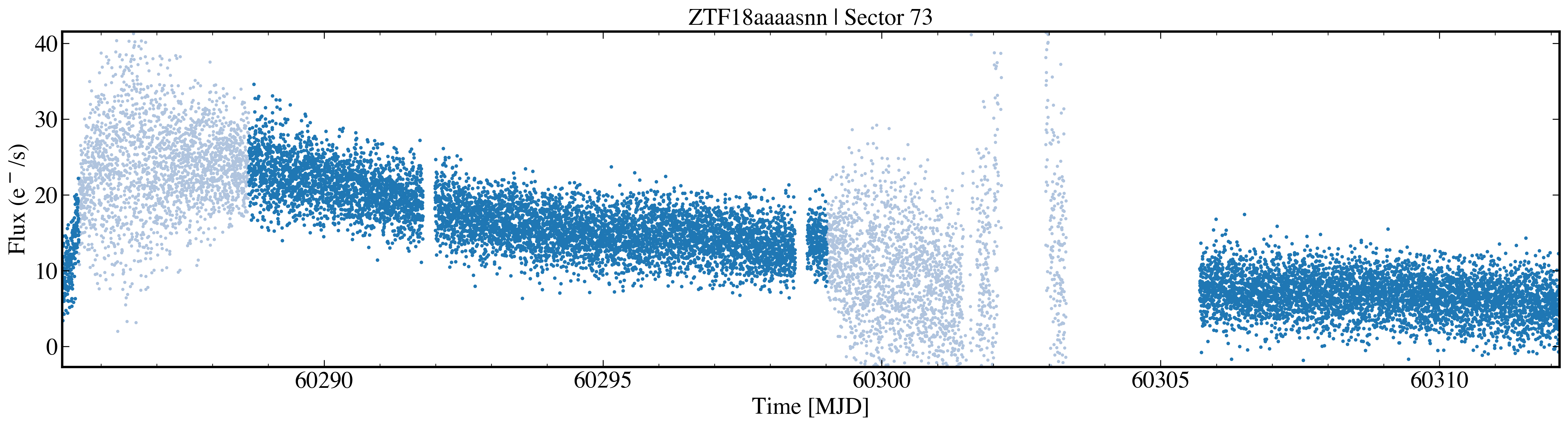}
    \includegraphics[width=0.89\linewidth]{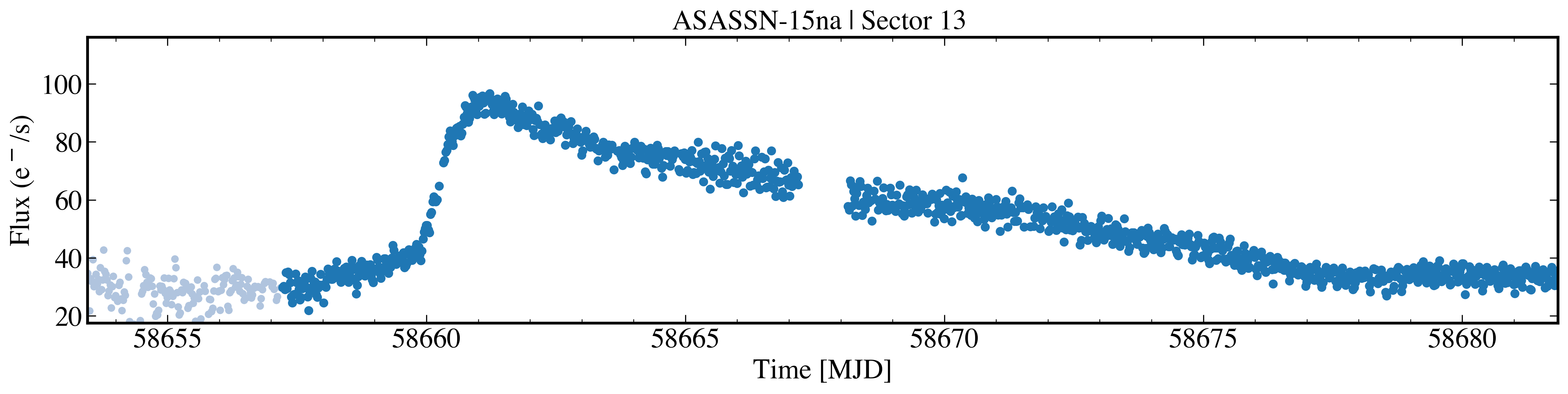}

    \caption{TESS light curves of four targets showing various types of outburst behaviour and quiescent states. Different superoutbursts phases of ASASSN-19ct are highlighted by green, orange, and brown colours and correspondingly labelled, 
    measurements with low-quality flags are shown in light blue colour.}
    \label{TESS:LC}
\end{figure*}

\begin{figure*}
    \centering
    \includegraphics[width=0.99\linewidth]{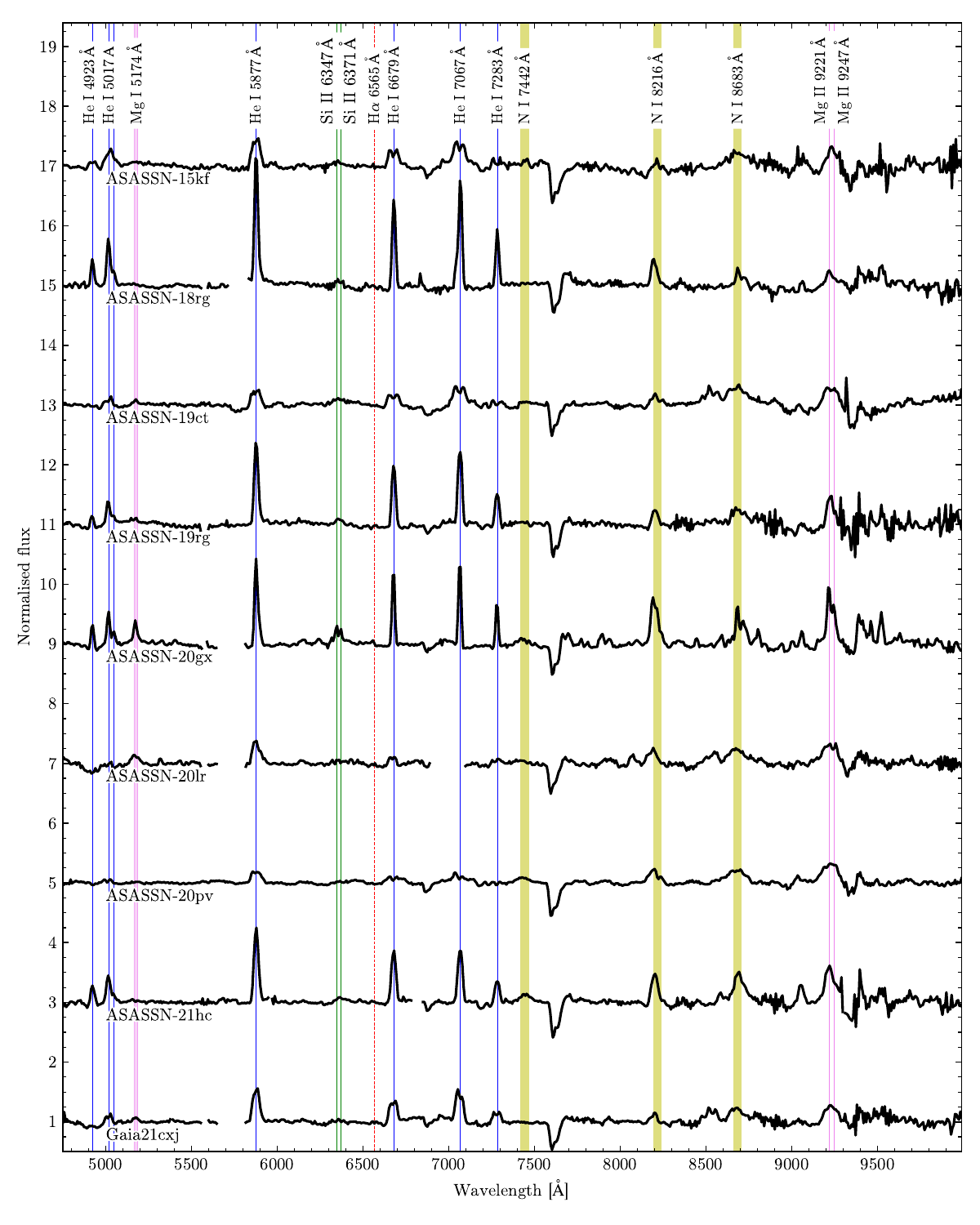}

    \caption{normalised spectra of AM~CVn stars. Positions of prominent spectral lines are marked by vertical lines, blends of multiple lines are marked by thick lines. The dashed red vertical line shows the potential position of H$\alpha$ line which is absent in all of the presented spectra. }
    \label{F:SPEC:NORM:AM}
\end{figure*}

\begin{figure*}
    \centering
    \includegraphics[width=0.99\linewidth]{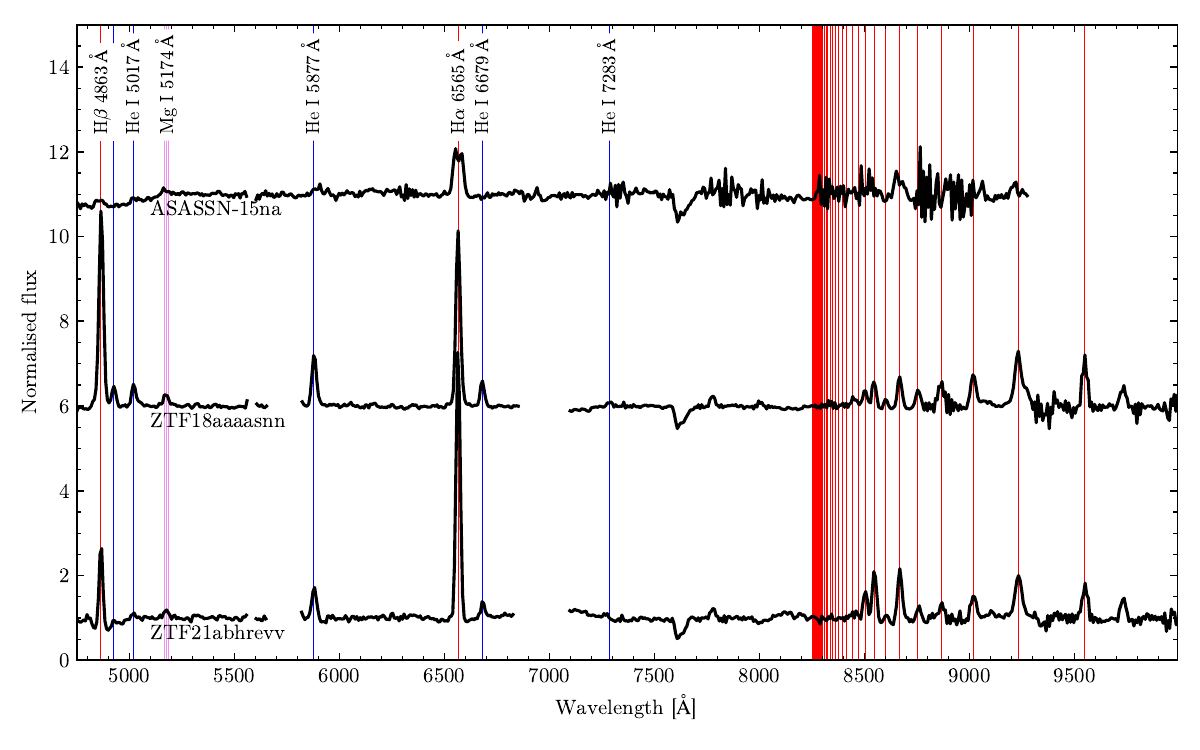}

    \includegraphics[width=0.99\linewidth]{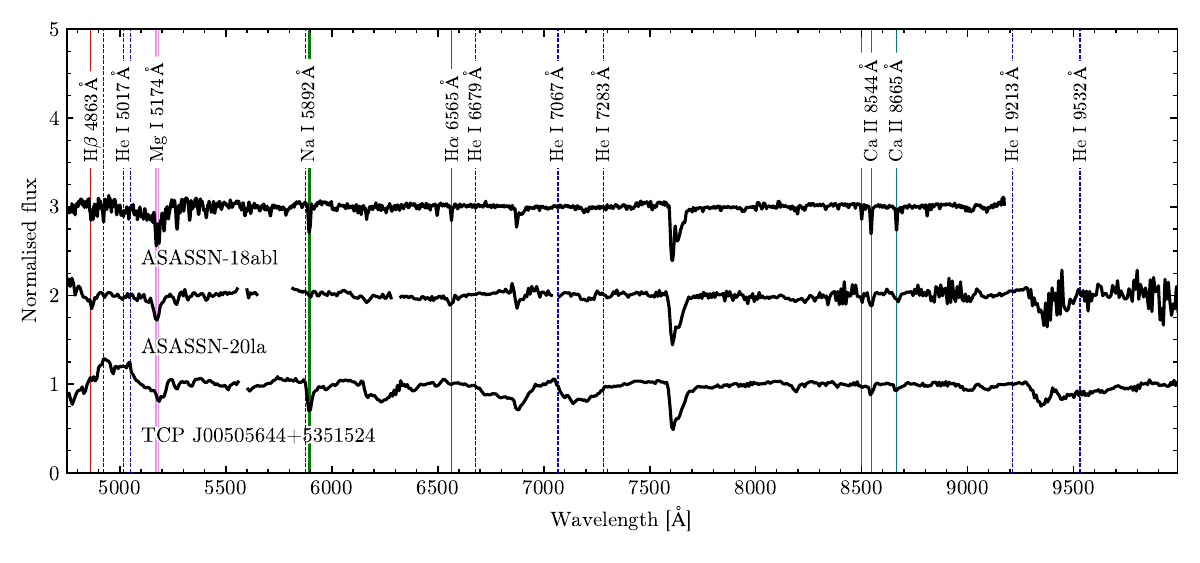}

    \caption{Top: normalised spectra of CVs. Positions of prominent spectral lines are marked by vertical lines. All presented spectra show hydrogen emission line from the Balmer series, ZTF18aaaasnn and ZTF21abhrevv show also hydrogen emission lines from the Paschen series. Wavelengths of Paschen lines are marked by vertical red lines. H$\alpha$ line of ASASSN-15na shows a clear double-peaked profile.
    Bottom: normalised spectra of evolved CV candidates. Positions of prominent spectral lines are marked by vertical lines, position of potential helium lines are marked by dashed blue lines. None of the presented spectra show emission lines typical for accretions discs.}
    \label{F:SPEC:NORM:CV_OTHER}
\end{figure*}

\section{Analysis}

\subsection{Blackbody fitting}
Given the low resolution of our spectra and relatively low signal-to-noise, 
we fitted the continuum of the flux-calibrated spectra with a blackbody model to characterise its slope instead of a more complicated model. As the flux calibration was problematic in some parts of the spectra, especially near the blue end of the observed range, we excluded these wavelength ranges for the fitting. 
Before fitting, we corrected the spectra for reddening. We used the three-dimensional map of dust reddening created by \cite{2019ApJ...887...93G} as the primary source, which allowed us to estimate the reddening correction according to the target's position on sky as well as its distance. As this map covers only sky north of declination of $-30\degree$, four of our targets are not included in its region. Therefore, for ASASSN-20pv and ASASSN-21hc we used reddening provided by the three-dimensional map created by \cite{2025arXiv250302657Z} which covers the southern Galactic plane. In cases of ASASN-15na and ASASSN-19ct, which are not located within the region of this map, we used full Galactic reddening from \cite{1998ApJ...500..525S}.

The temperatures corresponding to the best blackbody fit are listed in Table~\ref{tab:Params}. The optical flux in AM~CVn systems consists of contributions from different components, namely the primary star, the donor, the accretion disc, and the bright spot. Therefore, while the temperature derived by fitting the spectrum with a single blackbody spectrum can reflect the temperature of the WD primary to an extent, its value can be affected by radiation from the other components.

We also tested fitting a blackbody spectrum to the spectral energy distribution (SED) of the targets using the web tool VOSA \citep{2008A&A...492..277B} with which we obtained similar values of blackbody temperature. 

\subsection{Widths and peak separations of selected emission lines}

\begin{table*}[t]
    \centering
    \caption{FWHM for selected He I lines of AM~CVn systems}
    \label{tab:FWHM}
    {\tablefont\begin{tabular}{@{\extracolsep{\fill}}lcccccc}
    \toprule
     Target               & \multicolumn{2}{c}{He I 5876 \AA}        &         \multicolumn{2}{c}{He I 6678 \AA}         &         \multicolumn{2}{c}{He I 7065 \AA} \\
                     &      [\AA] &                    [km/s] &      [\AA] &                    [km/s] &      [\AA] &                    [km/s]\\ 
\hline
ASASSN-15kf          & $60.3\pm 6.6$    &    $3076.7\pm336.8$    &    $ 87.0\pm33.4$    &    $3905.6\pm1499.4$    &    $113.0\pm8.4$    &    $4794.9\pm356.4$ \\
ASASSN-18rg          & $22.9\pm 0.6$    &    $1168.4\pm 30.6$    &    $ 23.4\pm 0.4$    &    $1050.5\pm  18.0$    &    $25.0\pm1.7$    &    $1060.8\pm72.1$ \\
ASASSN-19ct          & $78.2\pm15.4$    &    $3990.0\pm785.8$    &    $ 97.3\pm28.8$    &    $4367.9\pm1292.9$    &    $108.3\pm9.5$    &    $4595.4\pm403.1$ \\
ASASSN-19rg          & $26.9\pm 0.9$    &    $1372.5\pm 45.9$    &    $ 26.1\pm 0.6$    &    $1171.7\pm  26.9$    &    $27.0\pm1.2$    &    $1145.7\pm50.9$ \\
ASASSN-20gx          & $23.8\pm 1.6$    &    $1214.4\pm 81.6$    &    $ 17.7\pm 0.3$    &    $794.6\pm  13.5$    &    $17.9\pm0.8$    &    $759.5\pm33.9$ \\
ASASSN-20lr          & $52.3\pm 8.1$    &    $2668.5\pm413.3$    &    $ 48.7\pm 8.2$    &    $2186.2\pm 368.1$    &                    &                 \\
ASASSN-20pv          & $78.0\pm13.5$    &    $3979.8\pm688.8$    &    $115.9\pm11.5$    &    $5202.9\pm 516.3$    &                    &                 \\
ASASSN-21hc          & $30.2\pm 0.9$    &    $1540.9\pm 45.9$    &    $ 29.8\pm 1.4$    &    $1337.8\pm  62.8$    &    $30.8\pm0.9$    &    $1306.9\pm38.2$ \\
Gaia21cxj            & $46.1\pm 2.6$    &    $2352.2\pm132.7$    &    $ 56.2\pm 7.8$    &    $2522.9\pm 350.2$    &    $51.0\pm5.3$    &    $2164.0\pm224.9$ 
    \botrule \\
    \end{tabular}}
\end{table*}

\begin{table*}[t]
    \centering
        \caption{Separation of the double-peaked profiles for selected He I lines of AM~CVn systems}
    \label{tab:2PS}
    {\tablefont\begin{tabular}{@{\extracolsep{\fill}}lcccccc}
    \toprule
     Target               & \multicolumn{2}{c}{He I 5876 \AA}        &         \multicolumn{2}{c}{He I 6678 \AA}        &         \multicolumn{2}{c}{He I 7065 \AA} \\
                     &      [\AA] &                    [km/s] &      [\AA] &                    [km/s] &      [\AA] &                    [km/s]\\
\hline
ASASSN-15kf          & $29.8\pm1.5$    &    $1520.5\pm76.5$    &    $38.3\pm1.2$    &    $1719.3\pm53.9$    &    $40.5\pm1.0$    &    $1718.5\pm42.4$ \\
ASASSN-19ct          & $35.2\pm1.6$    &    $1796.0\pm81.6$    &    $42.8\pm1.1$    &    $1921.4\pm49.4$    &    $44.9\pm1.8$    &    $1905.2\pm76.4$ \\
ASASSN-20lr          &                   &                      &    $26.5\pm1.9$    &    $1189.6\pm85.3$    &                      &                   \\
ASASSN-20pv          & $35.6\pm1.0$    &    $1816.4\pm51.0$    &                      &                      &                      &                   \\
Gaia21cxj            &                   &                      &    $28.0\pm1.5$    &    $1257.0\pm67.3$    &    $27.0\pm1.0$    &    $1145.7\pm44.1$ 
    \botrule
    \end{tabular}}
\end{table*}

The profiles of emission lines originating in the accretion disc are related to the disc velocities and inclination of the system. Usually, they may appear double-peaked, which is a typical profile for a disc viewed at an angle, and their separation can be used to infer the projection of velocity at the outer edge of the disc \citep{1981AcA....31..395S, 2016ApJ...822...99C}.

We measured full widths at half maximum (FWHMs) for selected He I emission lines in normalised spectra of 
our candidates. To determine FWHM of the lines, we fitted each line with a Gaussian profile using a Python package {\it SciPy} \citep{2020SciPy-NMeth}, the values of FWHM of the best fits are listed in Table~\ref{tab:FWHM}.

For those spectra that show double-peaked emission lines we measured separation of the peaks, we used the same set of He I lines as in the case of FWHM measurements. To determine the separation, we fitted each line with a model consisting of two identical Gaussian profiles which were shifted from the central wavelength by a value corresponding to half of the separation. Results of the best fits are given in Table~\ref{tab:2PS}.

\subsection{Analysis of TESS photometry}

For selected sources exhibiting apparent outburst activity, we visually isolated the different phases of the outburst to understand their periodic behaviour. In our periodicity analysis, we used the Lomb-Scargle technique \citep{1976Ap&SS..39..447L, 1982ApJ...263..835S} implemented by \textit{Astropy} Python packages. This method allows us to find the periodic signal and generate phase-folded light curves using the best-fit period. To evaluate the significance of the periodic signal, we used the False Alarm Probability (FAP) computed by the \textit{astropy module}. We calculated the threshold levels of 10\%, 0.1\%, and 0.001\% of FAP using the bootstrap method in  and included them in our periodograms to distinguish the real signal from noise. 
We fitted the peaks in periodograms with a Gaussian function to determine their positions, which we used as the best period. We estimated the uncertainties as half-widths at half maximum of the Gaussian functions.

\section{Results}

Nine of the observed targets exhibit spectral features typical for AM~CVns, their normalised spectra are shown in Figure~\ref{F:SPEC:NORM:AM}. All of these spectra show strong He I emission lines and no detectable hydrogen lines. Four of these targets exhibit single-peaked emission line profiles and five of them show double-peaked profiles. The spectra also exhibit blends of emission associated with other elements, such as Mg II or N I.

Three targets show hydrogen emission lines which classifies them as CVs, their normalised spectra are shown in Figure~\ref{F:SPEC:NORM:CV_OTHER}. Three targets show spectra that could be consistent with evolved cataclysmic variables, their normalised spectra are shown in Figure~\ref{F:SPEC:NORM:CV_OTHER}.

As all of the targets were previously reported as transients, their long-term light curves show at least one instance of outburst activity. Table~\ref{tab:SO_prop} lists superoutburst properties of the studied systems which we derived from the long-term light curves presented in Figure~\ref{F:LC:LT_ALL}.

Our analysis of TESS light curves identified outburst activity or periodic signals in four targets, each observed across multiple TESS sectors. Figure~\ref{TESS:LC} presents the light curves of these targets, the identified periods are listed in Table~\ref{tab:tesstargetslist_v2}. Periodic signals of three targets was identified during superoutbursts and were caused by superhumps. The periodicity of ASASSN-20pv was detected during quiescence and we interpret it as orbital period of this system.

Detailed descriptions of our results for each target is presented in Section~\ref{AP:TARGETS} of the Appendix.

Seven of our targets lie in the region covered by eROSITA catalogue of X-ray sources \citep{2024A&A...682A..34M}, five of which have X-ray positions located within $6"$ of their Gaia coordinates. The two systems without an X-ray detection are ASASSN-15na and ASASSN-17fp, which belong to the faintest systems in our sample. 
Observed X-ray fluxes for selected eROSITA energy bands are listed in Table~\ref{tab:XRay}, which also lists X-ray fluxes of Gaia21cxj listed in the Chandra Source catalogue \citep{2024ApJS..274...22E}. All eROSITA observations were obtained during quiescence phases of the systems. Gaia21cxj lacks photometric observations coinciding with the Chandra observations, therefore the activity phase during the observations can't be determined. However, considering the long recurrence times of its outbursts, they were likely obtained during a quiescence phase as well. 
Seven of our targets have UV photometry in the GALEX catalogue \citep{2017ApJS..230...24B},
which provides photometry in two bands: far-UV (FUV, $\lambda_\mathrm{eff} \sim 1528\,\text{\AA}$, 1344–1786 \AA) and near-UV (NUV, $\lambda_\mathrm{eff} \sim 2310\,\text{\AA}$, 1771–2831 \AA).
The fluxes derived from the photometry given in the catalogue are listed in Table~\ref{tab:UV}. 

\begin{table*}
    \caption{Properties of the studied systems}\label{tab:Params}
    \begin{center}{\tablefont\begin{tabular}{lccccccc}
    \toprule
     Name              &             Type  &     Period [min] & $T_{\mathrm{BB}}$ [K]   &          G [mag] &      BP-RP [mag] &                Parallax       & Distance [pc]   \\
\hline
ASASSN-15kf            &           AM~CVn &             $27.7$(sh)$^{b}$ & $     10\,324 \pm   86$  &             $19.4$ &             $0.18$ & $            1.91\pm 0.42$ & $  647_{-165}^{+265}$ \\
ASASSN-15na            &               CV &             $91.9$(sh)$^{c}$ & $      10\,304 \pm  297$  &                  &                  &                                &   \\
ASASSN-17fp            &           AM~CVn &             $51.0$(sh)$^{b}$   &         &                  &                  &                              &  \\
ASASSN-18abl           &          Evolved CV candidate &               $87.2$(sh) & $      5039 \pm   11$  &            $14.73$ &             $1.24$ & $            2.17\pm 0.03$   &        $  448_{-6}^{+6}$   \\
ASASSN-18rg            &           AM~CVn &               $46.0$(sh)$^{b}$ & $     10\,085 \pm  177$  &            $20.51$ &             $0.07$ & $            4.18\pm 1.31$ &   $  322_{-97}^{+337}$   \\
ASASSN-19ct            &           AM~CVn &               $30.9$(sh)$^{c}$ & $     14\,977 \pm   167$  &            $17.44$ &             $0.01$ & $            4.20\pm 0.08$ &   $  238_{-5}^{+5}$   \\
ASASSN-19rg            &           AM~CVn &             $43.9$(sh)$^{b}$  & $     13\,456 \pm  212$  &            $20.38$ &             $0.04$ & $            1.11\pm 0.63$&     $ 1160_{-374}^{+518}$   \\
ASASSN-20gx            &           AM~CVn &                  & $      8356 \pm  103$  &            $20.19$ &             $0.49$ & $            1.40\pm 0.52$   &              $  840_{-254}^{+393}$   \\
ASASSN-20la            &          Evolved CV candidate &                  & $      4999 \pm   34$  &                  &                  &                                &  \\
ASASSN-20lr            &           AM~CVn &                  & $      9824 \pm  133$  &            $19.66$ &             $0.46$ & $            2.19\pm 0.40$   &              $  514_{-87}^{+140}$  \\
ASASSN-20pv            &           AM~CVn &            $27.28$$^{c}$ & $     11\,884 \pm  143$  &            $17.11$ &             $0.22$ & $            3.78\pm 0.05$ &              $  262_{-3}^{+3}$  \\
ASASSN-21hc            &           AM~CVn &            $35.8$(sh)$^{b}$  & $      11\,606 \pm   130$  &            $19.45$ &             $0.35$ & $            2.37\pm 0.47$   &     $  458_{-97}^{+122}$  \\
Gaia21cxj              &           AM~CVn &             $37.6$$^{b}$ & $      8080 \pm  137$  &            $19.92$ &             $0.01$ & $            1.74\pm 0.43$   &        $  701_{-199}^{+273}$  \\
TCP J00505644+5351524  &          Evolved CV candidate &                  & $      3434 \pm   21$  &            $18.31$ &             $1.82$ & $            0.84\pm 0.18$   &              $ 1300_{-234}^{+417}$  \\
ZTF18aaaasnn           &               CV &            $91.21$(sh)$^{c}$ & $      6033 \pm   60$  &            $20.37$ &             $0.11$ & $            0.51\pm 0.82$   &        $ 2350_{-1483}^{+1593}$  \\
ZTF21abhrevv           &               CV &                  & $      8391 \pm  171$  &            $20.18$ &             $0.37$ & $            1.12\pm 0.55$   &              $ 1201_{-445}^{+746}$  
    \botrule
    \end{tabular}}\end{center}
    \begin{tabnote}
        Parallaxes are taken from the Gaia DR3 catalogue \citep{2022yCat.1355....0G, 2023A&A...674A...1G}, the distance correspond to geometric distances derived by \cite{2021AJ....161..147B} from Gaia EDR3 catalogue, superhump periods are marked by (sh).\\
 $^{(b)}$ Period listed in \cite{2025A&A...700A.107G}; $^{(c)}$ Period determined in this work\\
    \end{tabnote}
\end{table*}

\begin{table*}[]
\caption{Superoutburst properties of studied targets. Presented are recurrence times $T_\mathrm{SO}$, superoutburst durations $\tau_\mathrm{dur}$, amplitudes $A_\mathrm{SO}$, peak brightnesses, rise rates $\mu_\mathrm{r}$, and decline rates $\mu_\mathrm{d}$. The column of peak brightness lists also the photometric filter used to determine the peak value.  }
\label{tab:SO_prop}
\centering
{\tablefont\begin{tabular}{lccccccccc}
\toprule
Name                & Type   & SO & $T_\mathrm{SO}$ & $\tau_\mathrm{dur}$      & $A_\mathrm{SO}$     & Peak brightness        & $\mu_\mathrm{r}$   & $\mu_\mathrm{d}$   \\
                      &        &    & {[}d{]}         & {[}d{]}          & {[}mag{]}        &                        & {[}mag d$^{-1}${]} & {[}mag d$^{-1}${]} \\
\noalign{\smallskip}\hline
ASASSN-15kf           & AM CVn & 4  & $600$           & \textgreater{}8  &                  & 14.5 (V)               &                    & \textgreater{}0.19 \\
ASASSN-15na           & CV     & 2  & $\sim1500$      & \textgreater{}25 & \textgreater 2.4 & 15.4 (g)               &                    & 0.10               \\
ASASSN-17fp           & AM CVn & 1  & $>3000$         & 10               &                  & 15.7 (V)               &                    &                    \\
ASASSN-18abl          & Evolved CV candidate  & 1  & $>2200$         & 20               & 3.5              & 12.0 (g)               & 5.88               & 0.43               \\
ASASSN-18rg           & AM CVn & 1  & $>2400$         & 10               & 7.9              & 12.5 (g)               & \textgreater{}3.70 & 2.27               \\
ASASSN-19ct           & AM CVn & 14 & $180$ - $365$   & 7                & 4.0              & 13.3 (g)               & 2.27               & 2.17               \\
ASASSN-19rg           & AM CVn & 1  & $> 2000$        & 9                & 7                & 13.0 (g)               & 5.56               & 2.04               \\
ASASSN-20gx           & AM CVn & 2  & $\sim 1100$     & 7                & 5.9              & 14.4 (g)               & \textgreater{}2.22 & \textgreater{}0.80 \\
ASASSN-20la           & Evolved CV candidate  & 1  & $>1400$         & 7                & 4.9              & 16.3 (r)               & \textgreater{}4.35 & \textgreater{}1.28 \\
ASASSN-20lr           & AM CVn & 1  & $>1400$         & 6                & 5.4              & 14.7 (g)               & \textgreater{}0.91 & \textgreater{}1.02 \\
ASASSN-20pv           & AM CVn & 4  & $\sim500$       & 10               & 4.8              & 12.3 (g)               & 3.57               & 1.25               \\
ASASSN-21hc           & AM CVn & 1  & $>1800$         & 6                & 6.25             & 13.3 (g)               & \textgreater{}3.12 & 2.50               \\
Gaia21cxj             & AM CVn & 3  & $\sim2100$      & \textgreater{}5  & 5.91             & 14.1 (g)               &                    & 1.56               \\
TCP J00505644+5351524 & Evolved CV candidate  & 1  & $>1300$         & 4                & 4.5              & 15.0 (g)               &                    & 1.11               \\
ZTF18aaaasnn          & CV     & 6  & $\sim300$       & 16               & 4                & 16.8 (g)               &                    & 1.12               \\
ZTF21abhrevv          & CV     & 1  & $>1200$         & \textgreater{}22 & \textgreater{}7  & \textgreater{}12.7 (g) &                    & 1.06  
\botrule
\end{tabular}}
\end{table*}

\begin{table*}[]
\caption{X-ray fluxes $f$ of studied AM CVns and their and luminosities $L$ mass accretion rates $\dot{M}$ derived for the combination of listed energy bands. Periods listed  for ASASSN-20pv and Gaia21cxj are orbital periods, superhump periods are listed for the other systems. Critical mass accretion rates $\dot{M}_\mathrm{crit}^-$ and $\dot{M}_\mathrm{crit}^+$ were determined using Equation (A.2) of \cite{2012A&A...544A..13K}.}
\label{tab:XRay}
\centering
{\tablefont\begin{tabular}{llcccccc}
\toprule
Name                          &                                                                   & ASASSN-15kf            & ASASSN-19ct            & ASASSN-19rg            & ASASSN-20pv            & ASASSN-21hc            & Gaia21cxj              \\
Period                        & {[}min{]}                                                         & $27.7$                   & $30.9$                   & $43.9$                   & $27.3$                   & $35.8$                   & $37.6$                   \\
\noalign{\smallskip}\hline
$f_{(0.2 - 2.3)\mathrm{keV}}$ & {[}$10^{-13}\,\frac{\mathrm{erg}}{\mathrm{s}\,\mathrm{cm}^{2}}${]} & $1.40_{-0.37}^{+0.45}$ & $2.95_{-0.46}^{+0.51}$ & $1.07_{-0.27}^{+0.32}$ & $5.77_{-0.53}^{+0.57}$ & $1.31_{-0.31}^{+0.36}$ &                        \\
$f_{(0.2 - 0.5)\mathrm{keV}}$ & {[}$10^{-13}\,\frac{\mathrm{erg}}{\mathrm{s}\,\mathrm{cm}^{2}}${]} &  $0.00_{-0.00}^{+0.07}$  & $0.47_{-0.18}^{+0.24}$ & $0.32_{-0.14}^{+0.19}$ & $0.90_{-0.21}^{+0.26}$ & $0.12_{-0.08}^{+0.14}$ & $0.07_{-0.04}^{+0.04}$ \\
$f_{(0.5 - 1.0)\mathrm{keV}}$ & {[}$10^{-13}\,\frac{\mathrm{erg}}{\mathrm{s}\,\mathrm{cm}^{2}}${]} & $0.52_{-0.19}^{+0.25}$ & $0.79_{-0.19}^{+0.24}$ & $0.11_{-0.07}^{+0.11}$ & $1.54_{-0.24}^{+0.26}$ & $0.43_{-0.16}^{+0.19}$ &                        \\
$f_{(1.0 - 2.0)\mathrm{keV}}$ & {[}$10^{-13}\,\frac{\mathrm{erg}}{\mathrm{s}\,\mathrm{cm}^{2}}${]} & $0.70_{-0.25}^{+0.31}$ & $1.57_{-0.34}^{+0.39}$ & $0.74_{-0.22}^{+0.28}$ & $3.25_{-0.44}^{+0.43}$ & $0.66_{-0.22}^{+0.27}$ &                        \\
$f_{(2.0 - 5.0)\mathrm{keV}}$ & {[}$10^{-13}\,\frac{\mathrm{erg}}{\mathrm{s}\,\mathrm{cm}^{2}}${]} & $0.75_{-0.54}^{+1.05}$ & $3.10_{-1.32}^{+1.78}$ & $0.64_{-0.64}^{+1.21}$ & $5.63_{-1.58}^{+1.95}$ & $0.56_{-0.54}^{+1.03}$ &                        \\
$f_{(0.5 - 7.0)\mathrm{keV}}$ & {[}$10^{-13}\,\frac{\mathrm{erg}}{\mathrm{s}\,\mathrm{cm}^{2}}${]} &                        &                        &                        &                        &                        & $0.44_{-0.11}^{+0.11}$ \\
\noalign{\smallskip}\hline
$L_{(0.2 - 5.0)\mathrm{keV}}$ & [$10^{30}\,\frac{\mathrm{erg}}{\mathrm{s}}$] & $9.87_{-7.16}^{+12.19}$ & $4.01_{-1.39}^{+1.81}$ & $29.14_{-24.44}^{+41.07}$ & $9.32_{-2.05}^{+2.39}$ & $4.44_{-3.14}^{+4.93}$ & \\
$L_{(0.2 - 7.0)\mathrm{keV}}$ & [$10^{30}\,\frac{\mathrm{erg}}{\mathrm{s}}$] & & & & & & $2.98_{-1.94}^{+2.51}$ \\
\noalign{\smallskip}\hline
$\dot{M}_{(0.2 - 5.0)\mathrm{keV}}$ & [$10^{-13}\,\frac{\mathrm{M}_\odot}{\mathrm{year}}$] & $17.39_{-12.61}^{+21.48}$ & $7.07_{-2.44}^{+3.18}$ & $51.34_{-43.06}^{+72.36}$ & $16.41_{-3.62}^{+4.21}$ & $7.83_{-5.54}^{+8.68}$ & \\
$\dot{M}_{(0.2 - 7.0)\mathrm{keV}}$ & [$10^{-13}\,\frac{\mathrm{M}_\odot}{\mathrm{year}}$] & & & & & & $5.26_{-3.41}^{+4.43}$ \\
\noalign{\smallskip}\hline
$\dot{M}_\mathrm{crit}^-$ & [$10^{-13}\,\frac{\mathrm{M}_\odot}{\mathrm{year}}$] & $2.6$ & $2.6$ & $2.6$ & $2.6$ & $2.6$ & $2.6$ \\
$\dot{M}_\mathrm{crit}^+$ & [$10^{-10}\,\frac{\mathrm{M}_\odot}{\mathrm{year}}$] & $16.8$ & $20.4$ & $38.1$ & $16.4$ & $26.5$ & $29.0$ 
\botrule
\end{tabular}}
\end{table*}

\begin{table*}[]
\caption{UV fluxes of studied system obtained from the GALEX catalogue \citep{2017ApJS..230...24B}. Period listed in this table represent superhump periods of the systems.}
\label{tab:UV}
\centering
{\tablefont\begin{tabular}{llccccccc}
\toprule
Name         & Type    & Period    & $f_\mathrm{FUV}$                                                  & $f_\mathrm{NUV}$                                                  & $L_\mathrm{FUV}$                                 & $L_\mathrm{NUV}$                                 & \multicolumn{1}{c}{$\dot{M}_\mathrm{FUV+NUV}$ }                                                                                  \\
             &         & {[}min{]} & {[}$10^{-13}\,\frac{\mathrm{erg}}{\mathrm{s}\,\mathrm{cm}^{2}}${]} & {[}$10^{-13}\,\frac{\mathrm{erg}}{\mathrm{s}\,\mathrm{cm}^{2}}${]} & {[}$10^{30}\,\frac{\mathrm{erg}}{\mathrm{s}}${]} & {[}$10^{30}\,\frac{\mathrm{erg}}{\mathrm{s}}${]} & {[}$10^{-13}\,\frac{\mathrm{M}_\odot}{\mathrm{year}}${]} \\
             \noalign{\smallskip}\hline
ASASSN-18rg  & AM CVn  & $46.0$      & $0.20\pm0.10$                                                    & $0.94\pm0.15$                                                    & $0.25_{-0.19}^{+0.54}$                           & $1.16_{-0.73}^{+2.44}$                           & $2.49_{-1.62}^{+5.24}$             \\
ASASSN-19ct  & AM CVn  & $30.9$      & $22.53\pm1.25$                                                   & $29.03\pm0.81$                                                   & $15.24_{-1.03}^{+1.03}$                          & $19.63_{-0.94}^{+0.92}$                          & $61.43_{-3.48}^{+3.43}$            \\
ASASSN-20gx  & AM CVn  &           &                                                                    & $0.68\pm0.17$                                                    &                                                  & $5.77_{-3.76}^{+5.59}$                           & $10.17_{-6.63}^{+9.84}$             \\
ZTF18aaaasnn & CV      & $91.2$      &                                                                    & $0.38\pm0.16$                                                    &                                                  & $24.84_{-24.84}^{+35.28}$                        & $43.76_{-43.76}^{+62.15}$          \\
ZTF21abhrevv & CV      &           & $1.25\pm0.30$                                                    & $1.24\pm0.15$                                                    & $21.51_{-16.77}^{+27.21}$                        & $21.41_{-16.08}^{+26.72}$                        & $75.62_{-57.87}^{+95.02}$          \\
ASASSN-15na  & CV      & $91.6$      & $0.74\pm0.17$                                                    & $0.85\pm0.16$                                                    &                                                  &                                                  &                                           \\
ASASSN-18abl & Evolved CV candidate & $87.2$      & $0.79\pm0.25$                                                    & $1.48\pm0.21$                                                    & $1.91_{-0.60}^{+0.60}$                           & $3.56_{-0.51}^{+0.51}$                           & $9.62_{-1.95}^{+1.95}$             
\botrule
\end{tabular}}
\end{table*}

\section{Discussion}

\begin{figure*}
   \centering
   
   \includegraphics[width=0.99\linewidth]{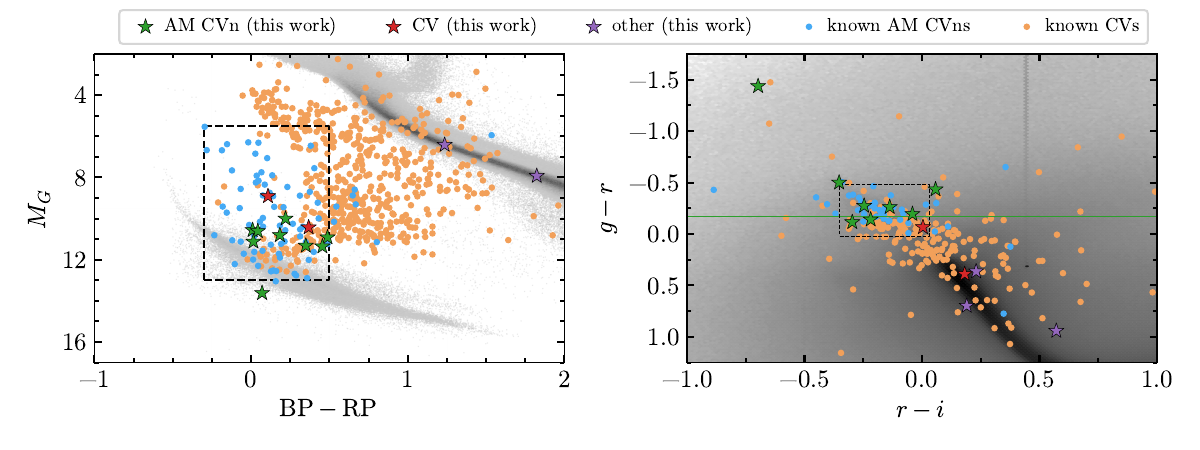}
   \caption{Left: Colour-magnitude diagram showing Gaia sources within 200 pc (grey), AM~CVns (blue), CVs (orange), AM~CVns and CVs from the studied sample (green stars and red stars, respectively), and targets from the studied sample of other type (purple star). No reddening correction was applied for the presented data. Right: Colour-colour diagram showing stars from the SDSS DR12 catalogue (grey), AM~CVns and CVs based on SDSS photometry, and targets from the studied sample based on ZTF,  SkyMapper, and PanStarrs1 photometry. Targets with only one available colour index are represented by a line. The dashed line represents the selection criteria used by \cite{2014MNRAS.439.2848C} for the identification of  AM~CVns.}
   \label{F:CMD}
\end{figure*}

\subsection{Colour-magnitude diagram}

Figure~\ref{F:CMD}, left panel, shows the positions of the observed target in a colour-magnitude diagram constructed using absolute Gaia magnitudes $M_G$ and colours based on Gaia's blue prism (BP) and red prism (RP) magnitudes. Targets ASASSN-15na, ASASSN-17fp, and ASASSN-20la are not included due to a lack of Gaia observations. The diagram also shows the positions of AM~CVn stars listed in the catalogue by \cite{2025A&A...700A.107G} and CVs from the catalogue by \cite{2003A&A...404..301R, 2011yCat....102018R}. 

All AM~CVn stars and CVs analysed in this paper are blue, grouped in a region close to the WD branch, where both of these types can be expected \citep{2020MNRAS.492L..40A, 2023MNRAS.524.4867I, 2025A&A...700A.107G}, indicating an important contribution from the accreting WD and the inner part of the accretion disc. All of them can be found slightly above the WD sequence, apart from one system, ASASSN-18rg, which is located below that sequence. However, parallax of this system was determined with a large error (see Table~\ref{tab:Params}), which might have affected its vertical position in the diagram. 

The diagram shows that AM~CVn stars tend to have bluer colours than CVs, even though there is a partial overlap, in which the targets analysed in this work are located. CVs found in this overlap are dominantly short-period CVs with orbital periods bellow 2 hours. All AM~CVns studied in this paper occupy the region for which \begin{equation}
    \mathrm{BP} - \mathrm{RP} < 0.5, 
\end{equation}which is true also for approximately $88\%$ of known AM~CVns with available Gaia photometry and for approximately $34\%$ of CVs from the catalogue by \cite{2011yCat....102018R}. 
Combining the properties of AM~CVns from our study and from literature, we can determine the boundaries of region in colour-magnitude diagram occupied by AM~CVns as \begin{equation}\begin{aligned}
    -0.3 <\mathrm{BP} &- \mathrm{RP} < 0.5, \\
    5.5< \;&M_G<13,
\end{aligned}\end{equation}which is occupied by $89\%$ of known AM~CVns, boundaries of this region are marked by black dashed line in Figure~\ref{F:CMD}.

Two clear outliers from the analysed sample are the two stars labelled as `other': ASASSN-18abl and TCP J00505644+5351524. The colour index of these stars shows that they are red objects and they are located on the main sequence, unlike most of the AM~CVn stars and CVs.
Given that these systems have shown superoutbursts, their binary nature is confirmed and the lack of emission lines indicate that the red component is clearly dominating the spectrum. 

\subsection{Colour-colour diagram}

Figure~\ref{F:CMD}, right panel, shows a de-reddened colour-colour diagram based on photometry in bands \textit{g}, \textit{r}, and \textit{i}. The photometry for the targets from this work was obtained primarily from ZTF. In cases where ZTF data were not available, we used photometry from SkyMapper and Pan-STARS. Only observation obtained at quiescence were considered. The photometry for known AM~CVns and CVs was obtained from SDSS. We adopted reddening values from \cite{1998ApJ...500..525S} and converted them to corresponding filters by relations derived by \cite{2011ApJ...737..103S}. We chose to use reddening derived by \cite{1998ApJ...500..525S} as it was also used in previous studies of AM~CVns \citep[for example][]{2009MNRAS.394..367R, 2013MNRAS.429.2143C, 2014MNRAS.439.2848C}, which allows us to easily compare our results with those of other authors.

AM~CVns from this paper show narrow spread in $g-r$ and they occupy the region for which \begin{equation}
    g-r < -0.11, 
\end{equation} which applies also to $84\%$ of known AM~CVns and $29\%$ of CVs from the catalogue of \cite{2011yCat....102018R}. This shows that $g-r$ colour index is a useful parameter for selection of AM~CVn candidates. The spread in $r-i$ is larger and the studied AM~CVns have values for which \begin{equation}
    r-i < 0.06, 
\end{equation} which is true also for $88\%$ of known AM~CVns and $60\%$ CVs used in this study for comparison.

A similar colour-colour diagram was also presented by \cite{2013MNRAS.429.2143C}, 
their selection criteria
for the identification of  AM~CVns are marked in the colour-colour diagram by a black dashed line. Five of the AM~CVns from this study lie inside of the region fulfilling these criteria, while three lie outside. Two of the outliers,  ASASSN-20pv ($r-i=-0.35$, $g-r=-0.50$) and ASASSN-21hc ($r-i=0.06$, $g-r=-0.43$), lie close to the boundary, while the third one, ASASSN-20lr ($r-i=-0.70$, $g-r=-1.44$), is a clear outlier. However, the reddening given by \cite{1998ApJ...500..525S} for the position of ASASSN-20lr is $E(B-V)=1.47$ which strongly affects the position of the star in the colour-colour diagram and given the fact that \cite{1998ApJ...500..525S} provides full Galactic reddening, it is likely that the reddening of ASASSN-20lr is overestimated. The three-dimensional extinction map by \cite{2019ApJ...887...93G} gives for the position of ASASSN-20lr and its Gaia distance $E(g-r)=0.35$. This value is considerably lower than the one of full Galactic reddening and leads to colours $r-i=-0.09$, $g-r=-0.35$, which lie within the region occupied by AM~CVns.

\begin{figure}
   \centering
   \includegraphics[width=0.99\linewidth]{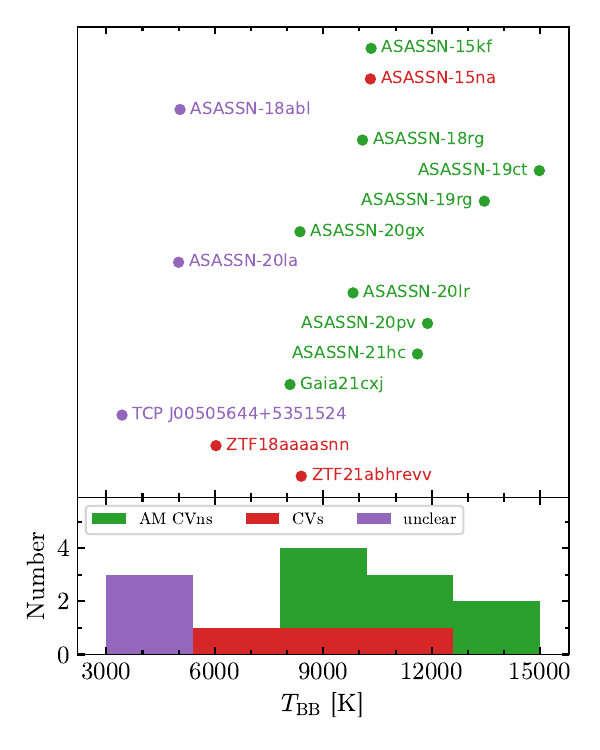}

   \includegraphics[width=0.99\linewidth]{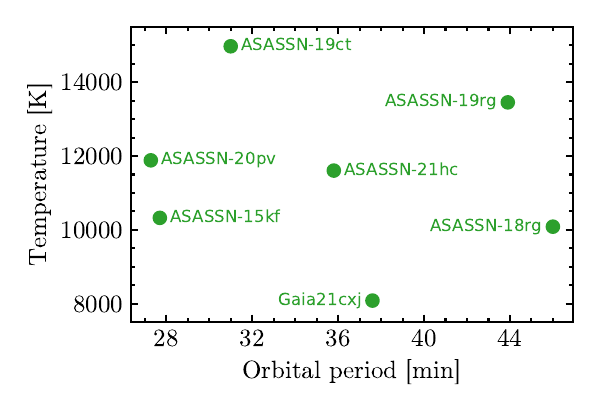}
   \caption{Top: Distribution of temperatures of the studied targets derived by fitting the continuum of spectra with a black-body model. Typical size of error is of the size of the symbols.
   Bottom: Relation between orbital periods of AM~CVns from this study and their temperatures determined from the black-body model. Typical size of error is of the size of the symbols.}
   \label{F:Temp_both}
\end{figure}

\subsection{X-ray and UV luminosities and accretion rates}

We computed the X-ray luminosities for our targets with available X-ray data using the Gaia distances determined by \cite{2021AJ....161..147B}, the resulting values are given in Table~\ref{tab:XRay}. 
All eROSITA observations were obtained during quiescence, which is also likely the case of Gaia21cxj.
The luminosities lie in the range between $2.98\cdot10^{30}\mathrm{erg}\,\mathrm{s}^{-1}$ and $29.14\cdot  10^{30} \, \mathrm{erg}\,\mathrm{s}^{-1}$, which agrees with the X-ray luminosities of AM CVn studied by \cite{2023JAVSO..51..227B}, who reported that short-period AM CVns show X-ray luminosities smaller than those predicted by models \citep{2012A&A...537A.104V} likely due to the boundary layer of short-period systems being optically thick. 
We computed GALEX-UV luminosities for our targets using an analogous approach, the obtained luminosities are listed in Table~\ref{tab:UV}.

The maximal possible luminosity of boundary layer $L_\mathrm{BL}$ is according to \citet[][equation 6.6]{2002apa..book.....F}\begin{equation}
    L_\mathrm{BL} = \frac{G\,M_1\,\dot{M}}{2\,R_1}\left[ 1-\frac{\Omega_1}{\Omega_\mathrm{K}} \right]^2,
\end{equation} where $R_1$, $M_1$, and $\Omega_1$ are the radius, mass, and surface angular velocity of the primary star, $G$ is the gravitational constant, $\dot{M}$ is the mass accretion rate, and $\Omega_\mathrm{K}$ is the Keplerian angular velocity of the boundary layer. If we assume $\Omega_1 \ll \Omega_\mathrm{K}$ and that $L_\mathrm{BL}$ is equal to the X-ray luminosity $L_\mathrm{X}$, we can estimate the mass accretion rate
as \begin{equation}
    \dot{M}=\frac{2\,R_1\,L_\mathrm{X}}{G\,M_1}.
\end{equation} However, this equation provides only lower limit on the mass accretion rate, as the assumptions $\Omega_1 \ll \Omega_\mathrm{K}$ and $L_\mathrm{BL} = L_\mathrm{X}$ lead to underestimating the value of $\dot{M}$. 
For the calculations, we assumed $M_1=0.85\,\mathrm{M}_\odot$ and we calculated corresponding radius $R_1 = 0.009\,\mathrm{R}_\odot$ using the relation given by \cite{1988ApJ...332..193V}.
The resulting  mass accretion rates derived from X-ray and UV luminosities are listed in Table~\ref{tab:XRay} and Table~\ref{tab:UV}, respectively. Their values are of orders $10^{-13} - 10^{-12}\,\frac{\mathrm{M}_\odot}{\mathrm{year}}$. 

For the accretion disc to be in an unstable state, in which outburst can occur, its accretion rate needs to be between critical values $\dot{M}_\mathrm{crit}^{-}$ and $\dot{M}_\mathrm{crit}^{+}$.
Using the same parameters for $M_1$ and $R_1$ and assuming the disc's inner radius is the size of accretor and the disc viscosity $\alpha_\mathrm{cold}=0.1$ we can estimate the critical mass accretion rate $\dot{M}_\mathrm{crit}^{-}=1.7\,\cdot10^{-13}\,\mathrm{M}_\odot\,\mathrm{year}^{-1}$ from the Equation (A.2) of \cite{2012A&A...544A..13K} for an AM CVn with $2\%$ of metals. All of the mass accretion rates obtained from the X-ray observations lie above this limit. By assuming a mass ratio $q=0.03$ \citep[as determined from studies of eclipsing systems,][]{2022MNRAS.512.5440V} and $\alpha_\mathrm{hot}=0.2$ we can estimate $\dot{M}_\mathrm{crit}^{+}$ for each target, the estimated values are listed in Table~\ref{tab:XRay}. 

All estimated $\dot{M}_\mathrm{crit}^{+}$ are about four orders of magnitude larger than the mass accretion rates obtained from the X-ray observations. This places the studied AM CVns in the unstable disc region, consistent with their observed transient behaviour.  Yet,  the derived accretion rates are below the values predicted by evolutionary models \citep{2021ApJ...923..125W} for mass transfer rates of the accretion discs. 
Since our mass accretion rates were determined  using only X-ray luminosities, the obtained values can be underestimated if most of the emission comes in the UV, which can be expected from previous studies of AM CVns \citep[e.g.][]{2006A&A...457..623R}. Therefore, the derived values can serve only as a lower limit for the mass accretion rates. 
This applies especially to short-period systems, for which the accretion rates derived from observed X-ray luminosities are underestimated due to the presence of a likely optically thick boundary layer \citep{2023JAVSO..51..227B}. 

Limitations of using only X-ray observations for accretion rate estimation can be seen on the case of ASASSN-19ct for which we determined $\dot{M}_\mathrm{FUV+NUV}=61.43\,\cdot10^{-13}\,\mathrm{M}_\odot\,\mathrm{year}^{-1}$ which is about ten times larger value than the one obtained from X-rays. However, due to lack of photometric monitoring coinciding with the GALEX observations, we cannot determine if the UV observations were obtained in quiescence or during a superoutburst. As the recurrence times of superoutbursts of ASASSN-19ct are shorter than one year, it is possible that the large UV flux was caused by a superoutburst.

The mass accretion rate of ASASSN-20gx lies above the limit $\dot{M}_\mathrm{crit}^{-}=2.6\,\cdot\,10^{-13}\,\mathrm{M}_\odot\,\mathrm{year}^{-1}$, but ASASSN-18rg shows accretion rate $\dot{M}_\mathrm{FUV+NUV}=2.49_{-1.62}^{+5.24}\,\cdot\,10^{-13}\,\mathrm{M}_\odot\,\mathrm{year}^{-1}$ which lies just beneath $\dot{M}_\mathrm{crit}^{-}$, however, the critical value is within the estimated uncertainty.
The mass accretion rates $\dot{M}_\mathrm{FUV+NUV}$ of ZTF18aaaasnn, ZTF21abhrevv, and ASASSN-18abl are above the predicted critical accretion rate for CVs $\dot{M}_\mathrm{crit}^{-} \approx 0.8\,\cdot\,10^{-13}\,\mathrm{M}_\odot\,\mathrm{year}^{-1}$ \citep{2011ApJS..194...28K}.

We note that mass transfer rates of a sample of AM CVns derived by \cite{2018A&A...620A.141R} from SEDs are of several orders higher and in agreement with evolutionary models. 
A difference between $\dot{M}$ derived from SED modelling and from X-rays can be expected, as SED modelling provides $\dot{M}$ which is characteristic for the mass transfer in the whole accretion disc, while X-rays and UV fluxes provide estimate of accretion onto the white dwarf. 
The systems analysed by \cite{2018A&A...620A.141R} also show smaller X-ray luminosities than is predicted by models, as was shown by \cite{2023JAVSO..51..227B}, which are of the same order as the X-ray luminosities derived in our study.

\begin{figure*}
   \centering
   \includegraphics[width=0.49\linewidth]{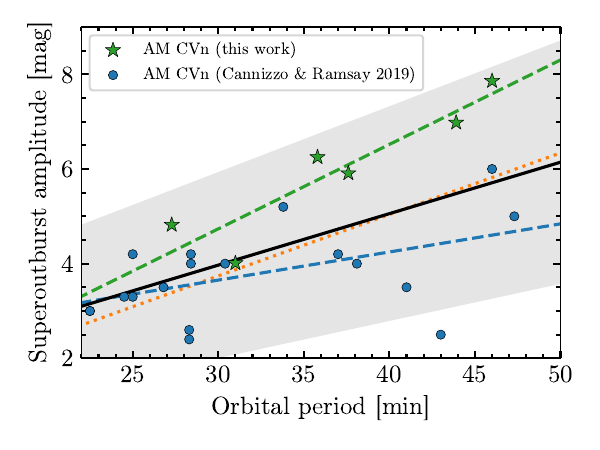}
   \includegraphics[width=0.49\linewidth]{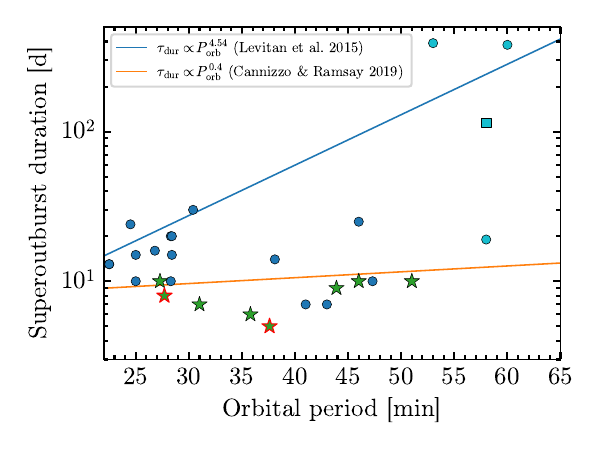}
   \caption{Left: Relation between orbital period and amplitude of superoutbursts for AM~CVns from this study (green stars) and from the study by \cite{2019AJ....157..130C} (blue circles). The dashed lines represent linear fits of individual samples, the orange dotted line represents the linear relation derived by \cite{2015MNRAS.446..391L}. The black solid line represents linear fit of data from both samples, the grey area represents $1\sigma$ error. Typical errors are smaller than the symbols.
   Right: Relation between orbital period and duration of superoutbursts for AM~CVns from this study (green stars), from the study by \cite{2019AJ....157..130C} (blue circles), and from \cite{2020ApJ...900L..37R, 2021MNRAS.505..215R, 2022ApJ...926...10R} (cyan circles). The cyan square marks the length of superoutburst and initial increase of brightness of ASASSN-21au as derived by \cite{2022ApJ...926...10R}. Systems with lower limits on the durations are marked by red outline. The blue line shows the empirical relation derived by \cite{2015MNRAS.446..391L}, the orange line shows the theoretical relation obtained by \cite{2015ApJ...803...19C} for disc instability model where we adopted value of disc viscosity $\alpha_\mathrm{hot}=0.2$ from \cite{2012A&A...545A.115K} and we used median values $M_1=0.85$ and $q=0.04$ of confirmed AM~CVns from \cite{2025A&A...700A.107G}.}
   \label{F:Per_Amp_Dur}
\end{figure*}

\subsection{Amplitudes of superoutbursts of AM~CVn and CV stars}

Figure~\ref{F:Per_Amp_Dur} shows the relation between orbital periods and amplitudes of superoutbursts for AM~CVns from this study and from the literature presented by \cite{2019AJ....157..130C}. The amplitudes of superoutbursts were determined from ground-based observations presented in 
Figure~\ref{F:LC:LT_ALL}, the values are listed in Table~\ref{tab:SO_prop}.
The Figure also shows the linear relation derived by \cite{2015MNRAS.446..391L} for AM~CVns with orbital periods between $22$ and $37$ minutes.
While there is a correlation between the amplitude of superoutbursts and the orbital period (Pearson correlation coefficient $c_\mathrm{P}=0.6$), there is a large dispersion especially for longer orbital periods. All but one of the AM~CVns from our study shown in the figure have amplitudes larger that it is predicted by the relation derived by \cite{2015MNRAS.446..391L} (orange dotted line) or the relation which can be derived from the sample presented by \cite{2019AJ....157..130C} (blue dashed line). The linear relation derived from all AM~CVns in this study is \begin{equation}
    A_\mathrm{SO}=(0.11\pm0.03)P_\mathrm{orb} + (0.72\pm1.04),
\end{equation} where $A_\mathrm{SO}$ is the amplitude and $P_\mathrm{orb}$ is the orbital period in minutes. It is possible that some of the previously published amplitudes could be underestimated due to incomplete coverage of the superoutbursts. 

The amplitudes of the confirmed CVs identified in this study are consistent with the amplitude limits established for their orbital period \citep{2016MNRAS.456.4441C, 2016MNRAS.460.2526O}. 

\subsection{Superoutburst duration for the AM~CVns}

By construction, our sample is formed of systems identified through their outbursts.  Figure~\ref{F:Per_Amp_Dur} shows the relations between orbital periods and durations of the superoutbursts $\tau_\mathrm{dur}$ for the AM~CVns from this work and from previous studies \citep{2019AJ....157..130C, 2020ApJ...900L..37R, 2021MNRAS.505..215R, 2022ApJ...926...10R}. All the AM~CVns from our study exhibited superoutbursts with durations between 6 and 10 days and they do not show any strong dependency on the orbital period. Similar short durations were already identified for systems with periods shorter than 35 min 
\citep{2021MNRAS.508.3275P}.
Interestingly, ASASSN-17fp (51 min) also showed a short duration superoutburst consistent with a disc instability origin, which contrasts with those systems with orbital periods longer than 50 min for which high state activity has been observed to last for years \citep{2020ApJ...900L..37R, 2021MNRAS.505..215R} and the origin of which is attributed to EMT. This supports the existence of a dichotomy as  observed in ASASSN-21au \citep{2022ApJ...926...10R}. The dichotomy is likely linked to different mass-transfer rates and even disc truncation \citep{2022ApJ...926...10R}.

It is possible that some of previously published durations of superoutbursts might be overestimated due to poor sampling of the light curves, which could cause echo outbursts and fading tails to appear as a part of the superoutburst.
The flat superoutburst duration distribution observed in figure \ref{F:Per_Amp_Dur} is in agreement with the expected dependency derived for a traditional disc instability model by \cite{2015ApJ...803...19C}, who predicts only a relatively modest dependency of $\tau_\mathrm{dur} \propto P_\mathrm{orb}^{\,0.4}$. However, as discussed for the case of ASASSN-19ct, it is possible that EMT is also present.

\begin{figure*}
   \centering
   \includegraphics[width=0.99\linewidth]{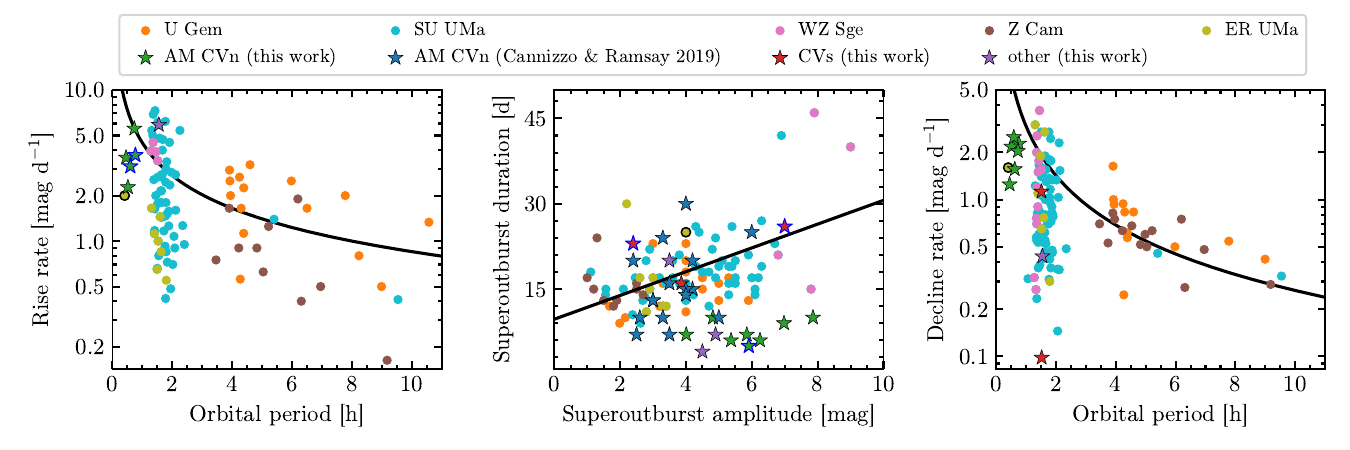}
   \caption{Left: Relation between orbital periods and rise rates of superoutbursts for AM~CVns and CVs. Centre: Relation between amplitudes and durations of superoutbursts of AM~CVns and CVs. Right: Relation between orbital periods and decline rates of superoutbursts for AM~CVns and CVs. Properties of CVs are taken from catalogue by \cite{2016MNRAS.460.2526O}, black lines represent their best fits of shown relations. Yellow point with black outline represents a known AM~CVn star CR Boötis, which is classified by \cite{2016MNRAS.460.2526O} as an ER~UMa system, star symbols with blue outline mark lower limits of the superoutburst properties.}
   \label{F:OH_comparison}
\end{figure*}

\subsection{Comparison of superoutburst properties of AM~CVns and CVs}

We determined the rise and decline rates of targets from this study from the corresponding phases of superoutbursts, following similar method as \cite{2016MNRAS.460.2526O}. 
Figure~\ref{F:OH_comparison}, left panel, shows relation between orbital periods and superoutburst rise rates $\mu_\mathrm{r}$ for AM~CVns from this study and for CVs analysed by \cite{2016MNRAS.460.2526O}, the right panel shows analogous relation for decline rates $\mu_\mathrm{d}$.  The central panel then shows the relation between superoutburst amplitudes and durations.

\cite{2016MNRAS.460.2526O} reported that CVs above the period gap ($P_\mathrm{orb} > 3\,\mathrm{h}$) show dependency of rise rates and decline rates of long outbursts and superoutbursts on their orbital periods. The CVs bellow the period gap ($P_\mathrm{orb} < 2\,\mathrm{h}$) show large dispersion in both rates with no dependency on orbital period.

All AM~CVns show fast changes of brightness during the decline and rise phases of their outbursts. The values of the rise rates are in the range $2.2\,\mathrm{mag}\,\mathrm{d}^{-1} \lesssim \mu_\mathrm{r}\lesssim5.6\,\mathrm{mag}\,\mathrm{d}^{-1}$ and the values of decline rates lie in the range $1.2\,\mathrm{mag}\,\mathrm{d}^{-1} \lesssim \mu_\mathrm{d}\lesssim2.5\,\mathrm{mag}\,\mathrm{d}^{-1}$. We note that the CV classified by \cite{2016MNRAS.460.2526O} as ER UMa, which lies in the same regions as AM~CVns, is a known AM~CVn star, CR Boötis.
All AM CVns show faster changes of brightness during the rise than during the decline, similarly as majority of CVs analysed by \cite{2016MNRAS.460.2526O}. 
However, they show much smaller dispersion in the both rates than the CVs bellow the period gap. WZ Sge stars also exhibit low dispersion of their rise rates, which is even smaller than for AM CVns, their decline rate dispersion is, however, larger and comparable with other type of CVs bellow the period gap. 

While the rise and decline  rates cannot be used to distinguish AM~CVns from CVs, the sample of AM CVns shown in Figure~\ref{F:OH_comparison} suggest  that selection criteria $\mu_\mathrm{r} \gtrsim 2\,\mathrm{mag}\,\mathrm{d}^{-1}$ and $\mu_\mathrm{d} \gtrsim 1\,\mathrm{mag}\,\mathrm{d}^{-1}$ can be used for identification of suitable candidates. 

The diagram displaying the relation between superoutburst amplitude and superoutburst duration shows that the AM CVns from our study exhibit superoutbursts of shorter durations than CVs with superoutbursts and long outbursts of similar amplitudes. However, the sample of AM CVns analysed by \cite{2019AJ....157..130C} shows large variety of superoutburst durations
comparable with the ones of CVs. As superoutbursts of AM CVns are typically followed by rebrightenings and long fading tails, it is possible that their durations might have been overestimated in some cases, as was shown by \cite{2021MNRAS.508.3275P}, which could explain the large dispersion.

\subsection{Black-body temperatures and spectra}

Figure~\ref{F:Temp_both} shows the distribution of black-body temperatures of the studied targets derived by fitting of the spectra and 
the relation between temperatures and orbital periods. Temperatures of AM~CVns vary between $8000\,\mathrm{K}$ and $15\,000\,\mathrm{K}$, CVs have temperatures between $6000\,\mathrm{K}$ and $10\,300\,\mathrm{K}$, and the evolved CV candidates exhibit temperatures between $3400\,\mathrm{K}$ and $5000\,\mathrm{K}$. Similarly as the colour indices, the temperature distribution displays that AM~CVns tend to show higher temperatures than CVs and the evolved CV candidates systems appear colder than both CVs and AM~CVns. 

The range of values determined for the AM~CVns seems in general to be in agreement with values reported by other authors \citep{2022MNRAS.512.5440V} and even with those obtained via simpler approaches, like fits to SEDs \citep{2024RNAAS...8..299M}. However, we note that for most systems, the black-body temperatures are well below the expected values for accretors in AM~CVns according to the models by \cite{2021ApJ...923..125W}. This is particularly evident for systems with periods below 35 min, something similar as observed for YZ Lmi by \citep{2022MNRAS.512.5440V}. This is not surprising as the cooler components of the binary are having an important contribution in the optical spectrum. 

We also noted that AM~CVn systems with similar periods, such as ASASSN-15kf and ASASSN-20pv (both with periods $\sim27.5$ min) have differences in temperature of around 1500 K. While distance can be a factor, the reddening towards these binaries is low and very similar ($E(B-V)=0.07$ and $0.08$). Both systems showed double peaked lines, which indicate high inclination. However, their He abundances are clearly different. ASASSN-15kf shows strong He emission lines, while the optical spectrum of ASASSN-20pv, which has a larger black-body temperature, shows only a broad He 5877 emission line. Having both systems large inclination and very similar periods, the difference in optical spectra could point to a different type of donor. This could also lead to different mass-transfer and mass-accretion rates (making the disc hotter in ASASSN-20pv and hence having a larger contribution) or even irradiation of the disc by the accretor. 

As the temperatures were determined by fitting de-reddened spectra, they could be overestimated in the cases for which we assumed full Galactic reddening. This is the case of ASASSN-15na,  ASASSN-19ct, and ASASSN-20la for which we used $E(B-V)=0.08$,  $E(B-V)=0.10$, and $E(g-r)=0.06$, respectively. 

The difference in values between CVs and AM~CVns can be attributed to the larger contribution of the donor star and accretion disc of the CVs as well as to the fact that AM CVns are expected to exhibit hotter primaries than short-period CVs \citep{2021ApJ...923..125W, 2017MNRAS.466.2855P}
The obtained temperature value for the CVs are well below the ones of the WDs in other SU UMa or WZ Sge with similar periods to our targets \citep{2017MNRAS.466.2855P, 2024AstBu..79..428S}, indicating that the donors and outer parts of the disc are clearly dominating. 

We note that the evolved CV candidate sources form a separate distribution from the emission line CVs and AM~CVns, suggesting to be dominated by the cold components in the binary. The temperature of the candidates is consistent with those determined for other systems \citep{2021MNRAS.508.4106E}.

\subsection{FWHMs and separations of emission lines}

\begin{figure}
   \centering
   \includegraphics[width=0.99\linewidth]{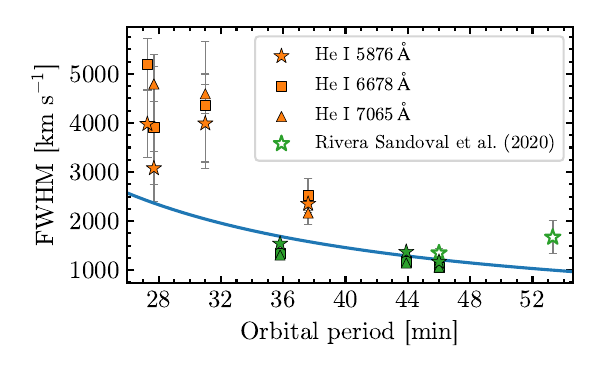}
   \caption{Relation between orbital periods of AM~CVns and FWHMs of their He I emission lines. Orange symbols represent double-peaked lines, green symbols represent single-peaked lines. The blue line represents the best fit of He I 5876 \AA\ data with an exponential function.}
   \label{F:FWHM_Per}
\end{figure}

\cite{2015ApJ...808...80C} found that that FWHMs of emission lines of accreting black holes are related to the radial velocity semi-amplitude of the donor star by relation \begin{equation}
    \label{eq:KvFWHM}
    K_2=0.233 \times \mathrm{FWHM},
\end{equation} where $K_2$ is the radial velocity semi-amplitude. 
\cite{2015ApJ...808...80C} also found similar relation for CVs and showed that FWHM and $K_2$ are tightly correlated for CVs above the period gap while the CVs inside and bellow the period gap show larger dispersion.
Here we decided to test this relation in the case of AM~CVns following a similar approach as 
\cite{2020ApJ...900L..37R}, who used FWHMs of disc emission lines to compare inclinations of systems with known orbital periods. 
Using the mass function \begin{equation} 
    \frac{M_1^3 \, \sin^3\left(i\right)}{\left(M_1 + M_2\right)^2}=\frac{P_\mathrm{orb}  K_2 ^3}{2\pi G}
\end{equation} where $M_1$ and $M_2$ are the masses of the primary and secondary star, $i$ is the inclination, and  $G$ is the gravitational constant, we can express $K_2$ as \begin{equation}
\label{eq:KvsPer}
    K_2=\left( 2 \pi G M_1 \right)^\frac{1}{3} \left(\frac{1}{1+q}\right)^\frac{2}{3} \sin(i)\, P_\mathrm{orb}^{-\frac{1}{3}}
\end{equation} where $q=\frac{M_2}{M_1}$ is the mass ratio.
This shows that $K_2$ is inversely proportional to $P_\mathrm{orb}^{\frac{1}{3}}$ and if a relation analogous to equation~\ref{eq:KvFWHM} holds also for AM~CVns, we can expect that FWHMs are proportional to $P_\mathrm{orb}^{\frac{1}{3}}$.

Figure~\ref{F:FWHM_Per} shows the relation between orbital periods  and FWHMs of He I  emission lines for  AM~CVns from this study and from \cite{2020ApJ...900L..37R}. The diagram shows that the largest FWHMs were measured for systems with short orbital periods, as can be expected from the equation~\ref{eq:KvsPer}. 
From the sample shown in the diagram, ASASSN-21hc with its $P_\mathrm{orb}=35.8\,\mathrm{min}$ shows relatively small FWHM when compared to Gaia21cxj, which has orbital period $P_\mathrm{orb}=37.6\,\mathrm{min}$. 
This suggests that ASASSN-21hc might have smaller inclination than Gaia21cxj, which is consistent with the detection of a single-peaked  profile of its emission lines

We note that while a dependency of FWHMs on orbital periods is present in all studied He I lines, there is as difference in measured FWHMs of the lines within each system. Specifically, $6678\,\mathrm{\text{\AA}}$ and $7065\,\mathrm{\text{\AA}}$ lines  show smaller FWHMs than $5876\,\mathrm{\text{\AA}}$ line in the case of of single-peaked profiles while they tend to be larger in the case double-peaked profiles.

We fitted the data of He I 5876 \AA\ with an exponential function to test the dependency of FWHM on $P_\mathrm{orb}$, the derived dependency is \begin{equation}
    \mathrm{FWHM} \propto P_\mathrm{orb}^{-1.32\pm0.40},
\end{equation} which does not agree with the expected $\mathrm{FWHM} \propto P_\mathrm{orb}^{-\frac{1}{3}}$. This discrepancy could be caused by small number of systems used for fitting as well as dependency on other properties of the systems, such as their inclination or the size of the accretion disc. A larger sample of systems is necessary to fully test the relation. 

ASASSN-20lr, which lacks any estimate of orbital period, shows large FWHMs and also double-peaked profile, which suggests that it might be a short-period system. Its FWHMs and  the separation of the double-peaked profile are similar to the values we derived for Gaia21cxj, which has a known orbital period $P_\mathrm{orb}=37.6\,\mathrm{min}$. Therefore, ASASSN-20lr might have similar orbital period or even shorter, depending on the inclination of the system.

Another AM~CVn which lacks orbital period estimation is ASASSN-20gx, which shows single-peaked emission lines which have the smallest FWHMs from the studied sample. This could be caused by either a long orbital period, a low inclination angle, or a combination of both.

\subsection{Evolved cataclysmic variable candidates}

Three systems from the studied sample, ASASSN-18abl, ASASSN-20la, and TCP J00505644+5351524 do not show spectral features typical for AM~CVn systems or hydrogen-rich CVs. Their common characteristics are spectra reminiscent of late-type stars with slope corresponding to blackbodies of low temperature ($3000\,\mathrm{K}$-$5000\,\mathrm{K}$), lack of emission lines in spectra, and superoutbursts with amplitudes between $3.5$-$4.7$ mags. The long term light curve show only one superoutburst for each system which suggests that their recurrent times are relatively long.

The temperatures of ASASSN-18abl and ASASSN-20la are higher than the typical effective temperatures of secondary components in CVs, for which $T_\mathrm{eff} \lesssim 3900\,\mathrm{K}$ \citep{2011ApJS..194...28K}. This suggests that ASASSN-18abl and ASASSN-20la might be CVs with evolved secondaries. A sample of such systems was studied by \cite{2021MNRAS.508.4106E} who determined the effective temperatures of their secondaries to be between $4700\,\mathrm{K}$ and $<8000\,\mathrm{K}$, which is true also for ASASSN-18abl and ASASSN-20la. Some objects studied by \cite{2021MNRAS.508.4106E} also show similar spectral features as ASASSN-18abl and ASASSN-20la, such as absence of emission lines and presence of absorption lines. Nevertheless, \cite{2023MNRAS.524..740Y} reported that the whole sample of CVs with evolved secondaries shows enhanced sodium abundances, which is difficult to quantify for our sample considering we are not fitting the same model to the Na I line as these authors did. 
However, ASASSN-20la shows very week absorption feature corresponding to the Na I doublet at $\lambda=5892\,\text{\AA}$, and while the doublet is one of the most prominent absorption lines of ASASSN-18abl, its equivalent width $\mathrm{EW=3.4\,\text{\AA}}$ is consistent with solar-like sodium abundances.

One possible explanation of the nature of TCP J00505644+5351524 could be an evolved CV with contamination by a neighbouring star. Gaia catalogue shows a nearby star which is about $1''$ from TCP J00505644+5351524 and about 2 magnitude dimmer. This dimmer star could be the actual transient which was wrongly associated with the brighter target. The spectrum and photometry obtained during quiescence could be then dominated by the light from this brighter target.

\subsection{Periodicity found in the TESS light curves}

Light curves of all four objects with TESS observations 
show detectable periodicity in their Lomb-Scargle periodograms. ASASSN-19ct shows clear superhump variation during the plateau phase of its superoutbursts with period $P_\mathrm{sh}=30.9\pm0.2\,\mathrm{min}$ which is in agreement with the superhump period previously reported for this object. The TESS light curve of this target is similar to the one of SDSS J1043+5632, though, the rebrightenings in ASASSN-19ct are asymmetric and resemble more to those of Gaia-16all \citep{2021MNRAS.508.3275P}. 
The TESS light curve of ASASSN-19ct reveals the whole sequence of echo outbursts for the first time.
This sequence had a duration of around 20 days, around 5 days longer than that of Gaia16-all. No quiescent period between the several echoes is observed. The amplitude and duration of each individual echo outburst is variable, but on average, each individual echo outbursts has a duration of $\sim 1$ day.
\cite{2021A&A...650A.114H} explain a similar behaviour in WZ Sge stars invoking the enhanced-mass-transfer model (EMT). 
Under this scenario, the decay phase of the superoutburst in  ASASSN-19ct would result from a reduced mass transfer rate.  However, the mass transfer from the donor must remain steady over some days to explain the sequence of echo outbursts. 
The lack of quiescent period between echoes would imply that the cooling front moving inwards from the outer disc is reflected by a heating front, thereby maintaining the disc in a hot state. To achieve this, a small inner disc radius would be needed (meaning no disc truncation) and perhaps a massive, hot accretor \citep{2001A&A...373..251D, 2021A&A...650A.114H}

ASASSN-20pv shows a period $P=27.280\pm0.008\,\mathrm{min}$ which was detected from  a light curve obtained in TESS sector 63 in quiescence, during which superhump variation are not expected to occur. Therefore, this period is likely related to orbital variations of this system. While the TESS light curve does not show large variations, this period is close to the reported superhump period $P_\mathrm{sh}=27.5\,\mathrm{min}$ \citep{vsn-a-26565} and therefore the periodicity is very likely real.
We also note that the same period was detected in TESS sector 90, even though at lower significance. If we assume this periodicity is indeed the orbital period, we can use the relations between period excess $\epsilon=\frac{P_\mathrm{sh}-P_\mathrm{orb}}{P_\mathrm{orb}}$ and mass ratio $q=\frac{M_2}{M_1}$ derived by \cite{2019MNRAS.486.5535M} to estimate the mass ratio of the system. If we assume the superhump period relates to stage B superhumps, we get $q_\mathrm{B}=0.044(5)$ and $q_\mathrm{C}=0.052(4)$ for the case of stage C superhumps, both of which are values typical for the known AM~CVns \citep{2025A&A...700A.107G}. If we follow the approach of \cite{2025A&A...700A.107G} and assume the mass of the primary to be  the average mass of CVs $M_1=0.83\pm0.17\,\mathrm{M}_\odot$derived by \cite{2020MNRAS.494.3799P}, we can estimate the mass of the secondary star to be either $M_{2, \mathrm{B}}=0.037(9)\,\mathrm{M}_\odot$ or $M_{2, \mathrm{C}}=0.043(9)\,\mathrm{M}_\odot$.

\section{Conclusion}

We used spectroscopic observations from Gemini to characterise a sample of 15 transients. Two targets had previous spectroscopic observations, but we further explore their properties during quiescence. Eight AM CVns were spectroscopically analysed for the first time, which represent increases of the number of spectroscopically analysed AM CVns by about 10\%. 

We analysed TESS photometry of the targets which shows outbursting activity of one AM CVn star, ASASSN-19ct and two CVs (ASASSN-15na and ZTF18aaaasnn.)
We determined the orbital period for one AM CVn for which only the superhump period was known. We also determined the superhump period for one CV with no previous periodicity known. 

Six of the AM~CVns from our sample show X-ray emission. We estimated their X-ray luminosities and mass accretion rates which agree with those reported for other AM CVn systems. The mass accretion rates are above $\dot{M}_\mathrm{crit}^-$ predicted by the accretion disc models, and they are therefore in agreement with expectations from the disc instability model and the observed transient behaviour. However, due to limitation of the analysis, they can serve only as lower limits of the actual mass accretion rates.

We can summarise our findings as follows:
\begin{itemize}
\item 
Four of the 15 targets were previously classified as AM~CVns based on their photometric properties. Here we provide the optical spectrum that fully confirms their nature. They show spectral features typical for AM~CVn stars, namely He I emission lines and absence of H lines.

\item 
We identified four new members of AM CVn stars using Gemini spectra.

\item 
Spectra of 3 of the candidates show H emission lines which classifies them as H-rich CVs.  

\item 
Spectra of 3 of the candidates did not show any signs of emission lines typical for accretion discs. Further, the spectra show slopes corresponding to relatively cold blackbody radiation ($3000\,\mathrm{K}$-$5000\,\mathrm{K}$) which disfavours the presence of a hot WD as a dominant component in these systems. We discard a symbiotic nature based on the lack of emission lines in their spectra. 
None of these three systems is a known X-ray source, which disfavours a magnetic CV, symbiotic nature, and accreting neutron star or black hole. 
The presence of outbursts and their spectra suggests that these systems might be CVs with evolved secondaries.

\item 
We found brightness variation in the TESS light curve of ASASSN-20pv with period $P=27.280(8)\,\mathrm{min}$. As this period was derived from a quiescent light curve, it is most likely an orbital period of this system. We used this period to estimate the mass ratio of the system $q$, which is between $0.044$ and $0.052$. 

\item 
The TESS light curve of ZTF18aaaasnn obtained during a superoutburst shows a brightness variation with period $P=91.2(2)\,\mathrm{min}$ which is most likely a superhump period.

\item 
The amplitudes of  the AM~CVn superoutbursts studied in this work are larger than predicted for a given orbital period by previous studies. Our sample confirms that systems with larger orbital periods tend to have superoutbursts with larger amplitudes. However, there is a large dispersion for a given orbital period.

\item 
The superoutburst duration of the studied AM~CVns range from 5 - 10 days, they show small dispersion in values and do not show strong dependency on the orbital period. This is in good agreement with theoretical predictions from the disc instability model and with durations determined by previous studies with TESS.

\item 
The widths of emission lines of AM~CVns show dependency on the orbital period, with the short-period system having larger widths and double-peaked profiles. 

\item 
The temperatures derived by fitting the spectra with a black-body model shows that AM~CVns tend to exhibit larger effective temperatures than CVs. However, the derived temperatures are lower than predicted for the accretors, as detections in the optical range are affected by contributions from other components, such as the accretion disc.

\item
We increased the number of know AM~CVn stars which brings us closer to understanding of these accreting systems.

\end{itemize}

Selection of AM~CVn candidates based on their photometric properties and subsequent follow-up analysis using low-resolution spectroscopy proved to be a viable method for identification of new AM~CVn systems. AM~CVn studied in this paper show colour indices $g-r < -0.11$, $r-i < 0.06$, and $\mathrm{BP} - \mathrm{RP}<0.5$, which applies also to majority of known AM~CVn systems. 
The Gaia magnitudes and colours of majority of AM~CVns lie in the ranges $5.5< \;M_G<13$ and -0.3 $<\mathrm{BP} - \mathrm{RP} < 0.5$.
Analysis of superoutburst properties revealed that the rise and decline rates of AM CVns have values for which $\mu_\mathrm{r} \gtrsim 2\,\mathrm{mag}\,\mathrm{d}^{-1}$ and $\mu_\mathrm{d} \gtrsim 1\,\mathrm{mag}\,\mathrm{d}^{-1}$, respectively.  While short period CVs of SU~UMa, WZ~Sge, and ER~UMa types can show similar values, they show larger spread with rise and decline rates extending down to $\mu_\mathrm{r} \sim 0.4\,\mathrm{mag}\,\mathrm{d}^{-1}$ and $\mu_\mathrm{d} \sim 0.2\,\mathrm{mag}\,\mathrm{d}^{-1}$
Using these parameters for selection of AM~CVn candidates has a large potential for identification of new systems using all-sky surveys such as LSST.

\section*{Acknowledgments}
We are grateful to the anonymous referee for providing
us with useful comments and suggestions that improved our manuscript.
JK and LRS acknowledge support from NASA grants NNH22ZDA001N-6152 and 80NSSC24K0638. MPM is partially supported by the Swiss National Science Foundation IZSTZ0\_216537 and by UNAM PAPIIT-IG101224. Based on observations obtained at the international Gemini Observatory, a program of NSF NOIRLab, which is managed by the Association of Universities for Research in Astronomy (AURA) under a cooperative agreement with the U.S. National Science Foundation on behalf of the Gemini Observatory partnership: the U.S. National Science Foundation (United States), National Research Council (Canada), Agencia Nacional de Investigaci\'{o}n y Desarrollo (Chile), Ministerio de Ciencia, Tecnolog\'{i}a e Innovaci\'{o}n (Argentina), Minist\'{e}rio da Ci\^{e}ncia, Tecnologia, Inova\c{c}\~{o}es e Comunica\c{c}\~{o}es (Brazil), and Korea Astronomy and Space Science Institute (Republic of Korea). 
The Gemini data were obtained from programs GN-2023B-Q-310 and GS-2024A-Q-311 (PI: Rivera Sandoval) and processed using DRAGONS (Data Reduction for Astronomy from Gemini Observatory North and South)
The Digitized Sky Surveys were produced at the Space Telescope Science Institute under U.S. Government grant NAG W-2166. The images of these surveys are based on photographic data obtained using the Oschin Schmidt Telescope on Palomar Mountain and the UK Schmidt Telescope. The plates were processed into the present compressed digital form with the permission of these institutions. The National Geographic Society - Palomar Observatory Sky Atlas (POSS-I) was made by the California Institute of Technology with grants from the National Geographic Society. The Second Palomar Observatory Sky Survey (POSS-II) was made by the California Institute of Technology with funds from the National Science Foundation, the National Geographic Society, the Sloan Foundation, the Samuel Oschin Foundation, and the Eastman Kodak Corporation. The Oschin Schmidt Telescope is operated by the California Institute of Technology and Palomar Observatory. The UK Schmidt Telescope was operated by the Royal Observatory Edinburgh, with funding from the UK Science and Engineering Research Council (later the UK Particle Physics and Astronomy Research Council), until 1988 June, and thereafter by the Anglo-Australian Observatory. The blue plates of the southern Sky Atlas and its Equatorial Extension (together known as the SERC-J), as well as the Equatorial Red (ER), and the Second Epoch [red] Survey (SES) were all taken with the UK Schmidt. Supplemental funding for sky-survey work at the ST ScI is provided by the European Southern Observatory.
Based on observations obtained with the Samuel Oschin Telescope 48-inch and the 60-inch Telescope at the Palomar Observatory as part of the Zwicky Transient Facility project. ZTF is supported by the National Science Foundation under Grants No. AST-1440341 and AST-2034437 and a collaboration including current partners Caltech, IPAC, the Oskar Klein Center at Stockholm University, the University of Maryland, University of California, Berkeley, the University of Wisconsin at Milwaukee, University of Warwick, Ruhr University, Cornell University, Northwestern University, and Drexel University. Operations are conducted by COO, IPAC, and UW.
This work has used data from the European Space Agency (ESA) mission
{\it Gaia} (\url{https://www.cosmos.esa.int/gaia}), processed by the Gaia Data Processing and Analysis Consortium (DPAC, \url{https://www. cosmos.esa.int/web/gaia/dpac/consortium}). 
Funding for the DPAC has been provided by national institutions, 
in particular, the institutions participating in the Gaia Multilateral Agreement.
We acknowledge with thanks the variable star observations from the AAVSO International Database contributed by observers worldwide and used in this research.
This paper includes data collected by the TESS mission.
Funding for the TESS mission is provided by the NASA Science Mission Directorate. 
Some of the data presented in this paper were obtained from the B. Mikulski Archive for Space Telescopes (MAST).
This research has made use of the SIMBAD database,
operated at CDS, Strasbourg, France.
This research has made use of "Aladin sky atlas" developed at CDS, Strasbourg Observatory, France.
This research has made use of the VizieR catalogue access tool, CDS,
 Strasbourg, France.

\bibliographystyle{aa_link}
\bibliography{References.bib}

\appendix
\counterwithin{figure}{section}

\section{Individual systems}
\label{AP:TARGETS}

\subsection{AM~CVn systems}
\subsubsection{ASASSN-15kf}

ASASSN-15kf was discovered during its superoutburst in 2015 by the ASAS-SN survey and was classified as an AM~CVn star based on its superhump period $P_{\rm sh}=27.7\,{\rm min}$ reported by \cite{vsn-a-18669}. \cite{2025A&A...700A.107G} includes it in their catalogue as a confirmed AM~CVn star due to its short period, however, no spectroscopic observations have been published for this system prior to this study. 

We analysed the long-term light curve presented in 
Figure~\ref{F:LC:LT_ALL},
which shows that
the first observed superoutburst occurred on $\mathrm{MJD}\;57\,169$ but the scarce photometric coverage does not allow us to estimate its duration. Another superoutburst occurred on $\mathrm{MJD}\;57\,794$ and lasted at least 8 days. It is difficult to determine the occurrence and properties of other superoutbursts due to scarce coverage of the light curve. One superoutburst likely occurred on $\mathrm{MJD}\;59\,254$ and was partly covered by the ASAS-SN survey. Another such instance occurred on $\mathrm{MJD}\;60\,340$, when ASAS-SN and ATLAS data show variations resembling a superoutburst followed by multiple rebrightenings. 
The first two detected superoutbursts imply a recurrence time is about $600$ d, however, the true value could be even shorter, as suggested by \cite{2025PASJ..tmp...92K} who list a 370 d recurrence time for this object.

The spectrum of ASASSN-15kf shows several He I emission lines with a double-peaked profiles and is devoid of any detectable hydrogen lines, confirming the classification of this target as an AM~CVn star. The spectrum also shows blends of N I and Mg II lines. The flux-calibrated spectrum shown in 
Figure~\ref{F:SPEC_ALL}
shows a blue continuum. The best black-body fit gives a temperature $T_\mathrm{BB}=10\,324\pm86 \,\mathrm{K}$.

\subsubsection{ASASSN-17fp}

ASASSN-17fp was discovered by the ASAS-SN survey in April 2017 during its superoutburst with peak magnitude $V=15.7$. \cite{2017ATel10334....1C} obtained a spectrum during the outburst and classified this object as a potential AM~CVn system due to the presence of He I absorption lines lines and absence of any hydrogen lines. \cite{2017ATel10354....1M} obtained multi-band photometric observation during the outburst and reported periodic variations with a period $P=51.0(1)\,{\rm min}$, which puts ASASSN-17fp among the long-period AM~CVn systems. 

The long-term light curve in 
Figure~\ref{F:LC:LT_ALL}
which we constructed from ground-based photometry shows only one superoutburst. It was first detected on $\mathrm{MJD}\;57\,871$ and its maximal brightness $V=15.7$ mags was observed on $\mathrm{MJD}\;57\,784$. The duration of the superoutburst was about 10 days. The long-term light curve shows one rebrightening which occurred on $\mathrm{MJD}\;57\,889$. Subsequent photometry consists of only occasional observations from ASAS-SN and ATLAS surveys which are too scarce for analysis of outbursting activity. If we assume that the superoutbursts observed in April 2017 was the most recent one, we can put a lower limit on the superoutburst recurrence time, which needs to be larger than about 8 years. 

The target's brightness during the Gemini spectroscopic observations proved to be too low for acquisition of usable spectra.

\subsubsection{ASASSN-18rg}

ASASSN-18rg was discovered by the ASAS-SN survey in 2018 during a superoutburst. This superoutburst lasted for about 10 days and was followed by a second superoutburst and multiple rebrightenings. \cite{vsn-a-22496} reported superhumps with amplitude of $0.1\,\mathrm{mag}$ and period $P_\mathrm{sh}=49\,\mathrm{min}$, making this system a candidate AM~CVn star. 

The long-term light curve presented in 
Figure~\ref{F:LC:LT_ALL}
shows that both superoutbursts are covered by photometry from ZTF, ASAS-SN, and AAVSO surveys. 
The first superoutburst started on $\mathrm{MJD}\;58\,337$ and peaked on $\mathrm{MJD}\;58\,338$ at magnitude $g=12.5$. The system stayed in plateau phase until $\mathrm{MJD}\;58\,346$ and the superoutburst ended on $\mathrm{MJD}\;58\,347$. Its amplitude was 7.9 magnitudes and its total duration was 10 days. The second superoutburst started on $\mathrm{MJD}\;58\,365$ and peaked on $\mathrm{MJD}\;58\,366$ at magnitude $g=13.7$. It also showed plateau phase which lasted until $\mathrm{MJD}\;58\,371$ and the superoutburst ended on $\mathrm{MJD}\;58\,372$. The second superoutburst was followed by a fading tail with multiple rebrightenings.
The system appears to be in quiescence since then, which implies a superoutburst recurrence time longer than 6 years.

The Gemini spectrum of this target shows strong single-peaked helium emission lines and no hydrogen lines, confirming the AM~CVn classification of this object. The system also shows blends of N I and Mg II lines. The flux-calibrated spectrum presented in 
Figure~\ref{F:SPEC_ALL}
shows blue continuum with slope corresponding to a black-body model with temperature $T_\mathrm{BB}=10\,085\pm177 \,\mathrm{K}$.

\subsubsection{ASASSN-19ct}

While ASASSN-19ct was discovered during its normal outburst in 2019 by the ASAS-SN survey, its long term light curve revealed multiple superoutbursts which occur with recurrence time of about one year \citep{vsn-a-22996}. A superoutburst, which occurred shortly after the system's discovery, showed a superhump variation with periodicity of $P_\mathrm{sh}=31\,\mathrm{min}$ \citep{vsn-a-23109}. The system was classified as an AM~CVn star based on this short periodicity and the presence of outbursts, however, no spectroscopic observations were available for this target previous to this work. 

The long-term light curve which we constructed from available ASAS-SN, AAVSO, and ATLAS photometry is presented in 
Figure~\ref{F:LC:LT_ALL}.
It spans about 13 years and shows 14 superoutburst, however, the light curve shows several gaps during which other superoutbursts could occur. This implies that the average recurrence time of superoutbursts is likely shorter than one year with the shortest time separation of two consecutive superoutburst of about 180 days.
\cite{2025PASJ..tmp...92K} list a recurrence time of 1 year for this target. Our light curve shows that the superoutburst typically lasts for 7 days, has amplitude of 4 magnitudes, reaches a peak magnitude $g=13.3$, and is then followed by a fading tail with multiple rebrightenings.

ASASSN-19ct was observed by TESS in five different sectors (see Table~\ref{tab:tesstargetslist_v2} for details), two of which show outbursting activity.  
The light curve obtained in Sector 10 (Figure~\ref{TESS:LC}, top), shows a superoutburst event, its onset is, however, not covered. The light curve exhibits an incomplete plateau phase between $\mathrm{MJD}\;58\,570$ - $58\,572$, which is followed by a rapid decay ($\mathrm{MJD}\;58\,572.354$ - $58\,573.23$) and then multiple rebrightenings occurring between $\mathrm{MJD}\;58\,573.23$ - $58\,593$. 
The rebrightenings exhibit an asymmetric shape with steep rise followed by a slower decline, which has been observed in other AM~CVns \citep{2021MNRAS.508.3275P}. Several rebrightenings between $\mathrm{MJD}\;58\,585$ - $58\,589$ show short plateaus upon reaching their maximal brightness. We also note that the system's brightness between individual rebrightenings was varying throughout the rebrightening phase and we found the same behaviour in the corresponding section of the long-term light curve based on ground-based photometry. The cadence of Sector 10 light curve is 30 min, which is very close to the reported superhump period $P_\mathrm{sh}=31\,\mathrm{min}$ and it is therefore not suitable for period analysis of superhump variations.

In TESS Sector 37, the system was initially detected in quiescence before experiencing a normal outburst with duration of 1.2 d starting on  $\mathrm{MJD}\;59\,325$. It was followed by a superoutburst which started on $\mathrm{MJD}\;59\,330$, and reached its peak brightness on $\mathrm{MJD}\;59\,331$.
The subsequent plateau phase is covered only partly, but it exhibits clear superhump variations with period $P_\mathrm{sh}=30.94\pm0.21$ min.
The outburst preceding the superoutburst is in fact a precursor outbursts, a similar feature was observed in CVs \cite[e.g.][]{2014PASJ...66...15O} as well as in other AM CVns \citep[e.g.][]{2021MNRAS.502.4953D}. 
For CVs with low mass ratios, \cite{2024A&A...689A.354J} explain these delayed superoutbursts as the result of a tidal instability that becomes active when a cooling wave propagates inwards, leading to the growth of the disc’s eccentricity.
An analogous mechanism could explain the same behaviour in AM CVns, as they are typically low-mass ratio systems.
Figure~\ref{F:tess_periodograms} shows a phase-folded light curve of the plateau phase and its Lomb-Scargle power spectrum.

The Gemini spectrum of ASASSN-19ct shows broad double-peaked He I emission lines confirming the AM~CVn classification. The spectrum further shows blends of N I, Mg I, and Mg II emission lines. The flux-calibrated spectrum in 
Figure~\ref{F:SPEC_ALL}
exhibits a continuum flux dominated at shorter wavelengths. The black-body fit of the continuum gives its temperature $T_\mathrm{BB}=14\,977\pm167 \,\mathrm{K}$ which is the largest among the targets in our sample. However, it is possible that the the temperature determination was affected by overestimation of the reddening, as the reddening correction of this target was based on the full Galactic reddening $E(B-V)=0.10$ given by \cite{1998ApJ...500..525S}. The target's distance $d  =238 \,\mathrm{pc}$, determined from the Gaia parallax, suggests that the actual reddening might be lower as well as the black-body temperature. The black-body temperature derived from the spectrum before dereddening was applied is $T_\mathrm{BB}=11\,559\pm99 \,\mathrm{K}$.

\subsubsection{ASASSN-19rg}

ASASSN-19rg was discovered by the ASAS-SN survey in 2019 during a superoutburst. The photometric observations obtained during the superoutburst showed variation with a period of either $P=109\,\mathrm{min}$ or its half \citep{vsn-a-23399}, later observations showed superhump variations with periodicity $P_\mathrm{sh}=44\,\mathrm{min}$ \citep{vsn-a-23432}, labelling this system as an AM~CVn candidate.

We constructed a long-term light curve 
(Figure~\ref{F:LC:LT_ALL})
which shows only one active period.
The initial superoutburst of this period started on $\mathrm{MJD}\;58\,669$ and reached its peak magnitude $g=13$ on $\mathrm{MJD}\;58\,670$, its amplitude was 7 magnitudes. The light curve shows a plateau phase which lasted until $\mathrm{MJD}\;58\,676$ and the superoutburst ended on $\mathrm{MJD}\;58\,678$. It was followed by a second superoutburst which started on $\mathrm{MJD}\;58\,686$, ended on $\mathrm{MJD}\;58\,692$, and was followed by a fading tail. As no other superoutburst was detected in the following observation, we can estimate a lower limit of 5 years in the superoutburst recurrence time.

The spectrum obtained at the Gemini observatory shows strong single-peaked He I emission lines, blends of N I emission lines, and Si II and Mg II emission lines. The spectrum does not show any signs of hydrogen lines, which classifies this object as an AM~CVn system.
The flux-calibrated spectrum shown in 
Figure~\ref{F:SPEC_ALL}
shows blue continuum which corresponds to a black-body model with a temperature $T_\mathrm{BB}=13\,456\pm212 \,\mathrm{K}$. The Gaia distance $d=1160_{-374}^{+518}\,\mathrm{pc}$ has large uncertainties which could affect the determination of the reddening of this targets and the subsequent temperature estimate. However, the three-dimensional map of \cite{2019ApJ...887...93G} predicts the same reddening $E(g-r)=0.05$ for all distances $d > 160\,\mathrm{pc}$ and therefore serves as a good estimation of reddening of ASASSN-19rg.

\subsubsection{ASASSN-20gx}

ASASSN-20gx was discovered during a superoutburst in June 2020 on $\mathrm{MJD}\;59\,016$ by the ASAS-SN survey. The available photometric observations do not cover the onset of the superoutburst, only the subsequent fading with several rebrightenings, which lasted for about 120 days.
\cite{2021PASJ...73.1375K} classified this target as an AM~CVn star based on its blue colour and rapid fading observed during rebrightenings. This system underwent an outburst also in December 2011 \citep{vsn-a-25860} which was observed by the Catalina Sky Survey. 

The long-term light curve presented in 
Figure~\ref{F:LC:LT_ALL}
which we constructed from available photometry shows that
another superoutburst occurred in June 2023, which suggests that the recurrence time of superoutbursts is about 3 years. The superoutbursts started on $\mathrm{MJD}\;60\,121$, peaked on $\mathrm{MJD}\;60\,123$ at magnitude $g=14.4$ with amplitude of 5.9 magnitudes, and lasted for about 7 days. It was followed by a fading tail which lasted for about 90 days.

The Gemini spectrum of this target shows strong single-peaked He I emission lines, Si II emission lines, and blends of Mg I, Mg II, and N I emission lines. Hydrogen lines are not present in the spectrum, which classifies this system as an AM~CVn star. The flux calibrated spectrum presented in 
Figure~\ref{F:SPEC_ALL}
shows a blue continuum for which we determined a black-body temperature $T_\mathrm{BB}=8356\pm103 \,\mathrm{K}$.

\subsubsection{ASASSN-20lr}

This object was discovered during its superoutburst in September 2020 by the ASAS-SN survey. The superoutburst occurred on $\mathrm{MJD}\;59\,102$, lasted for 6 days, had amplitude of 5.4 magnitudes with peak brightness $g=14.7$ and was followed by a fading tail with multiple rebrightenings. The system reached the quiescence level about 100 days after the onset of the superoutburst. \cite{2021PASJ...73.1375K} classified it as an AM~CVn candidate based on the short duration of the superoutburst and rapid fading of the observed rebrightenings. 

The long-term light curve presented in 
Figure~\ref{F:LC:LT_ALL}
does not show any additional superoutburst. The system remains in quiescence for  about 1400 days which suggests the superoutburst recurrence time is longer than $\sim 4$ years.

The spectrum of ASASSN-20lr shows multiple He I emission lines two of which (6679 \AA\ and 7283 \AA) show double-peaked profile. The spectrum also shows blends of Mg I, Mg II, and N I emission lines. Hydrogen lines are not present in the spectrum which confirms the classification of this target as an AM~CVn system. 
Figure~\ref{F:SPEC_ALL}
shows the flux-calibrated spectrum which exhibits blue continuum. The best black-body model fit gives temperature $T_\mathrm{BB}=9824\pm133 \,\mathrm{K}$.

\subsubsection{ASASSN-20pv}

ASASSN-20pv was discovered by the ASAS-SN survey in December 2020 during its superoutburst on $\mathrm{MJD}\;59\,198$. \cite{vsn-a-25144} determined the superhump period of $P_\mathrm{sh}=27.8\,\mathrm{min}$ based on AAVSO observations and classified the system as an AM~CVn candidate.

The long-term light curve which we constructed from ground-based data covers the time period between $\mathrm{MJD}\;57\,422$ and $\mathrm{MJD}\;60\,884$. While no superoutburst was detected during about 1800 days preceding the system's discovery, the light curve
clearly shows three additional superoutbursts which occurred on $\mathrm{MJDs}\;59\,599$, $60\,277$, and $60\,805$. The average recurrence time of the superoutbursts is 1.5 years. All of the recorded superoutbursts lasted about 10 days, had peak magnitude $g=12.3$ and were followed by a fading tail with multiple rebrightenings. The duration of the fading tails was between 30 and 45 days, apart from the one observed in 2022, which lasted for 180 days.

ASASSN-20pv was observed in 6 TESS sectors (see Table~\ref{tab:tesstargetslist_v2} for details) but it does not show outbursting activity in any of them. 
The power spectrum of the light curve from TESS Sector 63, presented in Figure~\ref{F:tess_periodograms}, shows a strong peak at period $\mathrm{P}=27.280\pm0.008\;\mathrm{min}$ well above the $0.001\%$ FAP. The power spectrum also shows a weaker peak corresponding to half of this period.  Given the fact that the light curve was obtained during quiescence and shows double-wave modulation, this periodicity is most likely caused by orbital variations. We detected the same periodicity in the TESS light curve from Sector 90  (see Figure~\ref{F:tess_periodograms}), even though at lower significance.

The Gemini spectrum shows He I emission lines. All helium lines detected in the spectrum show double-peaked profiles, but only the strongest one (He I 5877 \AA) is prominent enough for fitting a two-component Gaussian to determine the peak separation. The spectrum also shows blends of N I and Mg II emission lines and it does not show any hydrogen lines, confirming the classification of this system as an AM~CVn star. The flux-calibrated spectrum in 
Figure~\ref{F:SPEC_ALL}
shows a clear blue spectrum with a slope corresponding to a black-body model of temperature $T_\mathrm{BB}=11\,884\pm143 \,\mathrm{K}$.

\subsubsection{ASASSN-21hc}

This target was discovered in May 2021 by ASAS-SN survey during its superoutburst. \cite{vsn-a-25849} classified it as an AM~CVn candidate based on its rapid fading. \cite{vsn-a-25868}  determined a superhump period $P_\mathrm{sh}=35.8\,\mathrm{min}$ from AAVSO observations. 

The long-term light curve shown in 
Figure~\ref{F:LC:LT_ALL}
reveals only one activity period which started with a superoutburst.
The superoutburst occurred on $\mathrm{MJD}\;59\,338$ and its plateau phase lasted until $\mathrm{MJD}\;59\,344$
The whole superoutburst lasted for 6 days, had peak magnitude $g=13.3$, and was immediately followed by a second superoutburst on $\mathrm{MJD}\;59\,345$ which lasted for around 4 days. The second superoutburst was then followed by a fading tail with multiple rebrightenings.
The fact that only one activity period was covered by the long-term light curve suggests that the superoutburst recurrence time of this system is long. Using the quiescence period preceding the first superoutburst, we can estimate the recurrence time to be longer than $\sim 1800$ days.

The spectrum of ASASSN-21hc shows strong single-peaked He I emission lines, Si II emission lines, and blends of N I and Mg II emission lines. Any hydrogen emission lines are absent in the spectra which confirms the system's classification as an AM~CVn star.
Figure~\ref{F:SPEC_ALL}
shows the flux-calibrated spectrum of ASASSN-21hc which exhibits blue continuum. The best fit of the continuum with a black-body model gives a temperature $T_\mathrm{BB}=11\,606\pm130 \,\mathrm{K}$.

\subsubsection{Gaia21cxj (V744 And)}

Gaia21cxj (also known as V744 And, or SDSSJ0129+3842) is a known AM~CVn star which was discovered by \cite{2005AJ....130.2230A} in the SDSS spectral database based on its spectral features typical for an AM~CVn system -- He I emission lines and absence of hydrogen lines. \cite{2013MNRAS.432.2048K} obtained phase-resolved spectroscopy of this system and determined the orbital period $P_\mathrm{orb}=37.6\,\mathrm{min}$.

Figure~\ref{F:LC:LT_ALL},
second panel, shows the long-term light curve we constructed using the available ground based photometry.
It shows two partly covered superoutburst on $\mathrm{MJD}\;55\,165$ and $\mathrm{MJD}\;57\,026$. Another superoutburst occurred on $\mathrm{MJD}\;59\,382$, peaked at magnitude $g=14.1$, lasted for at least 5 days, and was followed by multiple rebrightenings. 
The average superoutburst recurrence time derived from the three observed superoutbursts is about 2100 days.  \cite{2025PASJ..tmp...92K} lists a recurrence time of 1250 days which is much shorter to the one we determined. Since no details are given by these authors, we assume that their value is based on additional private data.

The Gemini spectrum shows double-peaked He I emission lines and emission line blends of Mg I, Mg II, and N I. The flux-calibrated spectrum which we present in 
Figure~\ref{F:SPEC_ALL}
shows a continuum with a large blue excess which makes the spectrum difficult to model by a single black-body. Because of this, we fitted only the wavelengths larger than $6500\,\mathrm{\text{\AA}}$, for which a suitable fit can be obtained. The derived temperature is $T_\mathrm{BB}=8080\pm137 \,\mathrm{K}$. The temperature is low as it does not reflect the blue excess. Fitting such blue part of the spectrum  would require a more complex model.

\begin{figure*}
    \includegraphics[width=0.49\linewidth]{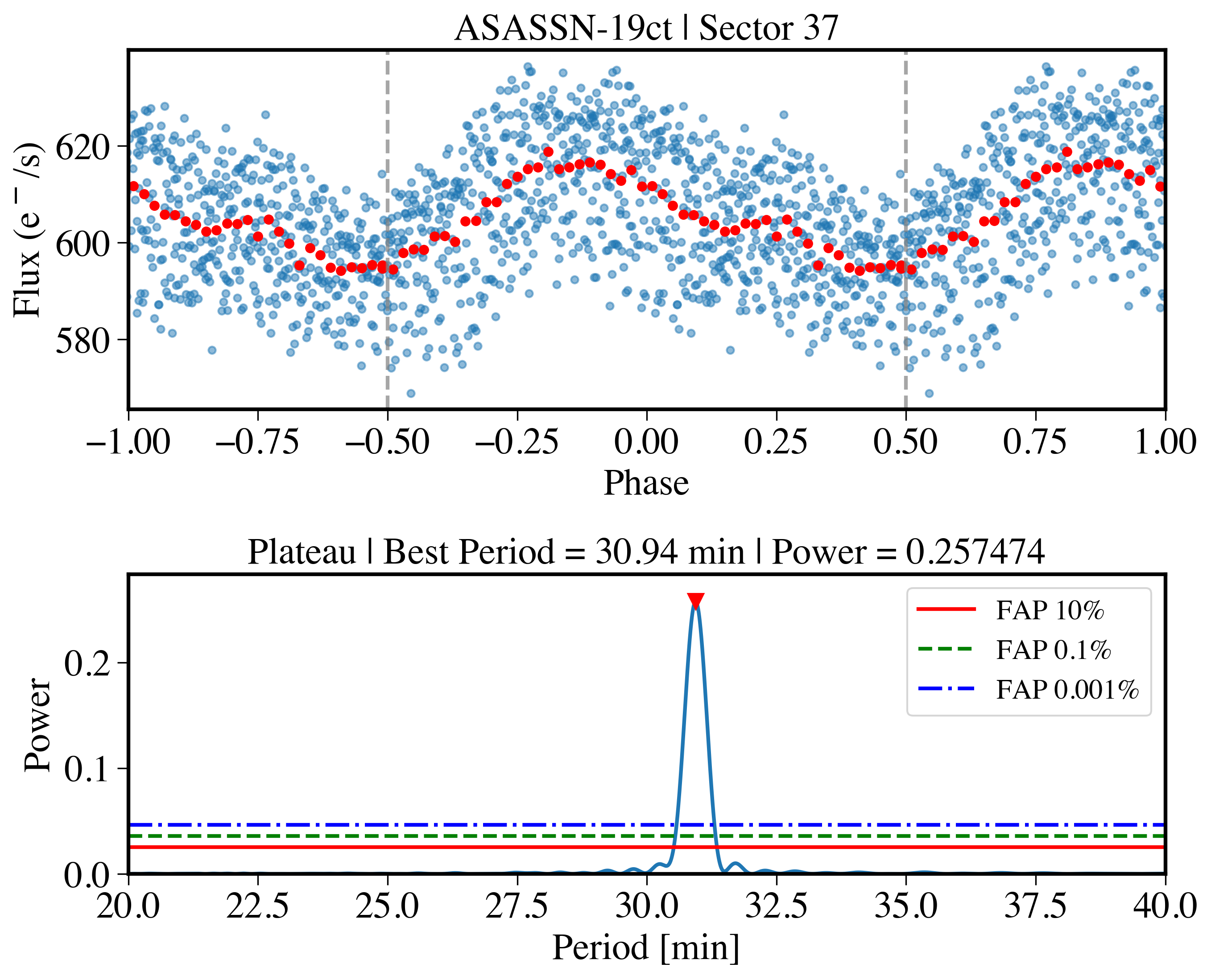}
    \includegraphics[width=0.49\linewidth]{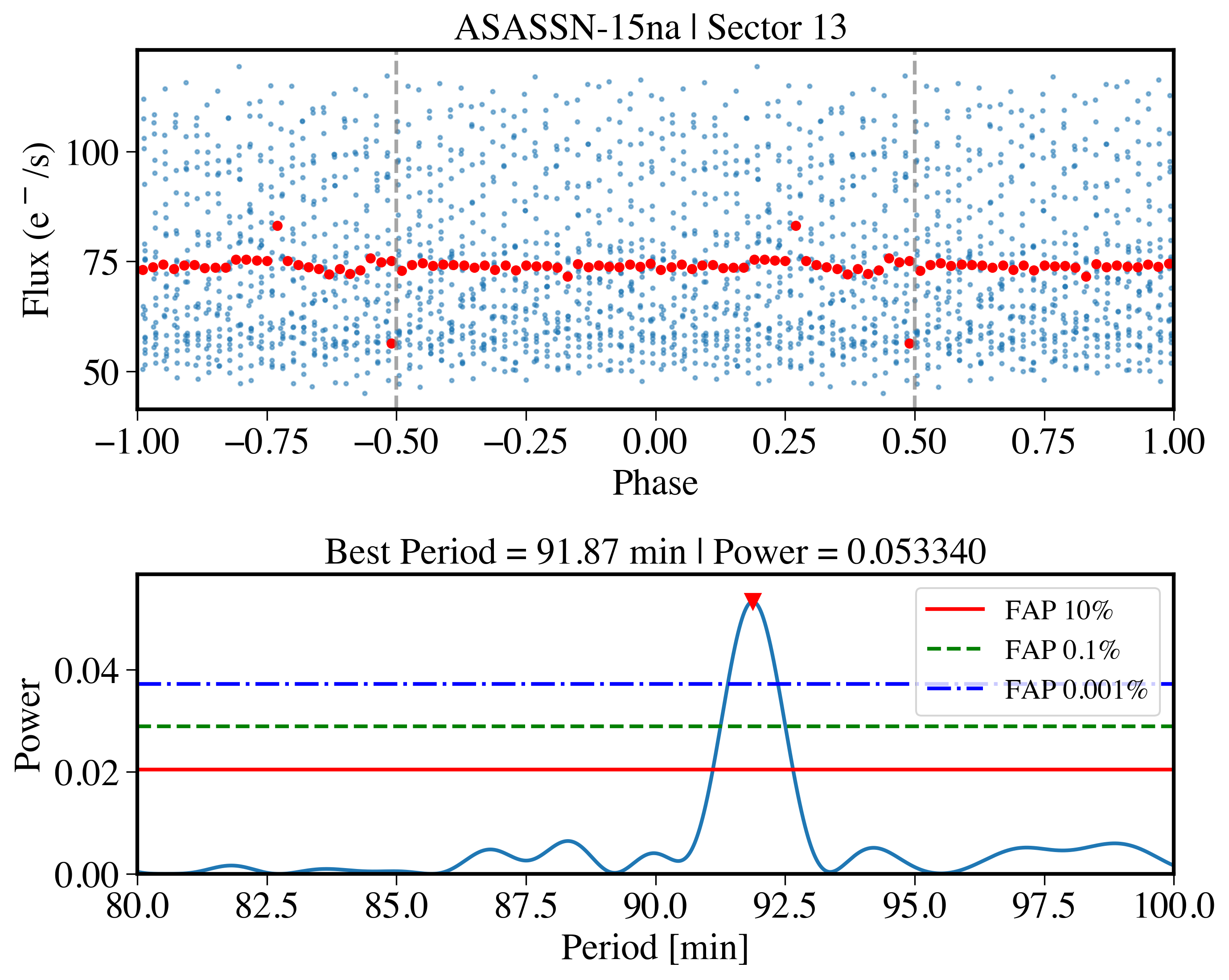}
    
    \includegraphics[width=0.49\linewidth]{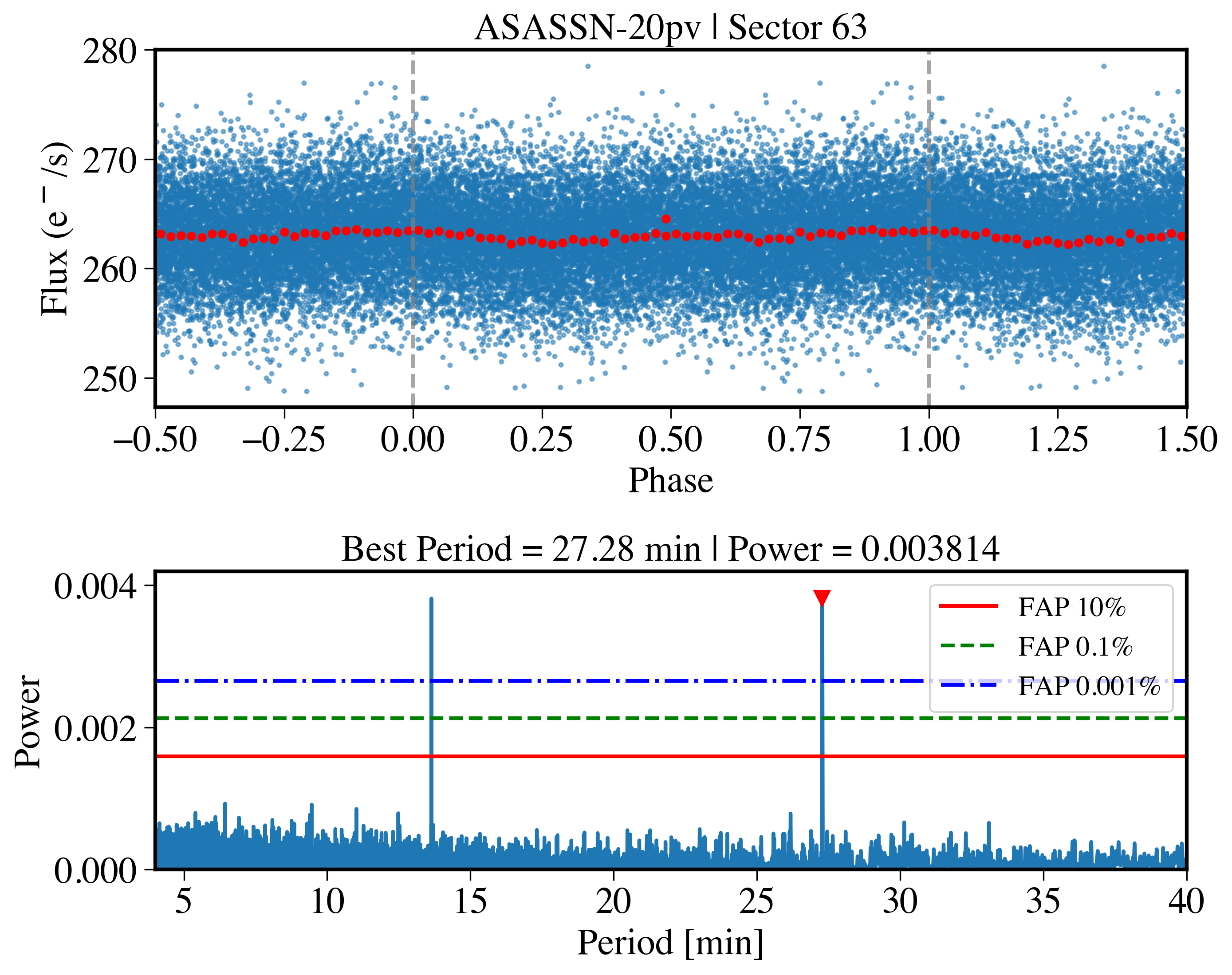}
    \includegraphics[width=0.49\linewidth]{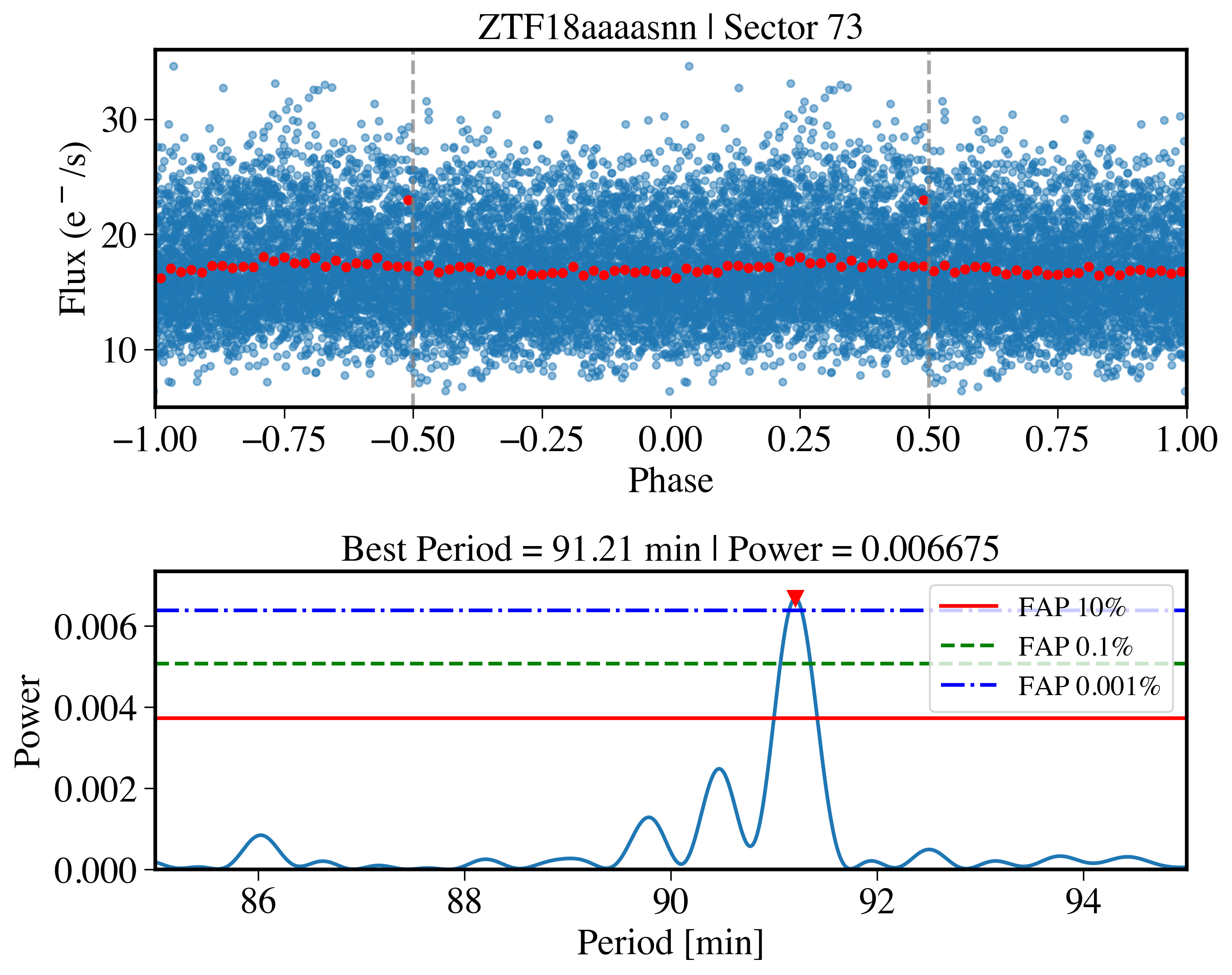}
    
    \includegraphics[width=0.49\linewidth]{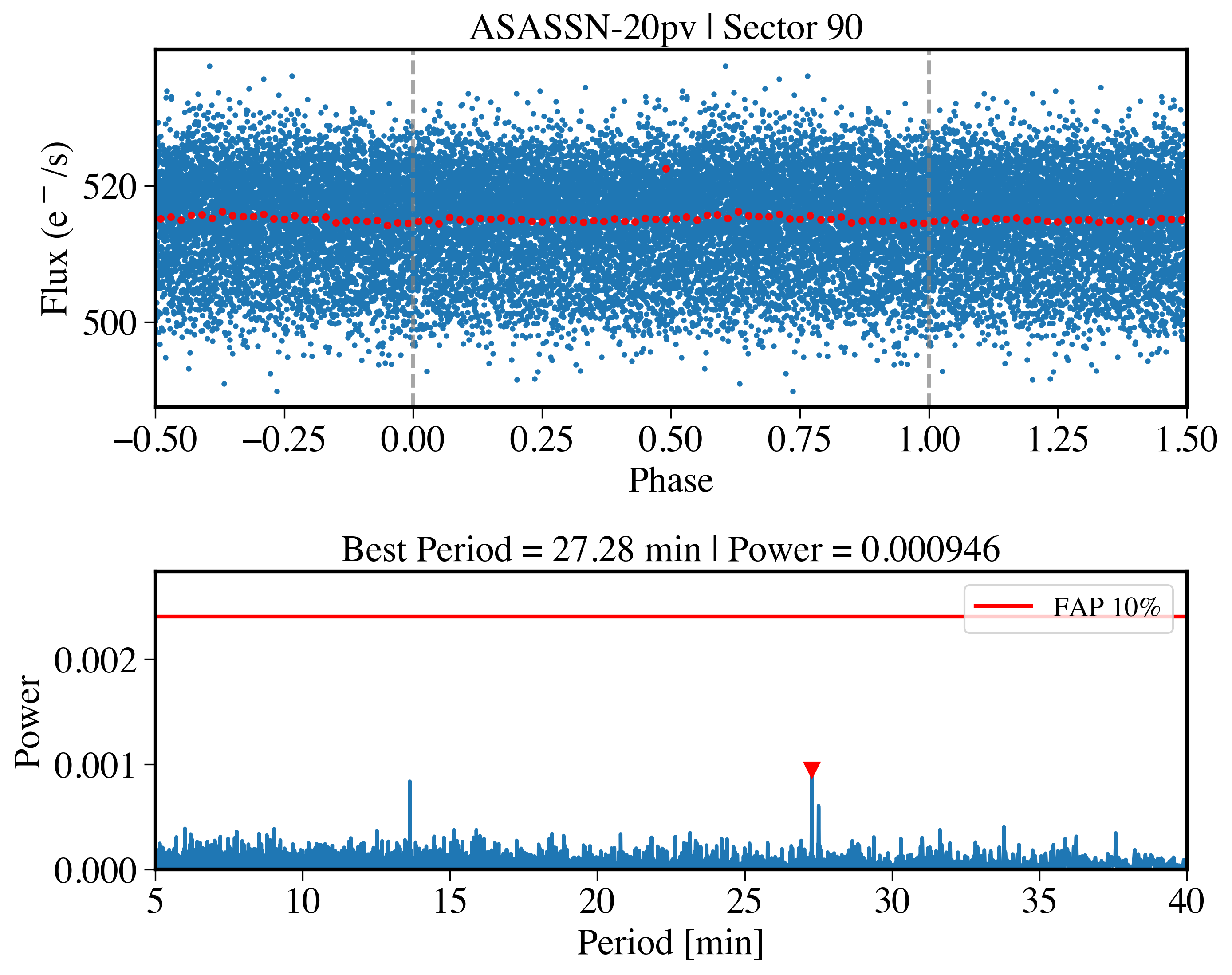}

    \caption{Phase folded light curves and periodograms of TESS observations. The red circles show binned light curve with 50 bins per phase. Both periodograms of ASASSN-20pv show peaks corresponding to the first and second harmonics of orbital variations. The large dispersion in light curve of ASASSN-15na is primarily caused by change of the system's brightness during the observed superoutburst, contamination by a nearby eclipsing binary produces variation with amplitude of about $10 \;e^{-}/s$. }
    \label{F:tess_periodograms}
\end{figure*}

\subsection{Cataclysmic variables}
\subsubsection{ASASSN-15na}

This system was discovered by the ASAS-SN survey in 2015 on $\mathrm{MJD}\;57\,223$ and \cite{vsn-a-18884} reported on a double wave modulation with amplitude of 0.1 mag and period of 91 min. It was suggested to be an AM~CVn system due to an apparent rapid fading and short superhump period of $P_{\rm sh}=45\,{\rm min}$ \citep{vsn-a-18899, vsn-a-18910}, but these findings were later corrected after analysis of additional photometry \citep{vsn-a-18923, vsn-a-18933} and the object was classified as a CV. \cite{2016PASJ...68...65K} analysed the superhumps of ASASSN-15na and derived a superhump period $P_{\rm sh}=93.5(2)\,{\rm min}$ which they used to determine the mass ratio $q=0.081(5)$.

The long-term light curve based on available photometry is shown in 
Figure~\ref{F:LC:LT_ALL}.
The superoutburst on $\mathrm{MJD}\;57\,223$ was not fully covered, the light curve shows only a fading tail which lasted for about 25 days. Another superoutburst was observed on $\mathrm{MJD}\;58\,660$ and lasted for at least 16 days.

The second superoutburst is covered by the TESS light curve from Sector 13. We inspected this portion of the light curve (Figure~\ref{TESS:LC}), to search for superhump variability. We note that the strongest peak in the Lomb-Scargle periodogram corresponds to a period $\mathrm{P}=223.598\pm0.517\;\mathrm{min}$ which is caused by contamination by a known nearby eclipsing binary CRTS J191904.8-494534. Therefore, we narrowed down the analysed period interval to periods between 80–180 minutes and identified a best-fit period of $\mathrm{P}=91.87\pm0.65\;\mathrm{min}$ (see Figure~\ref{F:tess_periodograms}), which is well above the significance threshold based on the FAP, and is consistent with the previously reported superhump period of this system.
The ASAS-SN light curve shows detections between the superoutbursts reaching $\sim15$ mags, which could correspond to normal outbursts in between the superoutbursts. Unfortunately, the TESS light curve does not cover these observation times. Considering the light curve behaviour and mass ratio we classify this target as a SU UMa star.

The spectrum of ASASSN-15na shows a double-peaked H$\alpha$ emission line confirming its classification as a CV. The separation of the two peaks is $29.6(6)\,\mathrm{\text{\AA}}$. Other features present in the spectrum are H$\beta$ emission line and a blend of Mg I emission lines, but they are much less prominent than the H$\alpha$ line. The presence of Balmer lines with moderate width also supports its classification as a SU UMa star. 
The flux calibrated spectrum shown in 
Figure~\ref{F:SPEC_ALL}
exhibits a blue continuum for which we derived a temperature $T_\mathrm{BB}=10\,304\pm297 \,\mathrm{K}$. The spectrum was de-reddened using the full Galactic reddening $E(B-V)=0.08$ taken from \cite{1998ApJ...500..525S}, which could lead to overestimation of the temperature. When assuming no reddening for the spectrum, the best fit with a black-body model gives $T_\mathrm{BB}=8903\pm220 \,\mathrm{K}$.

\subsubsection{ZTF18aaaasnn}

ZTF18aaaasnn was discovered by the ZTF survey and was classified as a CV candidate by \cite{2021AJ....162...94S}. 
We constructed a long-term light curve based on the available photometry which is presented in 
Figure~\ref{F:LC:LT_ALL}.
Its light curve shows numerous episodes of activity which can be classified as normal outbursts and superoutbursts. The superoutbursts covered by the light curve occurred on $\mathrm{MJD}\;58\,210$, $\mathrm{MJD}\;58\,483$, $\mathrm{MJD}\;58\,734$, $\mathrm{MJD}\;59\,324$, $\mathrm{MJD}\;59\,634$, and $\mathrm{MJD}\;60\,286$. Their recurrence time determined from these timings is about $300$ days.
It is possible that the light-curve shows also observations from other superoutbursts, but the scarce coverage does no allow to distinguish them from normal outbursts. The only superoutbursts which is covered almost in its entirety occurred on $\mathrm{MJD}\;58\,734$. Its amplitude was $\mathbf{4}$ magnitudes, it lasted for about $15$ days, and it shows a plateau phase which was followed by a steep decline.
The normal outbursts occur more frequently and they last typically only about 3 days.

ZTF18aaaasnn was observed by TESS in four different sectors (see Table~\ref{tab:tesstargetslist_v2} for details), one of which (Sector 73) covers the superoutburst which occurred on $\mathrm{MJD}\;60\,286$. Figure~\ref{F:tess_periodograms} shows the Lomb-Scargle periodogram of light curve from Sector 73. The most prominent peak we detected corresponds to a period $\mathrm{P}=91.21\pm0.20\,\mathrm{min}$ and its amplitude is above the $0.001\%$ FAP threshold. As the period was detected during a superoutburst, it is likely a superhump period, which agrees also with the shape of the phase-folded light curve shown in Figure~\ref{F:tess_periodograms}.

The Gemini spectrum shows strong hydrogen emission lines from the Balmer and Paschen series and also He I and Mg I emission lines. All emission lines appear single-peaked. 
The flux-calibrated spectrum is presented in 
Figure~\ref{F:SPEC_ALL},
the temperature derived from the black-body fit of its continuum is $T_\mathrm{BB}=6033\pm60 \,\mathrm{K}$. 
Even though the Gaia distance $d=2350_{-1483}^{+1593}$ has large uncertainties, it does not affect the estimation of the reddening, as \cite{2019ApJ...887...93G} predicts $E(g-r)=0.07$ for all distance $d>610\,\mathrm{pc}$.

The properties of superoutbursts and normal outbursts combined with the strong hydrogen emission lines can be used to classify this target as a CV of SU UMa sub-type.

\subsubsection{ZTF21abhrevv}

This target was classified as a CV candidate by \cite{2021AJ....162...94S} based on ZTF survey photometry. 
The long-term light curve which we constructed from available photometry is shown in 
Figure~\ref{F:LC:LT_ALL}.
It shows a superoutburst which started on $\mathrm{MJD}\;59\,378$, had amplitude of 7 magnitudes, and lasted for 22 days. It was then followed by a fading tail which showed one rebrightening on $\mathrm{MJD}\;59\,406$. The system returned to quiescence level on $\mathrm{MJD}\;59\,450$. As the light curve shows only one superoutburst, we can only estimate the lower limit on the superoutburst recurrence time from the duration of quiescence as at least 1200 days.

The Gemini spectrum shows strong emission lines of Balmer and Paschen series, all of which appear to be single-peaked. It also shows emission lines of He I and Mg I. Lines H$\alpha$ and H$\beta$ also show shallow broad absorption, which is more dominant in the case of H$\beta$. 
Figure~\ref{F:SPEC_ALL}
shows the flux-calibrated spectrum of ZTF21abhrevv, the black-body fit of its continuum gives a temperature $T_\mathrm{BB}=8391\pm171 \,\mathrm{K}$.

The properties of the observed superoutburst and the presence of hydrogen lines in the spectrum show that ZTF21abhrevv is most likely a CV of WZ Sge sub-type. The presence of the broad absorption in H$\alpha$ and H$\beta$ indicate that the accretion disc in this system is optically thick.

\subsection{Evolved cataclysmic variable candidates}
\subsubsection{ASASSN-18abl}

ASASSN-18abl was discovered during its outburst by \cite{2018TNSTR1914....1S} as part of the ASAS-SN survey.
\cite{vsn-a-22836} reported a superhump period $P_\mathrm{sh}=41.5\,\mathrm{min}$ suggesting this target could be an AM~CVn star. \cite{vsn-a-22850} reported that the superhump variation seems to be double-peaked with a period  $P_\mathrm{sh}=86\,\mathrm{min}$.
ASASSN-18abl was classified as a CV by \cite{2018TNSCR1926....1L} based on a spectrum obtained one day after the discovery. The spectrum was obtained at the Three Hills 
Observatory\footnote{\url{http://www.threehillsobservatory.co.uk}} and it shows blue continuum with weak absorption in H$\beta$ and H$\gamma$.

\begin{figure}
    \centering
    \includegraphics[width=0.99\linewidth]{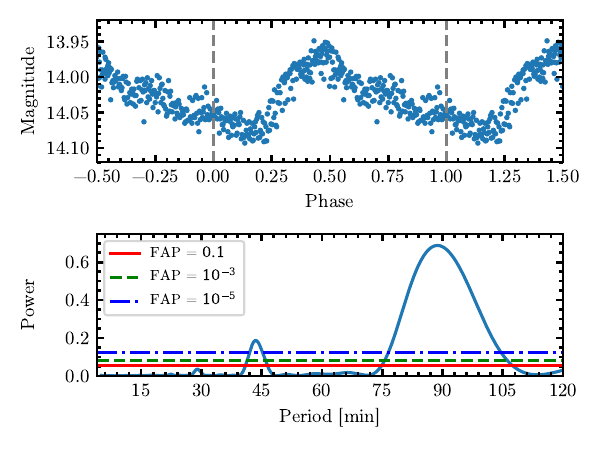}
    \caption{Phase-folded light curve and periodogram of AAVSO observations of ASASSN-18abl obtained between $\mathrm{MJD}\;58\,478.7$ and $\mathrm{MJD}\;58\,479.1$.}
    \label{F:LSP:18abl}
\end{figure}

The long-term light curve presented in 
Figure~\ref{F:LC:LT_ALL}
shows that only one outburst was detected in this system.
The outburst started on $\mathrm{MJD}\;58\,464$ and peaked on $\mathrm{MJD}\;58\,465$ with magnitude $g=12.0$. It lasted for 20 days and it is covered by photometry available in in the ASAS-SN, ZTF, and AAVSO databases. The light curve obtained during the fading tail shows single-peaked superhump variations. 
Figure~\ref{F:LSP:18abl} shows the periodogram of the variations observed between $\mathrm{MJD}\;58\,478.7$ and $\mathrm{MJD}\;58\,479.1$, which covers about four superhump periods. The periodogram shows two significant peaks, one corresponding to the superhump period ($P_\mathrm{sh}=89.5\pm9.3 \, \mathrm{min}$) and another corresponding to its half value ($P_\mathrm{sh}/2=43.6\pm2.0 \, \mathrm{min}$). As the second peak gives smaller uncertainty in the period determination, we adopted  $P_\mathrm{sh}=87.2\pm4.0 \, \mathrm{min}$ as the superhump period.
Given the fact that only one outburst was observed, we can estimate the lower limit of outburst occurrence to be about 2200 days.

The Gemini spectrum has a flat continuum with 
no emission lines and numerous absorption lines, the most prominent being Mg I, Na I, Ca II, and hydrogen lines.
The spectrum does not show any significant helium lines.
Flux-calibrated spectrum of ASASSN-18abl is shown in 
Figure~\ref{F:SPEC_ALL}.
It shows much redder continuum than the spectrum reported by \cite{2018TNSCR1926....1L} and fitting the continuum with a black-body model gives a temperature $T_\mathrm{BB}=5039\pm11 \,\mathrm{K}$. The relatively flat spectrum with prominent absorption lines shows similarities to the spectra of evolved cataclysmic variables identified by \cite{2021MNRAS.508.4106E}, which exhibit donors with higher temperatures and luminosities than typically CVs at corresponding orbital periods. However, the periods presented by these authors are longer than the one of ASASSN-18abl. The characteristics of the superoutburst, such as the duration, support this classification. 

\subsubsection{ASASSN-20la}

ASASSN-20la was discovered by ASAS-SN survey in August 2020 during outburst.  \cite{2021PASJ...73.1375K} classified the system as an AM~CVn star based on the properties of the outburst and the rebrightenings. 

We used ZTF r and i photometry to construct the long-term light curve presented in 
Figure~\ref{F:LC:LT_ALL}.
While ASAS-SN and ATLAS photometry is also available, it is reliable only during the superoutburst detection due to the low brightness of the target. The long-term light curve shows only one superoutburst which started on $\mathrm{MJD}\;59\,087$ and peaked on $\mathrm{MJD}\;59\,088$. It shows a plateau phase which ended on $\mathrm{MJD}\;59\,092$, the whole superoutburst lasted for about 7 days and was followed by multiple rebrightenings. The subsequent quiescence lasts for more than 1400 days, which can serve as the lower limit for the superoutburst recurrence time.

The Gemini spectrum shows broad absorption lines and no emission lines. The strongest absorption line corresponds to a blend of Mg I lines at $\lambda=5174\,\mathrm{\text{\AA}}$. H$\alpha$ and H$\beta$ absorption lines are also present in the spectra. The flux calibrated spectrum presented in 
Figure~\ref{F:SPEC_ALL}
shows a continuum whose flux increases towards shorter wavelengths, but the temperature determined by a black-body fitting $T_\mathrm{BB}=4999\pm34 \,\mathrm{K}$ which is much lower than those obtained for the AM~CVns and the CVs previously presented. The spectrum of ASASSN-20la was de-reddened using reddening $E(g-r)=0.06$ provided by \cite{2019ApJ...887...93G}. As the target's distance is not determined, we used a value which is predicted for distances $d > 480\,\mathrm{pc}$, which serves as the upper limit. However, the reddening has only a small effect on the temperature determination, as the value derived from the uncorrected spectrum is $T_\mathrm{BB}=4796\pm32 \,\mathrm{K}$.

The absorption lines, while shallower than the ones in ASASSN-18abl, especially the Na I, the cold temperature and presence of superoutburst suggest that this target could also be a CV with an evolved donor as those in \cite{2021MNRAS.508.4106E}.

\subsubsection{TCP J00505644+5351524}

This target was discovered by \cite{2021TNSTR2489....1K} in July 2021 during an outburst. \cite{vsn-a-26088} reported superhump observation with two possible periods detected ($39.6\,\mathrm{min}$ and $79.1\,\mathrm{min}$) and classified this target as either an AM~CVn or a WZ Sge star. 

\begin{figure*}
\section{Long-term light curves}
    \includegraphics[width=0.48\textwidth]{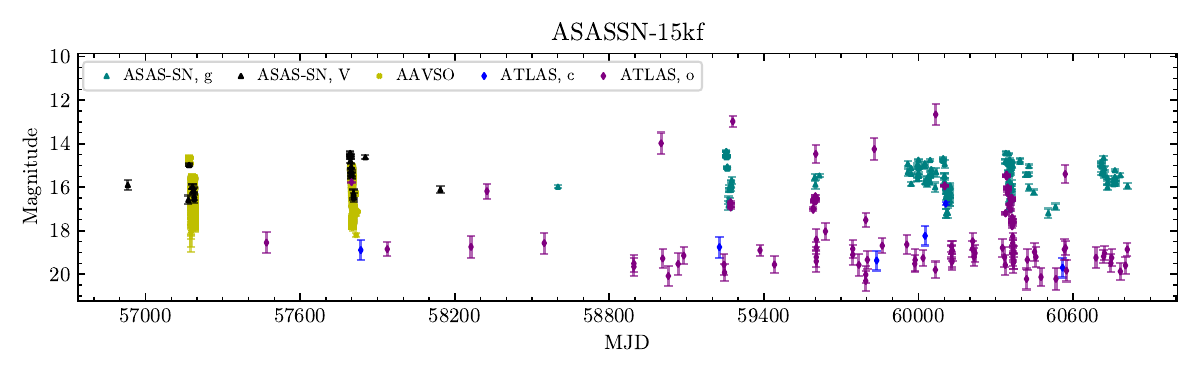}
    \includegraphics[width=0.48\textwidth]{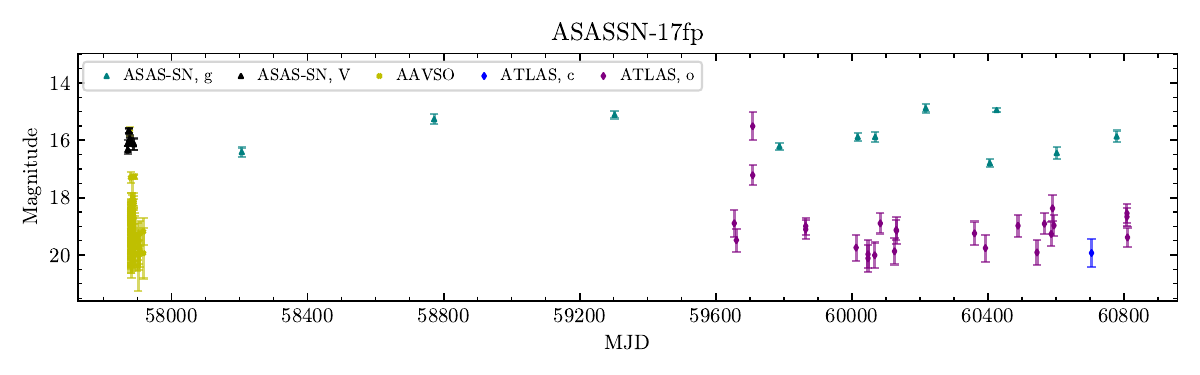}    
    
    \includegraphics[width=0.48\textwidth]{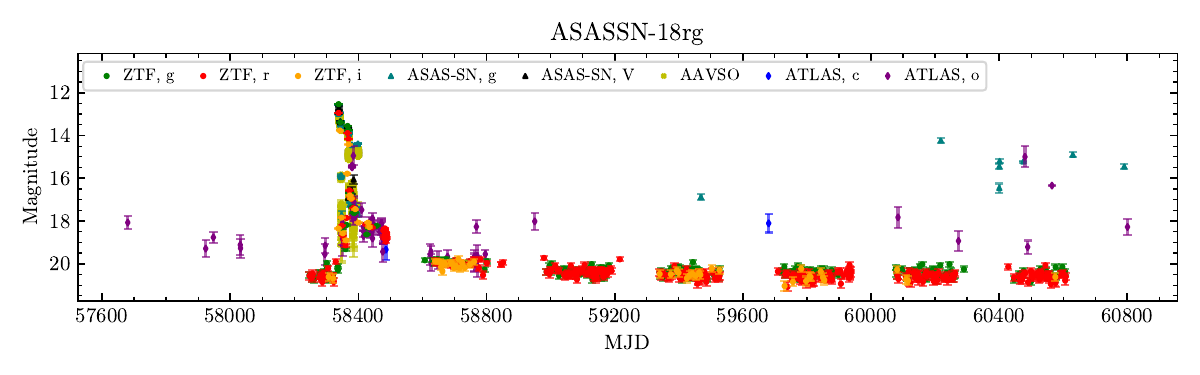}
    \includegraphics[width=0.48\textwidth]{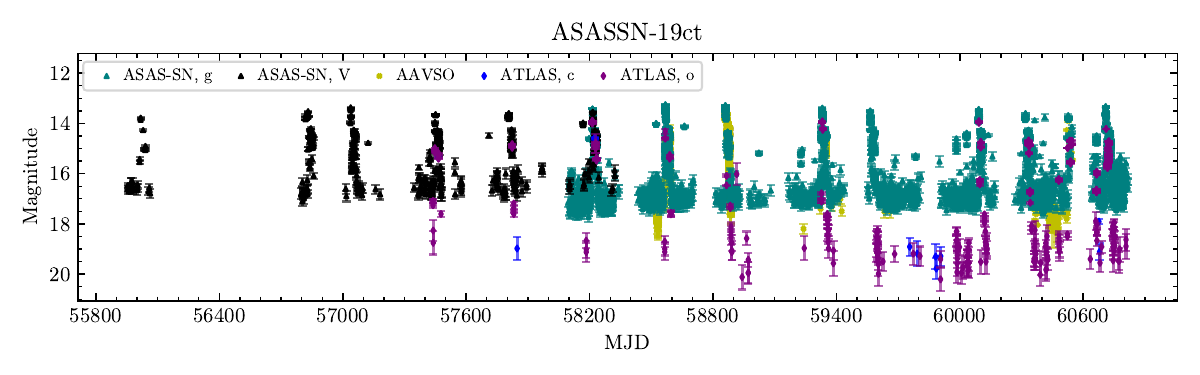}    
    
    \includegraphics[width=0.48\textwidth]{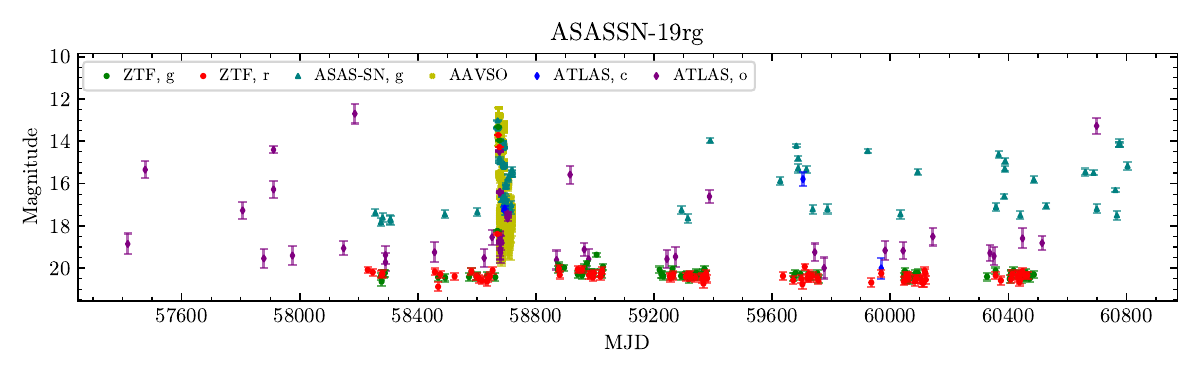}
    \includegraphics[width=0.48\textwidth]{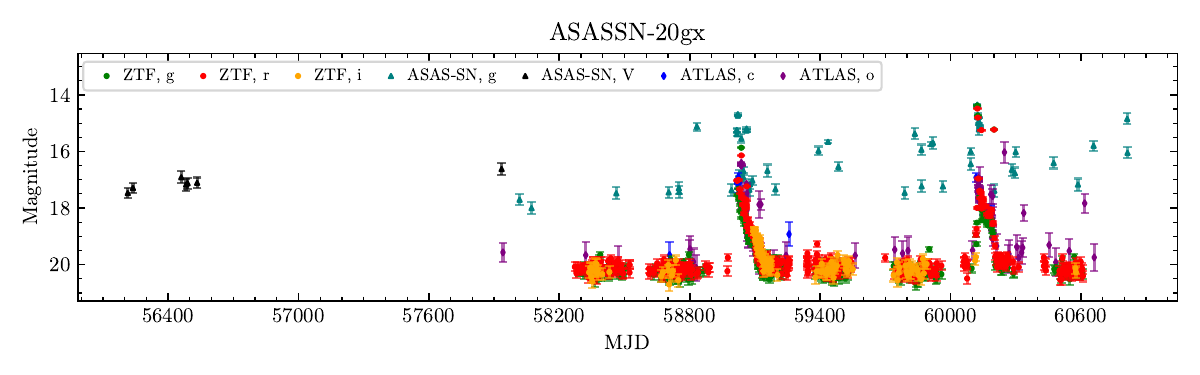}    
    
    \includegraphics[width=0.48\textwidth]{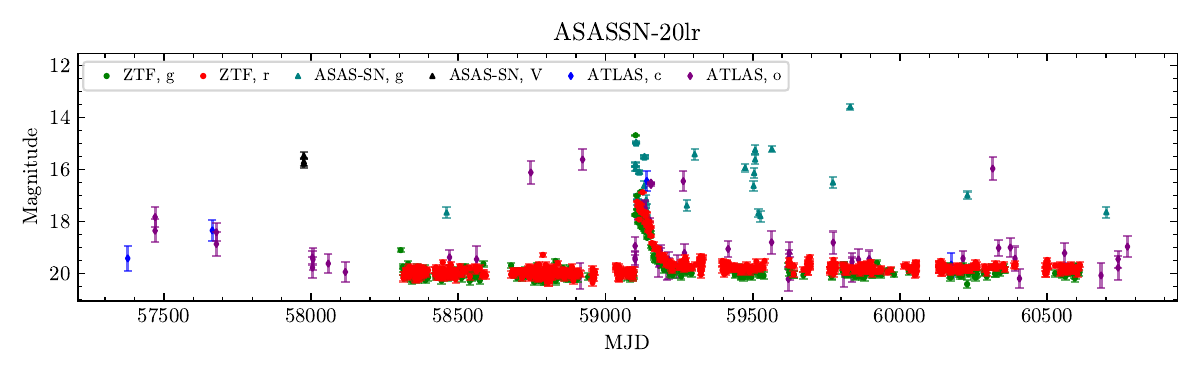}
    \includegraphics[width=0.48\textwidth]{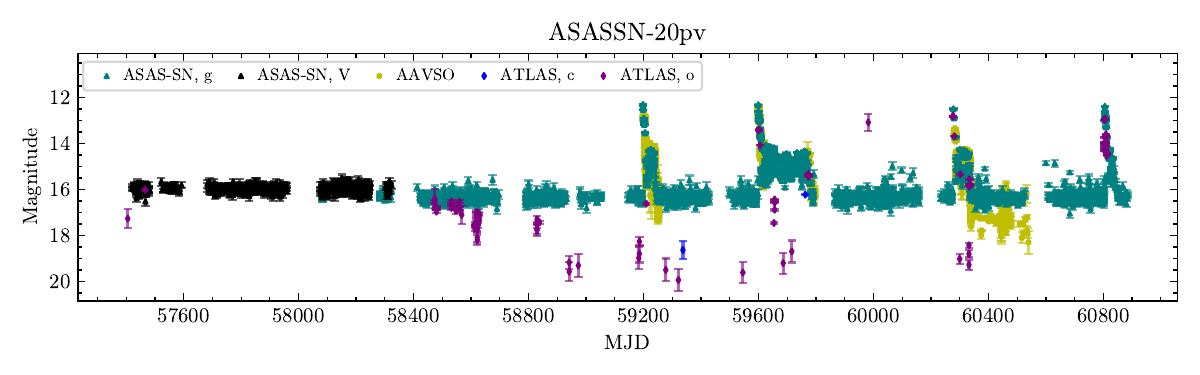}    
    
    \includegraphics[width=0.48\textwidth]{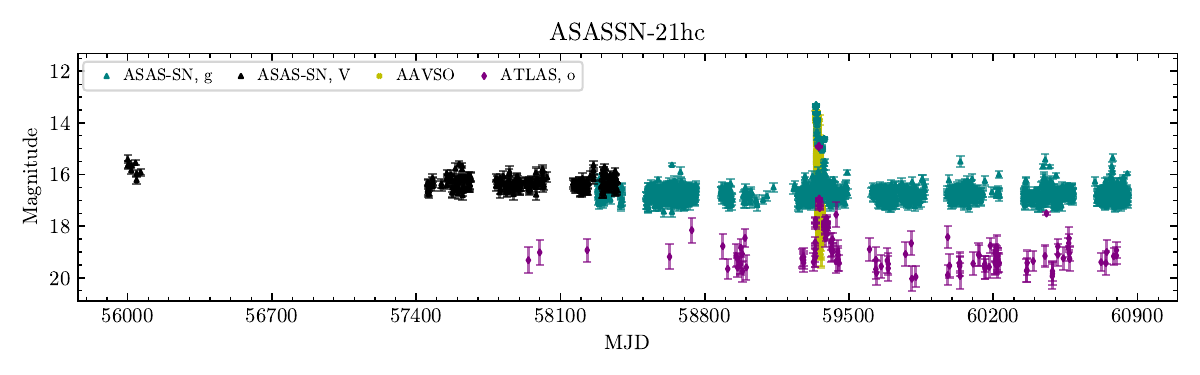}
    \includegraphics[width=0.48\textwidth]{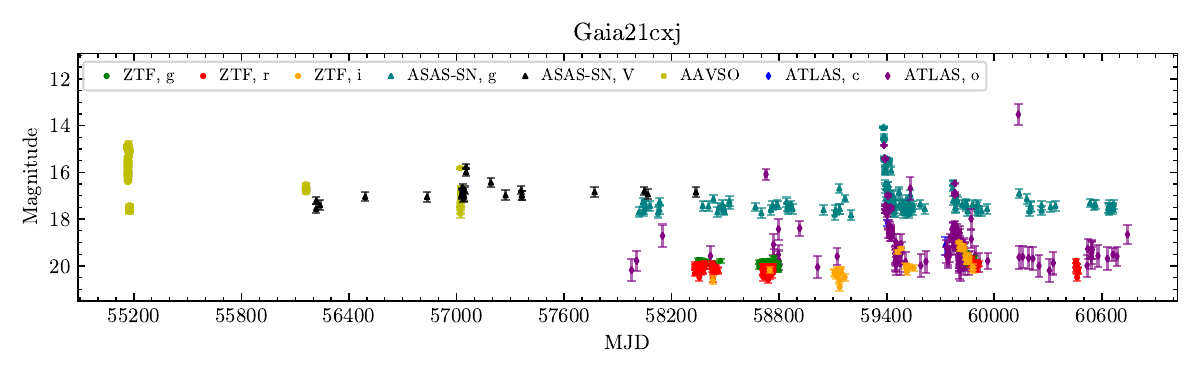}   
    
    \includegraphics[width=0.48\textwidth]{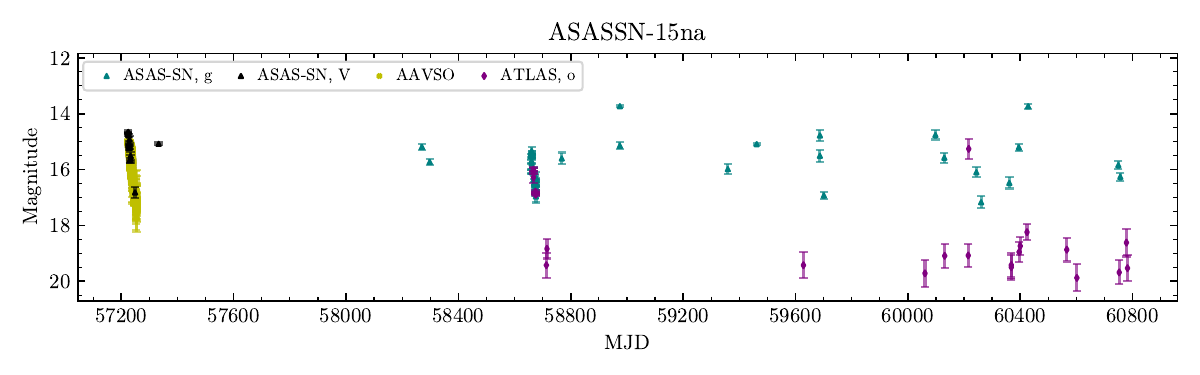}
    \includegraphics[width=0.48\textwidth]{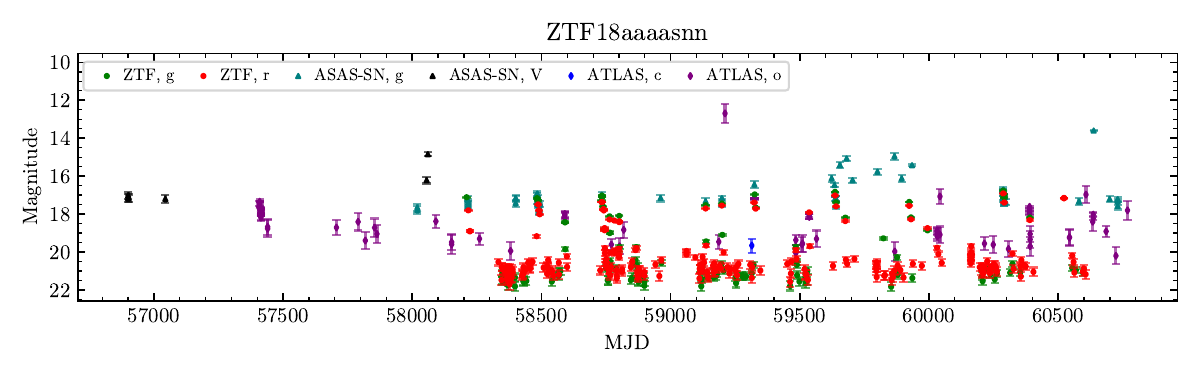}    
    
    \includegraphics[width=0.48\textwidth]{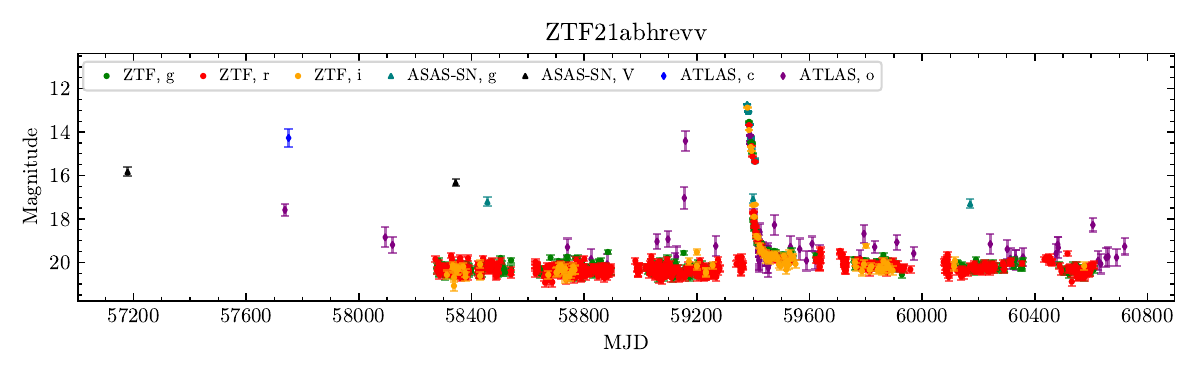}
    \includegraphics[width=0.48\textwidth]{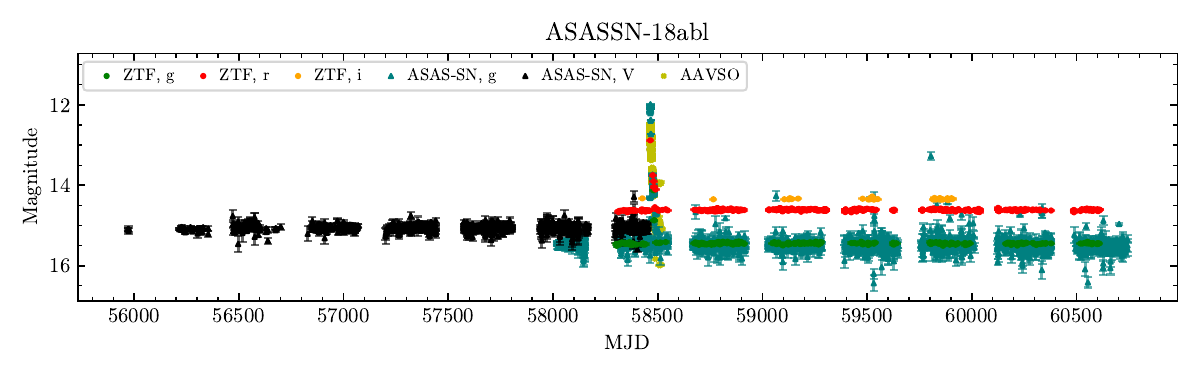}    
    
    \includegraphics[width=0.48\textwidth]{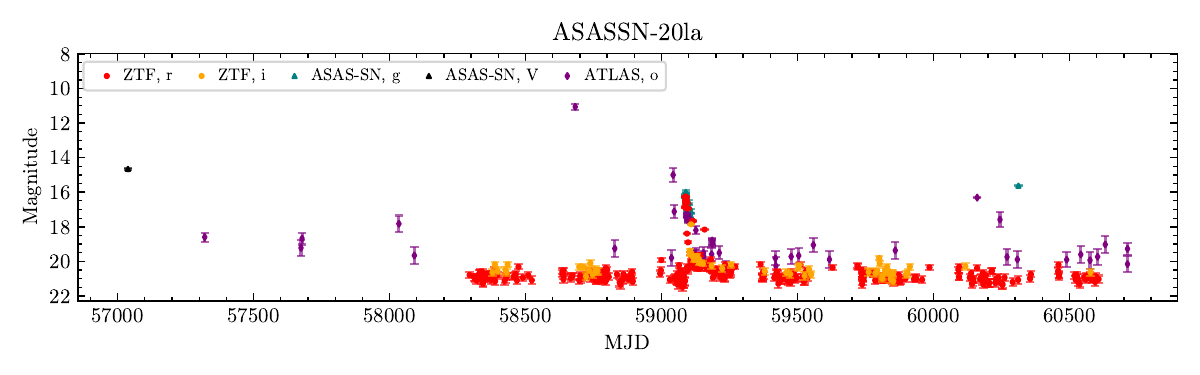}
    \includegraphics[width=0.48\textwidth]{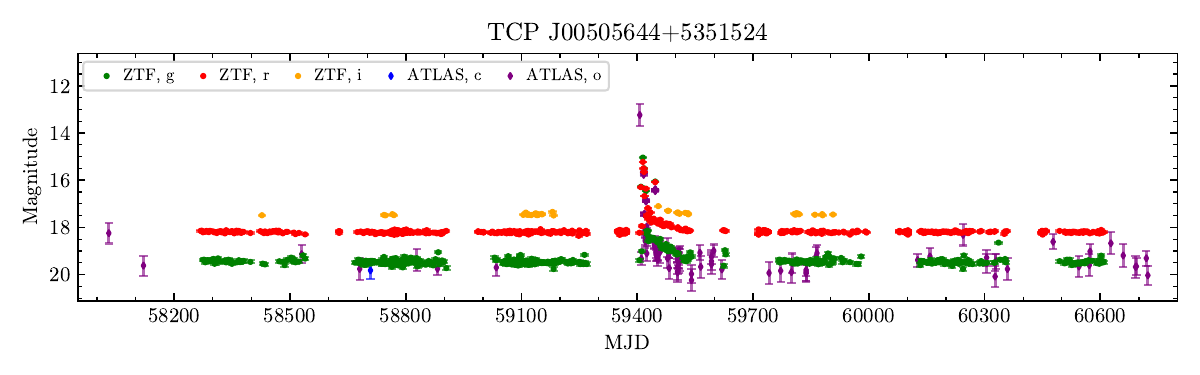}

    \caption{Long-term light curves based on ground based photometric observations.}
    \label{F:LC:LT_ALL}
\end{figure*}

We used the available photometry to construct a long-term light curve which is presented in 
Figure~\ref{F:LC:LT_ALL}.
The light curve shows that the first outburst was recorded on $\mathrm{MJD}\;59\,406.6$. This outburst lasted for $\sim 4$ days and was followed by multiple rebrightenings and a fading tail. The system reached its quiescence level on $\mathrm{MJD}\;59\,520$ at which it remains. We can only determine the lower limit on the recurrence time which is longer than about 1300 days.

\begin{figure*}
\section{Flux-calibrated spectra}
    \centering
    \includegraphics[width=0.47\linewidth]{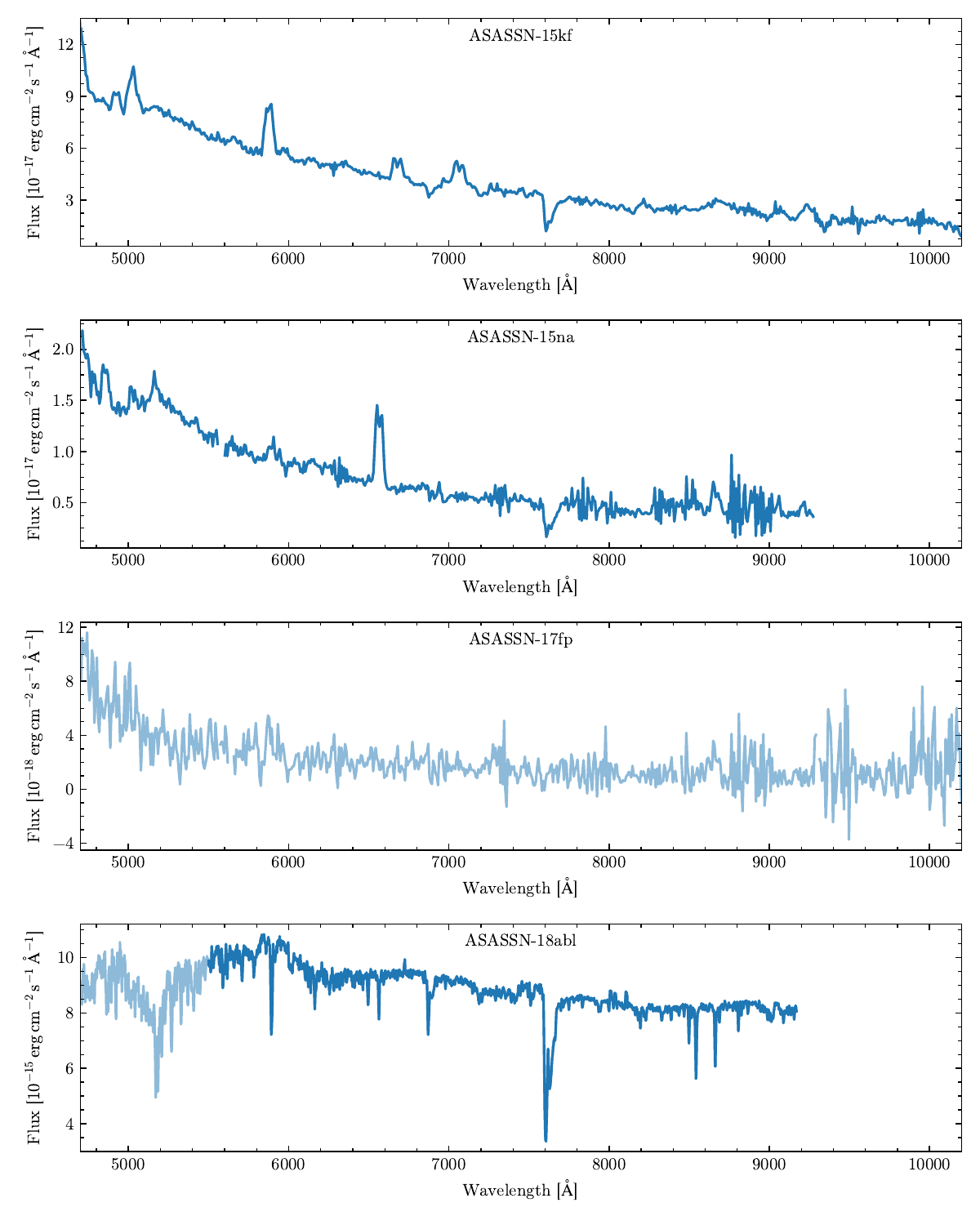}
    \includegraphics[width=0.47\linewidth]{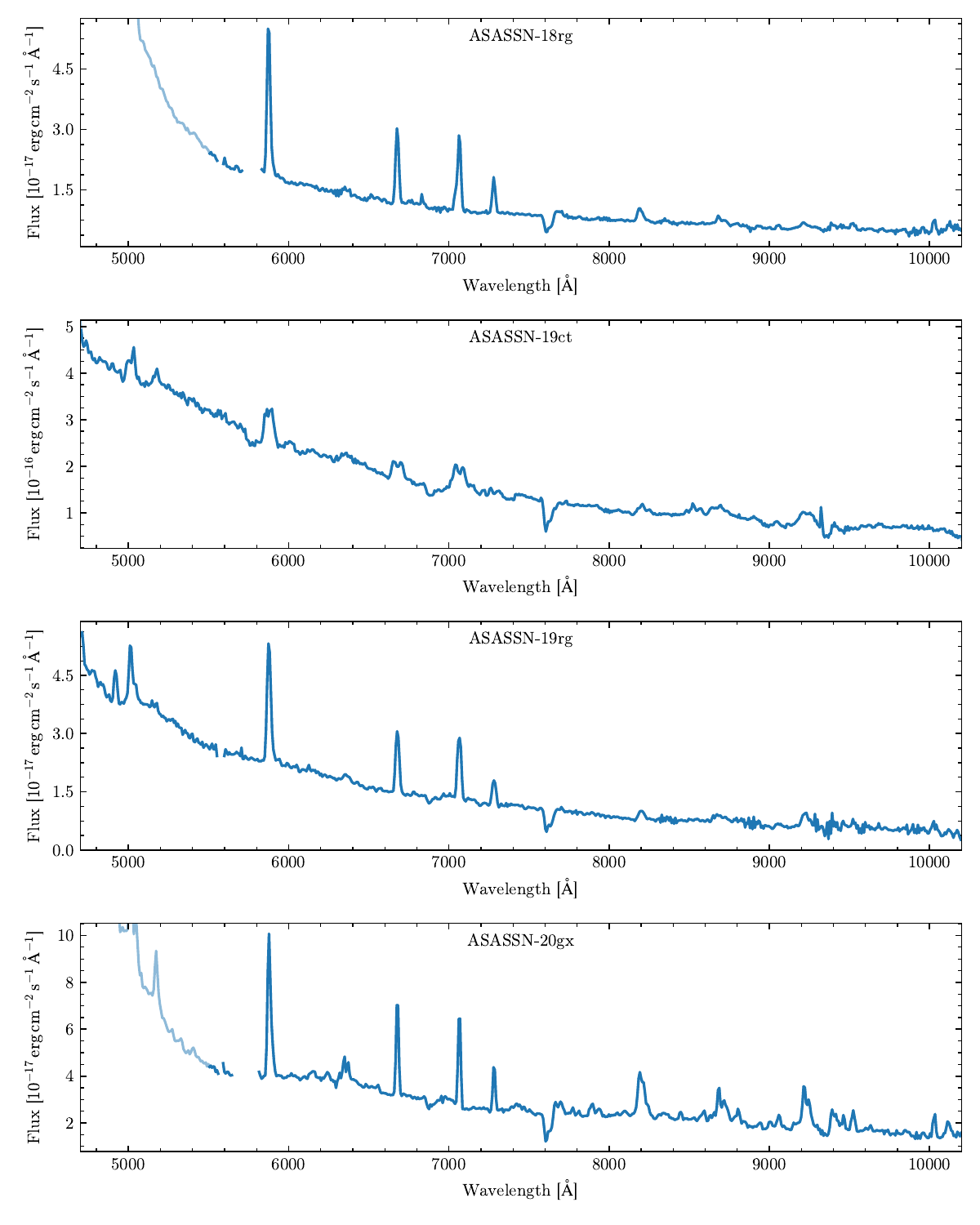}
    
    \includegraphics[width=0.47\linewidth]{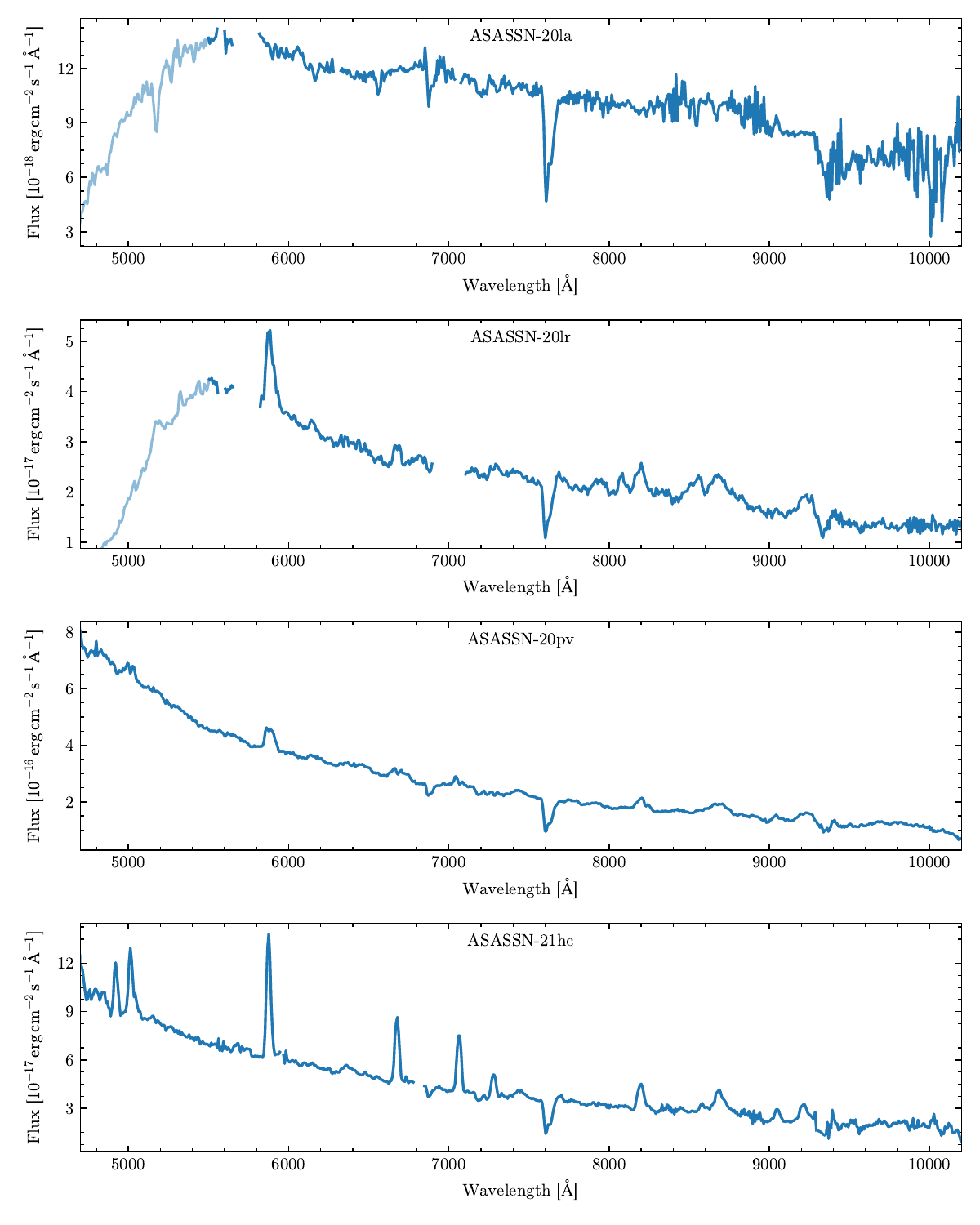}
    \includegraphics[width=0.47\linewidth]{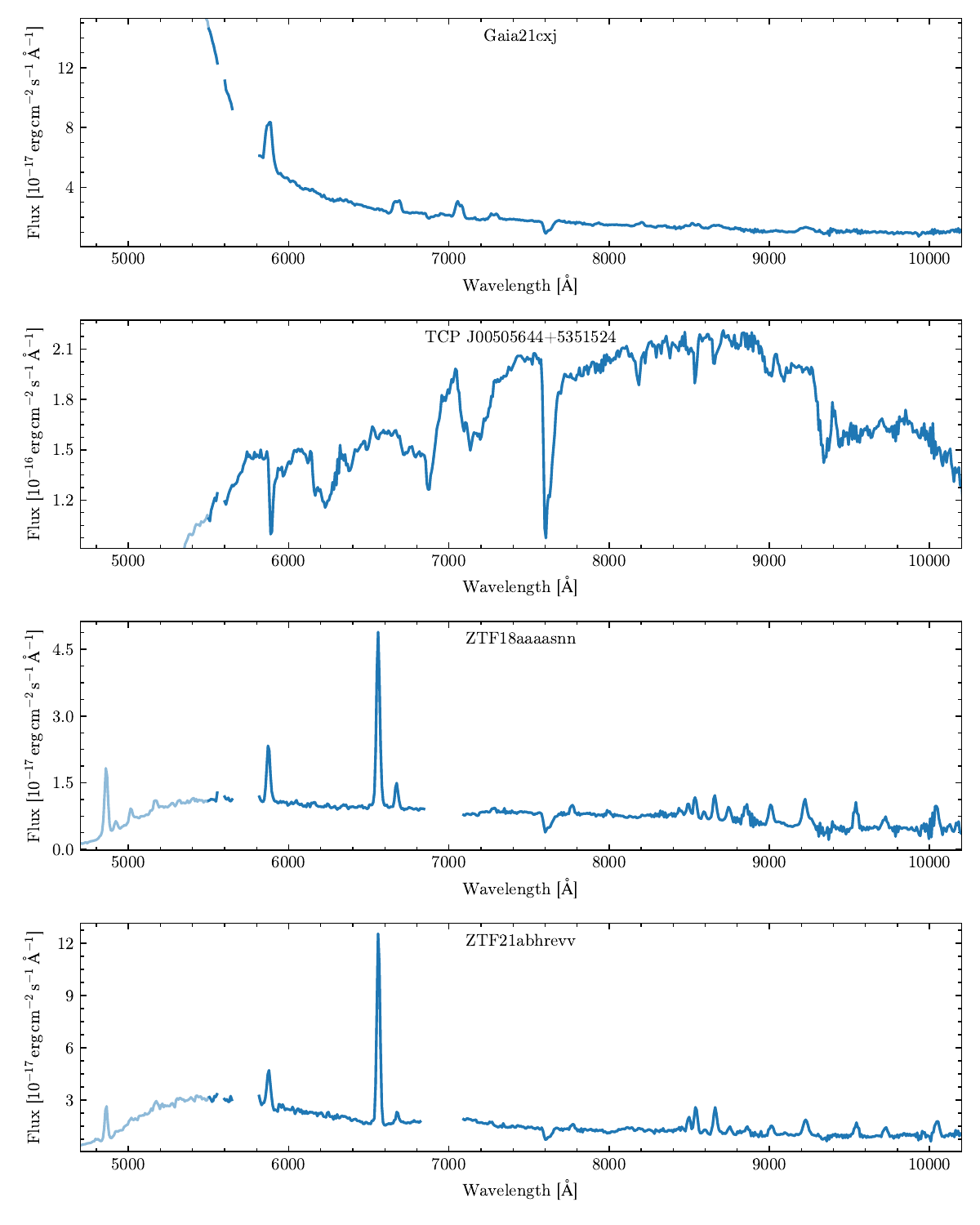}

    \caption{Flux-calibrated Gemini spectra. Regions with unreliable flux-calibration are presented in light colour.}
    \label{F:SPEC_ALL}
\end{figure*}

The Gemini spectrum exhibits broad absorption line bands typical for a late-type star. The spectrum shows H$\alpha$ absorption line and absorption lines of Na I and Mg I, but there is no evidence of spectral features which could be attributed to accretion or a presence of a white dwarf. 
Although it shows differences to the spectra of ASASSN-18abl and ASASSN-20la, it is possible that this system is also an evolved CV with contamination from nearby targets. 
The lack of X-ray emission agrees with this scenario. 
The flux-calibrated spectrum is shown in 
Figure~\ref{F:SPEC_ALL}
and it exhibits continuum with peak flux at wavelength $\lambda \sim 8800\,\text{\AA}$, the black-body fit gives a temperature $T_\mathrm{BB}=3434\pm21 \,\mathrm{K}$.

\end{document}